\documentclass[showpacs,superscriptaddress,nofootinbib]{revtex4}
\usepackage{epsfig}
\usepackage[english]{babel}
\usepackage{pgf}
\usepackage{amsfonts,amsmath,amssymb,enumerate,array,amssymb}
\usepackage{enumitem}
\usepackage{mathtools} 
\usepackage{multirow} 
\usepackage[latin1]{inputenc}
\usepackage{bm}
\usepackage{fixmath}
\usepackage{amsbsy}
\usepackage{graphicx}
\usepackage{capt-of}
\usepackage{dcolumn}
\usepackage{upgreek}
\usepackage{float}
\usepackage{subfig}
\usepackage[format=plain,justification=centerlast]{caption}
\usepackage{bbm}        

\usepackage{layouts}
\usepackage{natbib}
\usepackage{abrev-jour-long}

\usepackage{pdfpages}
\usepackage{color}
\usepackage{graphicx,epstopdf}
\usepackage[colorlinks=true,
 linkcolor=red,
 urlcolor=purple,
 citecolor=blue,hyperindex]{hyperref}

\newtheorem{theorem}{Theorem}

\begin{document}

	\title{(5+1)-Dimensional Analytical Brane-World Models: Intersecting Thick Branes}

	\author{Henrique Matheus Gauy}\email[]{henmgauy@df.ufscar.br}

	\author{Alex E. Bernardini}\email[]{alexeb@ufscar.br}

	\affiliation{Departamento de F\'isica, Universidade
		Federal de S\~{a}o Carlos, S\~ao Carlos, 13565-905 SP, Brazil}


	\begin{abstract}

		Two co-dimensional thick brane-worlds are investigated in quite general terms for two intersecting scalar fields generating the extra dimension defect. In general, when one considers two co-dimensional thick brane-worlds, the warp factor is constructed as a string-like defect. Considering a twofold-warp factor constructed from two intersecting warp factors, an alternative bulk configuration is examined. With the brane localization thus driven by two crossing scalar fields, the possible solvable models obtained from such a two co-dimensional setup are systematically discussed.
		The obtained solutions are classified as five different models organized into two subsets for which some of their physical properties are evaluated. For models $I$ and $II$, in the first subset, Einstein equation solutions are rigidly defined, up to some arbitrary constant. For models $III$, $IV$ and $V$, in the second subset, an additional degree of freedom not constrained by Einstein equations is admitted.
		The solutions are all obtained from a departure statement of assuming a conformally flat metric for the internal space, which is concomitant to the proper choice of coordinates. Eventual singularities in the curvature are identified, however, without affecting the physical appeal of the solutions described in terms of the stress energy tensor patterns, which are shown to be free of singularities for model $IV$, besides admitting straightforward reductions to $(4+1)$-dimensions.
		In particular, from the framework of models $III$ and $IV$, one is able to achieve brane-world solutions over two different geometries of $\mathbb{S}^{2}$ which, as demonstrated, can be reduced to trivial and non-trivial extensions of the well-known $(4+1)$-dimensional brane-worlds. Therefore, our $(5+1)$-dimensional results point to a consistent and expanded generalization of $(4+1)$-dimensional brane-world which naturally admits the possibility of an enlarged and maybe more accurate phenomenology. 
	\end{abstract}

	\date{\today}
	
	\maketitle

	\section{Introduction}

	In the last decades, inspired by the modern attempts of unifying all interactions, the idea of extra dimensions has been currently scrutinized. Admitting the possibility of extra dimensions playing some role in physics requires a deeper understanding of the evinced observation of three space dimensions. A naive explanation for the $(3+1)$-dimensional universe is based on the idea that extra dimensions can be compactified as a circle with a tiny radius of the order of the Planck length ($\approx10^{-33}$ cm). In this scenario, all the effects due to additional dimensions would be hidden to experimental measurements. In such a context, Arkani-Hamed--Dimopoulos--Dvali (ADD) \cite{ArkaniHamed1998a} and Randall--Sundrum (RS) \cite{RS-1,RS-2} seminal papers diffused the possibility of implementing large extra dimensions into realistic phenomenological contexts. In particular, as suggested by the ADD model \cite{ArkaniHamed1998a}, one of the most inspiring motivations for pursuing large extra dimensions in physics is the possibility of resolving the hierarchy problem \cite{ArkaniHamed1998a} in quantum field theories. While the ADD model was performed on a flat space, Randall-Sundrum (RS) models \cite{RS-1,RS-2} assume that the brane should gravitate, being spatially localized by an extra dimension warping effect so as to explain the field hierarchy. As a consequence, the RS model warped metric would admit an alternative to the ADD compactification \cite{RS-2}.
Once capturing the correct gravitational interaction in $(3+1)$-dimensions, such compactified models supported the construction of higher dimensional theories with infinitely extended extra dimensions, without affecting the success of Newton's theory of gravity. 
	
	Higher dimensional theories by themselves have also supported brane-world scenarios driven by topological defects \cite{Rubakov1983} where the fields of the Standard Model are hypothetically confined by brane-like regions of space \cite{Rubakov1983,Sundrum1999,ArkaniHamed1998a,Antoniadis1998,Kakushadze1999,Nussinov1999,Shiu1998,DONINI199959,ArkaniHamed1999,Cullen1999,Lukas1999}. The novel paradigm thus led to several spin-off models, including the now so-called thick brane-worlds, where the thin brane is replaced by a topological defect, an equivalent structure to those ones introduced for describing domain walls \cite{Rubakov1983}. The thick brane framework has thus been considered as an engendering tool for obtaining the configuration to the Randall-Sundrum model, by admitting some lump-like (non-topological) defect solution for the warp factor. In this case, the gravitational and matter fields should be localized in the brane, which may host some internal structure \cite{Bernardini2016,Almeida2009,Bazeia2004,Dzhunushaliev2010,Bazeia2002,DeWolfe2000,Ahmed2013,Gremm2000a,Chinaglia2016}.

	Besides working as a platform for the resolution of the hierarchy problem, thick brane-worlds in $(4+1)$-dimensions \cite{Bernardini2016,Almeida2009,Bazeia2004,Dzhunushaliev2010,Bazeia2002,DeWolfe2000,Ahmed2013,Gremm2000a,Kehagias2001,Kobayashi2002,Bronnikov2003,BAZEIA2009b,BarbosaCendejas2013,ZHANG2008,Melfo2003,Bazeia2009a,Koley2005,Bazeia2014,Chinaglia2016,Bernardini2013} have include the dark scalar field dynamics into their formulation. Notwithstanding the ferment in this field, theoretical and phenomenological connections with cosmology and astroparticle physics have also been evaluated \cite{Hall1999,Ahmed2014,Csaki2000a,Rubakov1983a,Binetruy2000,Binetruy2000a,Cline1999,Csaki1999,Csaki2000b,Flanagan2000,Kanti2000,Kanti1999,Bazeia2008,George2009,Kadosh2012,Kim2000,Binetruy2000b,Bowcock2000,Guha2018,Mukohyama2000,Chung1999,Ida2000,Chung2000,Mukohyama2000a,MERSINI2001,Bernardini2014,Casadio2014}. More recently, in the strict theoretical front, the possibility of humanly traversable wormholes in RS models \cite{Maldacena2021} has also been suggested.

	With the ultimate proposal of enlarging the phenomenology alternatives for thick brane scenarios, the present work intents to investigate six dimensional brane-worlds, an idea already explored through different facets which include, for instance, thin branes and string-like defects \cite{Csaki2000,ArkaniHamed2000,PARAMESWARAN200754,Liu2007,Dzhunushaliev2008,Koley2006,Singleton2004,Gherghetta2000,Park2003,Dzhunushaliev2009,Multamaeki2002,Gregory2000,Cohen1999}. However, instead of considering string-like defects as they are typically engendered from thick brane 1-dim warping mechanisms, the possibility of intersecting branes is here admitted. By considering a twofold-warp factor which is separable into two intersecting warp factors respectively driven by two intersecting scalar fields, several novel solutions for thick brane-world models are obtained. In fact, with respect to the featured compact internal structure, some of the resulting constructions here admitted shall contrast with the RS-2 paradigm. Otherwise, the involved scalar fields and self-gravity mechanisms shall consistently resemble the well-succeeded thick brane-world models in admitting lump-like defect solutions for the warp factors. Considering the eventual complexity of some $(5+1)$-dimensional metrics, our paper is constrained to finding and classifying classical solutions that define the corresponding brane-world scenarios so as to prepare the clean framework for describing the localization of gravitational and matter fields in next issues.

	Working with $(5+1)$-dimensions indeed is not only constrained to model building, but also to localizing spin $1/2$ particles without introducing additional fields, i.e. through the same warp factor features that results into the gravity localization \cite{PARAMESWARAN200754,Liu2007,Singleton2004,Gherghetta2000}. In some sense, this is not admitted in $(4+1)$-dimensions where some additional generating mass field mechanisms are required to achieve localization \cite{Bernardini2016,Almeida2009,Pal2008,ZHANG2008,Kehagias2001,Mendes2018,Liu2008,Liu2009,Liu2011}. Also, by placing brane-worlds over some novel topological spaces append the possibility of some new physics. While five dimensional setups have only two manifolds, $\mathbb{R}$ and $\mathbb{S}^{1}$, for the topology of the internal 1-dimensional space, six dimensional brane-worlds may exhibit a wide range of topologies from the 2-dimensional space. Due to the compact features of $\mathbb{S}^{2}$, our straightforward proposal lies in constructing the total space from a priori internal space $\mathbb{S}^{2}$, where, in particular, it is not regarded as the sphere, but as a set with a space topology homeomorphic to the sphere. Considering a departing topological manifold with a defined metric, and that $\mathbb{S}^{2}$, as a coupled structure, intrinsically carries several different metrics, models over two distinct geometries of $\mathbb{S}^{2}$, the sphere and the spheroid, can be solved and evaluated.

	A more specialized summary of the above procedure is provided by a departure metric given by $\boldsymbol{\sigma}=e^{-2f}\mathrm{d}u\otimes\mathrm{d}u+e^{-2h}\mathrm{d}v\otimes\mathrm{d}v$, which is nothing but the conformally flat metric \cite{Ray1992} $\boldsymbol{\sigma}=e^{-2B}\left(\mathrm{d}u\otimes\mathrm{d}u+\mathrm{d}v\otimes\mathrm{d}v\right)$ for any (pseudo-) Riemannian space of two dimensions, $(\mathbb{B}^{2},\boldsymbol{\sigma})$, although it is written in terms of different coordinates. Despite dealing with brane-world models with two co-dimensions, the choice of the coordinates implying into the conformally flat form of the metric is too restrictive. To find Einstein equation solutions for the warp factor, $A$, where the conformally flat approach would be intractable. When considering intersecting branes one assumes a twofold-warp factor $A$, $A=\tilde{A}+\hat{A}$, where $\tilde{A}$ and $\hat{A}$ depend on two different variables (i.e. $\tilde{A}(u)$ and $\hat{A}(v)$) with independent warping characteristics. The setup for the corresponding brane intersection is provided by two scalar fields, $\phi$ and $\zeta$, which also depend on the same two different variables, evidently with $\phi_{,v}=\zeta_{,u}=0$.

	Therefore, the scalar fields, $\phi$ and $\zeta$, shall drive the behavior of the warp factors, $\tilde{A}$ and $\hat{A}$, respectively. Such assumptions followed by the simplifying hypothesis of assuming the metric components associated with the co-dimensions to be separable, i.e. $f=\tilde{f}(u)+\hat{f}(v)$ and $h=\tilde{h}(u)+\hat{h}(v)$, shall result into two subset of sorted analytical solutions composing five different models: from $I$ to $V$, for which the physical properties and the reducibility to $(4+1)$ scenarios shall be evaluated.

	The paper is thus organized as follows. Sec.~\ref{preliminaries} presents the elementary introduction to the $(5+1)$-dimensional setup driven by two scalar fields and sets the equations to be solved. Sec.~\ref{intersecting} is devoted to the particular case of intersecting thick branes where the main assumptions of the proposed modeling is discussed.
The solutions for the so-called models from I to V is obtained.	Considering that only for models $I$ and $II$, in the above-mentioned first subset, Einstein equation solutions are rigidly defined, and that, for models $III$, $IV$ and $V$, in the above-mentioned second subset, an additional degree of freedom related to the coupled fields is not constrained by Einstein equations is admitted, and fixing the geometry of the internal space is mandatory for definitely determining all the fields.
Such aspects and their complete understanding are thus evaluated in sec.~\ref{predetermined}.
Our conclusions are drawn in sec.~\ref{Concl} so as to point to the possibility of an enlarged and maybe more accurate phenomenology.

	\section{(5+1)-Dimensional Brane-World Preliminaries}\label{preliminaries}
	
	The space-time is postulated to be a six dimensional manifold $\mathbb{E}^{6}$ that is, as a set, equivalent to the product space $\mathbb{M}^{4}\times \mathbb{B}^{2}$, where $\mathbb{M}^{4}$ is some four dimensional pseudo-Riemannian manifold and $\mathbb{B}^{2}$ is some two dimensional Riemannian manifold. The geometry of $\mathbb{E}^{6}$ is represented by the metric,
	\begin{equation}
	\boldsymbol{g}=e^{-2A}\omega_{\mu\nu}\mathrm{d}x^{\nu}\otimes\mathrm{d}x^{\mu}+ \sigma_{ij}\mathrm{d}x^{i}\otimes\mathrm{d}x^{j},\label{bundlemetric}
	\end{equation}
	where $A$ is the warp factor, $\boldsymbol{\omega}$ is the metric of the space-time $\mathbb{M}^{4}$ ($\boldsymbol{\omega}:\mathbb{M}^{4}\rightarrow\mathcal{T}^{\left(0,2\right)}\mathbb{M}^{4}$) and $\boldsymbol{\sigma}$ is the metric of the internal space $\mathbb{B}^{2}$ ($\boldsymbol{\sigma}:\mathbb{B}^{2}\rightarrow\mathcal{T}^{\left(0,2\right)}\mathbb{B}^{2}$). Here $A:\mathbb{B}^{2}\rightarrow \mathbb{R}$, which means that $A=A(u,v)$, with $u=x^{4}$ and $v=x^{5}$; $\omega_{\mu\nu}:\mathbb{M}^{4}\rightarrow \mathbb{R}$; and $\sigma_{ij}:\mathbb{B}^{2}\rightarrow \mathbb{R}$.
Clarifying the notation, Greek indices ($\mu$, $\nu$,...) are valued in the set $\{0,1,2,3\}$, uppercase Latin indices ($M$, $N$,...) are valued in $\{0,1,2,3,4,5\}$, lowercase Latin indices ($m$, $n$, $i$, $j$,...) are valued in $\{4,5\}$ (and represent the bulk co-dimensions) and the labels $x^{4}=u$ and $x^{5}=v$, represent the choice of coordinates for the co-dimensions $(\mathbb{B}^{2})$; the use of notation $T_{45}\equiv T_{uv}$ whenever suited, indicates that $"4"="u"$ and $"5"="v"$; derivatives, whenever suited, will be represented by a {\em comma}, i.e. $f_{,\mu}:=\partial f/\partial x^{\mu}$; finally, tensors when being referred to its (abstract) entirety will be in boldface, as $\boldsymbol{g}$, but its components will be cast in regular font, as $g_{\mu\nu}$.

	Let now one supposes that matter in this space are of scalar nature and it corresponds to two canonical real scalar fields minimally coupled to gravity. The action for gravity is the usual Einstein-Hilbert action in six dimensions so as to have
	$$
	S=S_{g}+S_{\phi},
	$$
	$$
	S_{g}=\int\mathrm{d}^{6}x\sqrt{-\mathrm{g}}\, 2{M}^{4}\, R,
	$$
	$$
	S_{\phi}=-\int\mathrm{d}^{6}x\sqrt{-\mathrm{g}} \left(\frac{g^{MN}}{2}\phi_{,M}\phi_{,N}+\frac{g^{MN}}{2}\zeta_{,M}\zeta_{,N}+\mathcal{V}\right),
	$$
	where $\phi:\mathbb{B}^{2}\rightarrow \mathbb{R}$ ($\phi{\equiv}\phi(u,v)$), $\zeta:\mathbb{B}^{2}\rightarrow \mathbb{R}$ ($\zeta{\equiv}\zeta(u,v)$), $\mathcal{V}$ is some function of $\phi$ and $\zeta$, and $\mathrm{g}=\det{\left(g_{MN}\right)}$. The equations of motion for the related fields ($\phi$, $\zeta$ and the metric $\boldsymbol{g}$) are obtained as
	\begin{align}
	\square\phi=\frac{1}{\sqrt{-\mathrm{g}}}\left[\sqrt{-\mathrm{g}}\,g^{MN}\phi_{,N}\right]_{,M}&=\mathcal{V}_{,\phi},\label{scalarfield1}
\\
	\square\zeta=\frac{1}{\sqrt{-\mathrm{g}}}\left[\sqrt{-\mathrm{g}}\,g^{MN}\zeta_{,N}\right]_{,M}&=\mathcal{V}_{,\zeta},\label{scalarfield2}
\\
	R_{MN}-\frac{1}{2}g_{MN}R&=\frac{T_{MN}}{4M^{4}},\label{einsteinfield}
	\end{align}
	where the stress energy tensor ($T_{MN}$) is defined as
	\begin{equation}
	T_{MN}:=\phi_{,M}\phi_{,N}+\zeta_{,M}\zeta_{,N}-g_{MN}\left(\frac{\phi^{,K}\phi_{,K}}{2}+\frac{\zeta^{,K}\zeta_{,K}}{2}+\mathcal{V}\right).\label{stresstensordefinition}
	\end{equation}
	The conservation of the stress energy tensor implies that
	$$
	\left(\nabla_{M}T\right)^{MN}=0\iff\left(\nabla_{M}G\right)^{MN}=0,
	$$
	which, for the two scalar fields, results into
	$$
	\left(\nabla_{M}T\right)^{MN}=\phi^{,N}\left(\square\phi-\mathcal{V}_{,\phi}\right)+\zeta^{,N}\left(\square\zeta-\mathcal{V}_{,\zeta}\right)=0.
	$$
	As previously mentioned, the accomplishment of the intersecting thick brane scenario admits scalar fields, $\phi=\phi(u)$ and $\zeta=\zeta(v)$, regarded as independent quantities one from each other, with $\phi_{,v}=\zeta_{,u}=0$. From such an assumption one has
	\begin{equation}
	\left(\nabla_{M}G\right)^{MN}=0\iff\begin{cases}
	\square\phi-\mathcal{V}_{,\phi}=0,
	\\
	\square\zeta-\mathcal{V}_{,\zeta}=0.
	\end{cases}\label{redundant}
	\end{equation}
	This means that any solution of Eq.~\eqref{einsteinfield} also satisfies Eqs.~\eqref{scalarfield1} and \eqref{scalarfield2} for the scalar fields. Nevertheless, this is only true for $\phi_{,v}=\zeta_{,u}=0$. Therefore, in this case, the scalar field Eqs.~\eqref{scalarfield1} and \eqref{scalarfield2} can be regarded as completely redundant\footnote{The analytical solutions must not only define the metric, but also the scalar fields as functions of $u$ and $v$, which does not necessarily implies into identifying $\mathcal{V}$ explicitly in terms of the scalar fields, i.e. $\mathcal{V}=\mathcal{V}(\phi,\,\zeta)$.}.
	
	More generically, to realize the field equations one first writes down the components of the Einstein tensor through a straightforward -- even if long and tedious -- process. To simplify the following steps, a rescaling of the metric given by $\boldsymbol{g}=e^{-2A}\boldsymbol{\hat{g}}$ can be used to remove the conformal factor, where one defines
	$$
	\boldsymbol{\hat{g}}=\omega_{\mu\nu}\mathrm{d}x^{\mu}\otimes\mathrm{d}x^{\nu}+\hat{\sigma}_{ij}\mathrm{d}x^{i}\otimes\mathrm{d}x^{j},
	$$
	and $\boldsymbol{\hat{\sigma}}=e^{2A}\boldsymbol{\sigma}$. Notice that the metric $\boldsymbol{\hat{g}}$ is factorable since $\omega_{\mu\nu}:\mathbb{M}^{4}\rightarrow \mathbb{R}$ and $\hat{\sigma}_{ij}:\mathbb{B}^{2}\rightarrow \mathbb{R}$, and the calculations that follow can be easily extended to any dimension.
	
	One can now write the relation between the operators compatible with $\boldsymbol{g}$ and $\boldsymbol{\hat{g}}$: the connection, the Riemann and the Einstein tensors compatible with $\boldsymbol{g}$, calling it $\nabla$, $R^{M}{}_{NPQ}$ and $G_{MN}$, respectively, and those ones compatible with $\boldsymbol{\hat{g}}$, calling it $\hat{\nabla}$, $\hat{R}^{M}{}_{NPQ}$ and $\hat{G}_{MN}$, respectively. The Einstein equations are thus re-defined in terms of the metric $\boldsymbol{\hat{g}}$ rather than $\boldsymbol{g}$. Meanwhile, the Einstein tensor of the metric $\boldsymbol{g}$ can be recast in terms of the metric $\boldsymbol{\hat{g}}$ and $A$, in the form of
	\begin{equation}
	{G}_{MN}=\hat{G}_{MN}+ 4\,\hat{\nabla}_{M}\hat{\nabla}_{N}A+4\,\hat{\nabla}_{M}A\hat{\nabla}_{N}A-4\,\hat{g}_{MN}\hat{\square}A+6\,\hat{g}_{MN}\hat{\nabla}^{P}A\hat{\nabla}_{P}A.\label{GMN}
	\end{equation}
	To compute $\hat{G}_{MN}$ in order to obtain the equations of motion, one firstly notices that the Riemann tensor $\mathbf{\hat{R}}$ is factorable,
	\begin{align*}
	\mathbf{\hat{R}}&=\hat{R}^{\rho}_{\;\;\alpha\mu\nu}(x^{\kappa})\;\frac{\partial\;\;}{\partial x^{\rho}}\otimes\mathrm{d}x^{\alpha}\otimes\mathrm{d}x^{\nu}\otimes\mathrm{d}x^{\nu}+\hat{R}^{j}_{\;\;klc}(x^{s})\;\frac{\partial\;\;}{\partial x^{j}}\otimes\mathrm{d}x^{k}\otimes\mathrm{d}x^{l}\otimes\mathrm{d}x^{c}
	\\
	&=\mathcal{R}^{\rho}_{\;\;\alpha\mu\nu}(x^{\kappa})\;\frac{\partial\;\;}{\partial x^{\rho}}\otimes\mathrm{d}x^{\alpha}\otimes\mathrm{d}x^{\nu}\otimes\mathrm{d}x^{\nu}+\hat{\Sigma}^{j}_{\;\;klc}(x^{s})\;\frac{\partial\;\;}{\partial x^{j}}\otimes\mathrm{d}x^{k}\otimes\mathrm{d}x^{l}\otimes\mathrm{d}x^{c},
	\end{align*}
	where $\hat{R}^{\rho}_{\;\;\alpha\mu\nu}$ encodes de curvature of $\left(\mathbb{M}^{4},\boldsymbol{\omega}\right)$, and which has been labeled by $\mathcal{R}^{\rho}_{\;\;\alpha\mu\nu}$, and $\hat{R}^{j}_{\;\;klc}$ encodes de curvature of $\left(\mathbb{B}^{2},\boldsymbol{\hat{\sigma}}\right)$, which has been labeled by $\hat{\Sigma}^{j}_{\;\;klc}$. From here on $\mathcal{R}$ and $\varDelta$ are the curvature and covariant derivative compatible with $\boldsymbol{\omega}$. Analogously, $\Sigma$ and $\triangle$ are compatible with $\boldsymbol{\sigma}$, with $\hat{\Sigma}$ being compatible with $\boldsymbol{\hat{\sigma}}$. Also, a shortened notation given in terms of $\boldsymbol{\varDelta}:=\omega^{\mu\nu}\varDelta_{\mu}\varDelta_{\nu}$ and $\triangle^{2}:=\sigma^{ij}\triangle_{i}\triangle_{j}$ shall be useful in the following steps.
	
	From the Riemann tensor, the set of expressions for Ricci tensors and Riccis scalar are given by
	\begin{align*}
	\hat{R}_{\mu\nu}&={\hat{R}^{M}}{}_{\mu M\nu}={\hat{R}^{\kappa}}{}_{\mu \kappa\nu}={\mathcal{R}^{\kappa}}{}_{\mu \kappa\nu}\left(x^{\rho}\right)=\mathcal{R}_{\mu\nu}\left(x^{\rho}\right),
	\\
	\hat{R}_{ij}&={\hat{R}^{M}}{}_{i M j}={\hat{R}^{m}}{}_{i m j}={\hat{\Sigma}^{m}}{}_{i m j}\left(x^{l}\right)=\hat{\Sigma}_{ij}\left(x^{l}\right),
	\\
	\hat{R}&=\hat{g}^{MN}\hat{R}_{MN}={\omega}^{\mu\nu}\mathcal{R}_{\mu\nu}+\hat{\sigma}^{ij}\hat{\Sigma}_{ij}=\mathcal{R}+\hat{\Sigma},
	\end{align*}
	which can be re-introduced into Eq.~\eqref{GMN} so as to return
	\begin{align}
	{G}_{\mu\nu}&=\mathcal{R}_{\mu\nu}-\frac{1}{2}{\omega}_{\mu\nu}\mathcal{R}-\frac{1}{2}{\omega}_{\mu\nu}\hat{\Sigma}-4\,{\omega}_{\mu\nu}\hat{\triangle}^{2}A+6\,{\omega}_{\mu\nu}\hat{\sigma}^{ij}A_{,i}A_{,j},\nonumber
	\\
	\nonumber
	\\
	{G}_{ij}&=\hat{\Sigma}_{ij}-\frac{1}{2}\hat{\sigma}_{ij}\hat{\Sigma}-\frac{1}{2}\hat{\sigma}_{ij}\mathcal{R}+4\,\hat{\triangle}_{i}\hat{\triangle}_{j}A+4A_{,i}A_{,j}-4\,\hat{\sigma}_{ij}\hat{\triangle}^{2}A+6\,\hat{\sigma}_{ij}\hat{\sigma}^{mn}A_{,n}A_{,m}.\nonumber
	\end{align}
	Finally, by substituting the above expressions into Einstein field equations decoupled from Eq.~\eqref{einsteinfield}, one finds
	\begin{align}
	&\mathcal{R}_{\mu\nu}-\frac{1}{2}{\omega}_{\mu\nu}\mathcal{R}={\omega}_{\mu\nu}\left[\frac{1}{2}\hat{\Sigma}+4\hat{\triangle}^{2}A-6\hat{\sigma}^{ij}A_{,i}A_{,j}-\frac{e^{-2A}}{4M^{4}}\left(\frac{\phi^{,K}\phi_{,K}}{2}+\frac{\zeta^{,K}\zeta_{,K}}{2}+\mathcal{V}\right)\right],\label{fieldequationfiber}
	\\
	\nonumber
	\\
	&\hat{\Sigma}_{ij}-\frac{1}{2}\hat{\sigma}_{ij}\left(\mathcal{R}+\hat{\Sigma}\right)+4\,\hat{\triangle}_{i}\hat{\triangle}_{j}A+4A_{,i}A_{,j}-4\,\hat{\sigma}_{ij}\hat{\triangle}^{2}A\nonumber
	\\
	&\qquad\qquad\qquad\qquad\qquad+6\,\hat{\sigma}_{ij}\hat{\sigma}^{mn}A_{,n}A_{,m}=\frac{1}{4M^{4}}\Bigg[\phi_{,i}\phi_{,j}+\zeta_{,i}\zeta_{,j}-e^{-2A}\hat{\sigma}_{ij}\left(\frac{\phi^{,K}\phi_{,K}}{2}+\frac{\zeta^{,K}\zeta_{,K}}{2}+\mathcal{V}\right)\Bigg].\label{fieldequationbase}
	\end{align}
	Since $\mathcal{R}_{\mu\nu}$ and $\mathcal{R}$ are functions of space-time, $\mathbb{M}^{4}$, and $\hat{\Sigma}_{ij}$ and $\hat{\Sigma}$ are functions of the internal space, $\mathbb{B}^{2}$, then one may separate variables at Eqs.~\eqref{fieldequationfiber} and \eqref{fieldequationbase}. Through a more familiar notation, one chooses a separation constant which can be interpreted as the so-called cosmological constant $\Lambda$. After some mathematical manipulation, one thus obtains
	\begin{equation}\label{fibergeometry}
	\mathcal{R}_{\mu\nu}=\Lambda{\omega}_{\mu\nu},
	\end{equation}
	\begin{align}\label{basegeometry1}
	&\frac{1}{2}\hat{\Sigma}+4\hat{\triangle}^{2}A-6\hat{\sigma}^{ij}A_{,i}A_{,j}-\frac{e^{-2A}}{4M^{4}}\left(\frac{\phi^{,l}\phi_{,l}}{2}+\frac{\zeta^{,l}\zeta_{,l}}{2}+\mathcal{V}\right)=-\Lambda,
	\end{align}
	\begin{align}\label{basegeometry2}
	&\hat{\Sigma}_{ij}-\frac{1}{2}\hat{\sigma}_{ij}\left(4\Lambda+\hat{\Sigma}\right)+4\,\hat{\triangle}_{i}\hat{\triangle}_{j}A+4A_{,i}A_{,j}-4\,\hat{\sigma}_{ij}\hat{\triangle}^{2}A\nonumber
	\\
	&\qquad\qquad\qquad\qquad\qquad+6\,\hat{\sigma}_{ij}\hat{\sigma}^{mn}A_{,n}A_{,m}=\frac{1}{4M^{4}}\left[\phi_{,i}\phi_{,j}+\zeta_{,i}\zeta_{,j}-e^{-2A}\hat{\sigma}_{ij}\left(\frac{\phi^{,l}\phi_{,l}}{2}+\frac{\zeta^{,l}\zeta_{,l}}{2}+\mathcal{V}\right)\right].
	\end{align}
	From the above results, Eq.~\eqref{fibergeometry} defines the geometry of space-time $\left(\mathbb{M}^{4},\boldsymbol{\omega}\right)$, and one can readily obtain some solutions summarized by
	\begin{enumerate}
		\item if $\Lambda=0$, a solution is a Minkowski space, $\boldsymbol{\omega}=\boldsymbol{\eta}$;
		\item if $\Lambda>0$, a solution is a de Sitter space of four dimensions ($\mathbbm{d}\mathbb{S}^{4}$), $\boldsymbol{\omega}=\boldsymbol{\omega}^{+}$;
		\item if $\Lambda<0$, a solution is an anti-de Sitter space of four dimensions ($\mathbb{A}\mathbbm{d}\mathbb{S}^{4}$); $\boldsymbol{\omega}=\boldsymbol{\omega}^{-}$;
		\item there are also FRW space-times solutions of these equations for all values of $\Lambda$ \cite{Ahmed2014}.
	\end{enumerate}
	
	Therefore one may fit each of these $(3+1)$ solutions in the model construction that follows. Again, to simplify the notation, whenever one is dealing with a space-time $\mathbb{M}^{4}$ where $\Lambda=0$, its metric will be labeled $\boldsymbol{\eta}$, while either for $\Lambda>0$ or for $\Lambda<0$, it will be labeled either by $\boldsymbol{\omega}^{+}$ or by $\boldsymbol{\omega}^{-}$, respectively.
	Hence the subsequent steps can be resumed by obtaining the solutions for Eqs.~\eqref{basegeometry1} and \eqref{basegeometry2}, which define the geometry of the internal space $\left(\mathbb{B}^{2},\boldsymbol{\hat{\sigma}}\right)$. However, since they are still expressed in terms of $\boldsymbol{\hat{\sigma}}$, it should be simpler to work with the started geometry preliminarily resumed by $\boldsymbol{\sigma}$. Turning back to such a departure metric, one firstly writes
	\begin{align*}
	\hat{\Xi}^{l}_{ij}&=\Xi^{l}_{ij}+A_{,j}\delta^{l}_{i}+A_{,i}\delta^{l}_{j}-A_{,s}\sigma^{ls}\sigma_{ij},
	\\
	\hat{\Sigma}_{ij}&={\Sigma}_{ij}-\sigma_{ij} \triangle^{2} A,
	\\
	\hat{\Sigma}&=e^{-2A}\left[{\Sigma}-2\triangle^{2} A\right],
	\\
	\hat{\triangle}_{i}\hat{\triangle}_{j}A&={\triangle}_{i}{\triangle}_{j}A-2A_{,i}A_{,j}+\sigma^{ls}\sigma_{ij}A_{,l}A_{,s},
	\end{align*}
which, once substituted into Eqs.~\eqref{basegeometry1} and \eqref{basegeometry2}, after some straightforward manipulations, lead to

\begin{equation}\label{potential1}
\frac{\mathcal{V}}{4M^{4}}=2\Lambda e^{2A}+2\triangle^{2}A-8A^{,m}A_{,m},
\end{equation}
\begin{equation}\label{basespaceequations}
\Sigma_{ij}-\sigma_{ij}\Lambda e^{2A}+4\,{\triangle}_{i}{\triangle}_{j}A-\sigma_{ij}\triangle^{2}A+4\,\sigma_{ij}A^{,m}A_{,m}-4A_{,i}A_{,j}=\frac{\phi_{,i}\phi_{,j}+\zeta_{,i}\zeta_{,j}}{4M^{4}},
\end{equation}
from which one can notice that some coordinate degree of freedom is still present.
 
Eqs.~\eqref{potential1} and \eqref{basespaceequations} encode the needed information to determine the geometry of space $\left(\mathbb{B}^{2},\boldsymbol{\sigma}\right)$, the warp factor, $A$, and the scalar fields, $\phi$ and $\zeta$. From a geometrical perspective, they clearly illustrate why the two co-dimensional problem is circumstantially more complicated then one co-dimensional analysis. The existence of curvature for the internal space $\mathbb{B}^{2}$ increases the complexity of the equations to be solved. For a one co-dimension problem, the equations are, up to some constants, equivalent, but the curvature is null. In addition, the complexity that arises solely from topological considerations is evinced: while for one co-dimension there only two possible topologies, $\mathbb{R}^{1}$ or $\mathbb{S}^{1}$, for two co-dimensions a vaster scenario can be explored.

Turning back to the systematic procedure for solving Eqs.~\eqref{potential1} and \eqref{basespaceequations}, one can state the following theorem \cite{Ray1992}, 
\begin{theorem}
	Every 2-dimensional (pseudo-) Riemannian space $(\mathbb{B}^{2},\boldsymbol{\sigma})$ is conformally flat.
\end{theorem}

This means that, without loss of generality, one can consider the metric of the space of co-dimensions to be conformally flat, i.e.
$$
\boldsymbol{\sigma}=e^{-2{B}(u,v)}\left(\mathrm{d}u\otimes\mathrm{d}u+\mathrm{d}v\otimes\mathrm{d}v\right).
$$

As previously argued, one has made a previous choice for the coordinates so as to be able to write the resulting expression for the metric. However, since Eq.~\eqref{basespaceequations} may not be analytically solvable, it would be counterproductive to keep that expressed in terms of conformal coordinates. A more treatable set coordinates for solving the resulting differential equations, that do not result in a conformally flat metric, can be identified by rewriting the system in terms of the following metric,
\begin{equation}
\boldsymbol{\sigma}=e^{-2{f}(u,v)}\mathrm{d}u\otimes\mathrm{d}u+e^{-2{h}(u,v)}\mathrm{d}v\otimes\mathrm{d}v.\label{assumptionmetric}
\end{equation}
This is not the most general metric choice\footnote{The most general one would allow non-diagonal terms}, but it does allow for some leeway when solving the equations. Naturally, this is equivalent to the conformally flat form, since one has just used a different set of coordinates. By substituting the metric choice from \eqref{assumptionmetric} into the field Eqs.~\eqref{potential1} and \eqref{basespaceequations}, it is straightforward to write
\begin{align}
\frac{\mathcal{V}}{8M^{4}}&=\Lambda e^{2A}+e^{2f}\left(A_{,uu}+A_{,u}f_{,u}-A_{,u}h_{,u}-4A_{,u}{}^{2}\right)+e^{2h}\left(A_{,vv}+A_{,v}h_{,v}-A_{,v}f_{,v}-4A_{,v}{}^{2}\right),\label{potential2}
\\
\frac{\phi_{,u}{}^{2}+\zeta_{,u}{}^{2}}{4M^{4}}&=e^{2h-2f}\left(f_{,vv}+f_{,v}h_{,v}-f_{,v}{}^{2}+4A_{,v}{}^{2}-3f_{,v}A_{,v}-A_{,vv}-A_{,v}h_{,v}\right)\nonumber
\\
&\qquad\qquad\qquad\qquad\qquad\qquad+h_{,uu}+f_{,u}h_{,u}-h_{,u}{}^{2}+3A_{,uu}+3f_{,u}A_{,u}+A_{,u}h_{,u}-\Lambda e^{2A-2f}\label{uu},
\\
\frac{\phi_{,v}{}^{2}+\zeta_{,v}{}^{2}}{4M^{4}}&=e^{2f-2h}\left(h_{,uu}+f_{,u}h_{,u}-h_{,u}{}^{2}+4A_{,u}{}^{2}-3h_{,u}A_{,u}-A_{,uu}-A_{,u}f_{,u}\right)\nonumber
\\
&\qquad\qquad\qquad\qquad\qquad\qquad+f_{,vv}+f_{,v}h_{,v}-f_{,v}{}^{2}+3A_{,vv}+3h_{,v}A_{,v}+A_{,v}f_{,v}-\Lambda e^{2A-2h}\label{vv},
\\
\frac{\phi_{,u}\phi_{,v}+\zeta_{,u}\zeta_{,v}}{4M^{4}}&=4A_{,uv}+4f_{,v}A_{,u}+4h_{,u}A_{,v}-4A_{,u}A_{,v}\label{uv}.
\end{align}

From Eqs.~\eqref{potential2}-\eqref{uv} one can determine the warp factor and scalar fields, and consequently obtaining the defect that generates the thick brane. 
From this point, different techniques must be employed for solving Eqs.~\eqref{potential2}-\eqref{uv} analytically. One may separate the techniques into two opposite categories:
\begin{enumerate}
	\item Starting from a predetermined internal space $(\mathbb{B}^{2},\boldsymbol{\sigma})$, which in some other words correspond to the preliminary knowledge of $f$ and $h$, one can thus calculate the warp factor $A$;
	\item Starting with no knowledge of the geometry of the internal space $(\mathbb{B}^{2},\boldsymbol{\sigma})$, i.e. of $f$ and $h$, one thus assume some simplifying hypothesis so as to solve the equations in order to find $A$, $f$ and $h$.\label{second}
\end{enumerate}

Our focus will be on the second technique, which can be later connected to the first one by using their solutions to fit them into predetermined geometries. 
Looking at the second technique, the equations will necessarily determine the metric, but the topology of $\mathbb{B}^{2}$ will still remain undetermined. This fact is true, since the metric does not have, in general, enough information to define the topological properties of space, with the exception of some of the compact characteristics of the latter, which is only possible because of the Bulk geometry\footnote{According to Refs.~ \cite{Gibbons2001,Leblond2001}, one can extract out of Einstein equations whether or not the space $\mathbb{B}^{2}$ is non-compact.}. Besides this special topological invariant, not many topological statements can be extracted about the spaces here within, unless it is imposed {\em a priori}. Such an indeterminacy will be advantageous to the model building, since the same solution may fit different topologies and thus configure distinctive space-times.

\section{Intersecting Thick Branes}\label{intersecting}

For branes regarded as the intersection between the defects generated by $\phi$, such that $\phi_{,v}=0$, and by $\zeta$, such that $\zeta_{,u}=0$, which are achieved through an appropriate choice of coordinates, $u$ and $v$, one can follow the strong supposition that the warp factor $A$ and the functions $f$ and $h$ will all be separable functions of $u$ and $v$,\begin{align*}
A&=\hat{A}(v)+\tilde{A}(u),
\\
f&=\hat{f}(v)+\tilde{f}(u),
\\
h&=\hat{h}(v)+\tilde{h}(u).
\end{align*}
which can be summarized by a metric restricted by \eqref{assumptionmetric}, as it does not have diagonal terms.

With such assumptions clearly identified, then Eq.~\eqref{uv} implies into
\begin{equation}
\hat{f}_{,v} \tilde{A}_{,u}+\tilde{h}_{,u} \hat{A}_{,v}-\tilde{A}_{,u} \hat{A}_{,v}=0,\label{intersec}
\end{equation}
which can be solved under two independent subliminar hypothesis.

Firstly, when either $\hat{A}_{,v}$ or $\tilde{A}_{,u}$ are set equal to zero, thus one has either $\hat{f}_{,v}=0$, if $\hat{A}_{,v}=0$, or $\tilde{h}_{,u}=0$, if $\tilde{A}_{,u}=0$, -- consequently, the most simplified scenario. It constrains either $u$ or $v$ to be compactified since there would be no way of localizing fields along the direction for which $A$ is null (constant). For instance, with $\hat{A}=0$, and arbitrarily setting $\hat{h}=0$, the resulting metric would be cast in the form of
\begin{equation}
\boldsymbol{g}=e^{-2\tilde{A}}\omega_{\mu\nu}\mathrm{d}x^{\mu}\otimes\mathrm{d}x^{\nu}+e^{-2\tilde{f}}\mathrm{d}u\otimes\mathrm{d}u+e^{-2\tilde{h}}\mathrm{d}v\otimes\mathrm{d}v,\label{stringlikedefect}
\end{equation}
which leads to a string-like defect for the warp factor. These set of solutions have already been widely investigated \cite{Koley2006,Gherghetta2000,Park2003,PARAMESWARAN200754,Singleton2004,Multamaeki2002} even when they are not driven by scalar fields. Considering our more general interest, such constructions will not be further pursued. However, it is worth to mention that several models that shall be more deeply understood also have, as limiting cases, string-like solutions for the warp factor.

Secondly, the most promising scenario emerges from considering non-vanishing values for both components, $\tilde{A}_{,u}$ and $\hat{A}_{,v}$. 
Following a simplified stratagem, from Eq.~\eqref{intersec}, one may write $\hat{f}=p\hat{A}$ and $\tilde{h}=\left(1-p\right)\tilde{A}$, where $p\in\mathbb{R}$, while $\hat{h}$ and $\tilde{f}$ are mapped by an aleatory correspondence with the coordinates $u$ and $v$. In this case, the metric is recast in the form of
\begin{equation}
\boldsymbol{g}=e^{-2\hat{A}}e^{-2\tilde{A}}\omega_{\mu\nu}\mathrm{d}x^{\mu}\otimes\mathrm{d}x^{\nu}+e^{-2p\hat{A}}e^{-2\tilde{f}}\mathrm{d}u\otimes\mathrm{d}u+e^{-2\hat{h}}e^{-2\left(1-p\right)\tilde{A}}\mathrm{d}v\otimes\mathrm{d}v,\label{intersectingmetric}
\end{equation}
which leads to a novel class of solutions which indeed is not covered by the metric from \eqref{stringlikedefect}.

As implicitly mentioned, from the metric Eq.~\eqref{intersectingmetric}, one can realize that the exchange of coordinates $u \leftrightarrow v$ (as well as $f \leftrightarrow h$), does not modifies the space-time, which is just re-labeled in terms of $u \leftrightarrow v$. This means that a model with $p=p_1$ is equivalent to a model with $p=1-p_{1}$, which can be mathematically expressed in terms of the equivalence relation, $\forall\;p\in \mathbb{R}:p\sim1-p$, i.e. for any $p$ value there is an equivalent model with $p$ replaced by $1-p$. Thus, the algorithm for solving the equations of motion can be constrained by choosing, for instance,
\begin{equation*}
p\in\mathbb{R}/\hspace{-3.5pt}\sim=\left\{p\in\mathbb{R}\;|\;p\geq1/2\right\},
\end{equation*}
such that the equations to be solved, \eqref{potential2}-\eqref{vv}, can be resumed by
\begin{align}
\frac{\mathcal{V}}{8M^{4}}=& e^{2 p\hat{A}}e^{2 \tilde{f}}\left[\left(p-5\right){\tilde{A}_{,u}}{}^{2}+\tilde{f}_{,u} \tilde{A}_{,u}+\tilde{A}_{,uu}\right]\nonumber
\\
&\qquad\qquad\qquad\qquad\qquad+e^{2 \hat{h}}e^{2 \left(1-p\right)\tilde{A}} \left[\hat{h}_{,v} \hat{A}_{,v}-\left(p+4\right){\hat{A}_{,v}}{}^{2}+ \hat{A}_{,vv}\right]+\Lambda e^{2 \hat{A}}e^{2 \tilde{A}},\label{einsteinsepV}
\\
\frac{e^{2 p\hat{A}}e^{2 \tilde{f}}}{4M^{4}}{\phi_{,u}}^{2}=& e^{2 p\hat{A}}e^{2 \tilde{f}} \left[p\left(1-p\right){\tilde{A}_{,u}}{}^{2}+\left(4-p\right)\tilde{f}_{,u}\tilde{A}_{,u}+\left(4-p\right)\tilde{A}_{,uu}\right]-\Lambda e^{2 \hat{A}}e^{2 \tilde{A}}\nonumber
\\
&\qquad\qquad\qquad\qquad\qquad+e^{2 \hat{h}}e^{2 \left(1-p\right)\tilde{A}}\left[\left(4-3p-p^{2}\right) {\hat{A}_{,v}}{}^{2}+\left(p-1\right)\hat{A}_{,v} \hat{h}_{,v}+\left(p-1\right)\hat{A}_{,vv}\right],\label{einsteinsep1}
\\
\frac{e^{2\hat{h}}e^{2\left(1-p\right)\tilde{A}}}{4M^{4}}{\zeta_{,v}}^{2}=& e^{2p\hat{A}}e^{2\tilde{f}} \left[p\left(5-p\right)\tilde{A}_{,u}{}^{2}- p\tilde{f}_{,u} \tilde{A}_{,u}-p\tilde{A}_{,uu}\right]-\Lambda e^{2\hat{A}}e^{2\tilde{A}}\nonumber
\\&
\qquad\qquad\qquad\qquad\qquad+e^{2\hat{h}}e^{2\left(1-p\right)\tilde{A}} \left[p\left(1-p\right){\hat{A}_{,v}}{}^{2}+\left(3+p\right) \hat{h}_{,v} \hat{A}_{,v}+\left(3+p\right)\hat{A}_{,vv}\right].\label{einsteinsep2}
\end{align}
Notice that Eq.~\eqref{einsteinsepV} just defines the potential as a function of $u$ and $v$. Unless one imposes to the potential $\mathcal{V}$ its analytical dependence on $\phi$ and/or $\zeta$, which would suppress some degrees of freedom from Eqs.~\eqref{einsteinsep1} and \eqref{einsteinsep2}, Eq.~\eqref{einsteinsepV} is redundant to the solutions from Eqs.~\eqref{einsteinsep1} and \eqref{einsteinsep2} when they are used to obtain $\mathcal{V}$.
Otherwise, the analytical solutions for Eqs.~\eqref{einsteinsep1} and \eqref{einsteinsep2} can be obtained under the following constraints.
\begin{enumerate}
	\item When $\Lambda=0$, thus the brane is flat;
	\item When $\Lambda\neq0$, but $p=0$ (or $p=1$).
\end{enumerate}
This happens because the term with the cosmological constant will necessarily contribute to a function that depends on both variables, unless $p=0$ (or $p=1$) or the brane is flat ($\Lambda=0$).

\subsection{The Flat Brane Case ($\Lambda=0$)}

After applying the separation of variables technique, Eqs.~\eqref{einsteinsep1} and \eqref{einsteinsep2} are written as
\begin{align}
& (4+p){\hat{A}_{,v}}^2-\hat{h}_{,v} \hat{A}_{,v}-\hat{A}_{,vv}=\frac{C_{1}}{1-p}e^{2 p\hat{A}}e^{-2 \hat{h}},\label{flatbrane1}
\\
& \frac{{\phi_{,u}}^{2}}{4M^{4}} -\left(4-p\right)\tilde{f}_{,u}\tilde{A}_{,u}-p\left(1-p\right){\tilde{A}_{,u}}^2-\left(4-p\right)\tilde{A}_{,uu}=C_{1}e^{-2 \tilde{f}}e^{2 \left(1-p\right)\tilde{A}},\label{flatbrane2}
\\
& \left(5-p\right) {\tilde{A}_{,u}}^2-\tilde{f}_{,u} \tilde{A}_{,u}-\tilde{A}_{,uu}=\frac{C_{2}}{p} e^{-2\tilde{f}}e^{2\left(1-p\right)\tilde{A}},\label{flatbrane3}
\\
& \frac{{\zeta_{,v}}^{2}}{4M^{4}}-\left(3+p\right) \hat{h}_{,v} \hat{A}_{,v}-p\left(1-p\right){\hat{A}_{,v}}^{2}-\left(3+p\right) \hat{A}_{,vv}=C_{2}e^{-2p\hat{A}} e^{2\hat{h}},\label{flatbrane4}
\end{align}
where $C_{i}\in\mathbb{R}$, $i\in\left\{1,2\right\}$, are the separation constants. To find solutions of Eqs.~\eqref{flatbrane1}-\eqref{flatbrane4} one needs to separate the $p=0$ (or $p=1$) case from the $p\neq 0$ (or $p\neq1$).

\subsubsection{The $p\neq0$ (or $p\neq1$) Case (Models $I$ and $II$)}

Essentially, the above introduced sequence of steps for preparing the equations of motion to be solved corresponds to some kind of suppression of unnecessary degrees of freedom. 
Looking at Eqs.~\eqref{flatbrane1}-\eqref{flatbrane4}, the coordinate freedom are represented by $\hat{h}$ and $\tilde{f}$. Again, the coordinate constraints, $\hat{h}=p\hat{A}$ and $\tilde{f}=\left(1-p\right)\tilde{A}$, are chosen in order to simplify the equation manipulability.
With the metric in the form of
\begin{equation}
\boldsymbol{g}=e^{-2\tilde{A}}e^{-2\hat{A}}\eta_{\mu\nu}\mathrm{d}x^{\mu}\otimes\mathrm{d}x^{\nu}+e^{-2\left(1-p\right)\tilde{A}}e^{-2p\hat{A}}\left(\mathrm{d}u\otimes\mathrm{d}u+\mathrm{d}v\otimes\mathrm{d}v\right)\label{metricflat}
\end{equation}
corresponds to the singular configuration for which a conformally flat approach simplifies the equation resolutions. From Eqs.~\eqref{flatbrane1} and \eqref{flatbrane3}, the solutions obtained are expressed by
\begin{align}
\hat{A}&=\hat{A}_{0}-\frac{1}{4} \ln \bigg\{\cosh \Big[2c_{v}\big(v+v_{0}\big)\Big]\bigg\}\label{warpfactorv},
\\
\tilde{A}&=\tilde{A}_{0}-\frac{1}{4} \ln \bigg\{\cosh \Big[2c_{u}\big(u+u_{0}\big)\Big]\bigg\}\label{warpfactoru},
\end{align}
where, without loss of generality, one set the boundary conditions as given by $\hat{A}_{0}=\tilde{A}_{0}=0$, with
\begin{align*}
c_{v}{}^{2}&=-\frac{C_{1}}{p-1},
\\
c_{u}{}^{2}&=\frac{C_{2}}{p},
\end{align*}
where $c_{v}$, $c_{u}$ $\in \mathbb{C}$, but either $Im(c_{i})=0$ or $Re(c_{i})=0$, since $C_{1}$, $C_{2}$ and $p$ are real constants.

To develop models which can ``localize'' fields on the brane, one may break this solution into two different configurations, one for $p\geq3$ and $Im(c_{u})=0$, and another for $p\leq3$ and $Re(c_{u})=0$. They correspond to the models that shall be further explored in appendix \ref{longermodelIandII}.

Starting with $p\geq3$ ($Im(c_{u})=0$), which is now labeled model $I$, one finds the metric ($u_{0}=v_{0}=0$),
\begin{equation}\label{metricI}
\boldsymbol{g}^{I}=\sqrt{\cosh \left(2c_{u}u\right)\left|\cos \left(\frac{n\varphi}{2}\right)\right|}\eta_{\mu\nu}\mathrm{d}x^{\mu}\otimes\mathrm{d}x^{\nu}+\sqrt{\frac{\left|\cos \left(\frac{n\varphi}{2}\right)\right|^{p} }{\cosh^{p-1} \left(2c_{u}u\right)}}\left(\mathrm{d}u\otimes\mathrm{d}u+r^{2}\mathrm{d}\varphi\otimes\mathrm{d}\varphi\right).
\end{equation}

Scalar fields and potential are resumed by
	\begin{align}
	\mathcal{V}^{I}&=-8M^{4}\operatorname{sech}^{\left(1-p\right)/2} \left(2c_{u}u\right)\operatorname{sec}^{p/2}
	\left(\frac{n\varphi}{2}\right)\left(c_{u}{}^{2}-\frac{n^{2}}{16r^{2}}\right),\label{potentialI}
	\\
	\phi^{I}&=\pm 2M^{2}\sqrt{{ }^{ }a_{\phi}}\Bigg\{u\sqrt{1+b_{\phi}}-\frac{\sqrt{b_{\phi}}}{2c_{u}} \operatorname{arcsinh}\left[\sqrt{b_{\phi}}\tanh\left(2c_{u}u\right)\right]\nonumber
	\Bigg.\\&\Bigg.
	\qquad\qquad\qquad\qquad\qquad\qquad\quad\,\,\,\,\,-\frac{\sqrt{1+b_{\phi}}}{4 c_{u}}\ln \left[\frac{\sqrt{1+b_{\phi}} \sqrt{1+b_{\phi} \tanh ^2\Big(2c_{u}u\Big)}+1-b_{\phi} \tanh \Big(2c_{u} u\Big)}{\sqrt{1+b_{\phi}} \sqrt{1+b_{\phi} \tanh ^2\Big(2c_{u}u\Big)}+1+b_{\phi} \tanh \Big(2c_{u} u\Big)}\right]\Bigg\},\label{scalarphiI}
	\\
	\zeta^{I}&=\pm M^{2}\cos \left(\frac{n\varphi}{2}\right)\frac{4r\sqrt{2a_{\zeta}} \sqrt{1-b_{\zeta} \tan ^2\left(\frac{n\varphi}{2}\right)}}{n \sqrt{1- b_{\zeta}+ (1+b_{\zeta}) \cos \left(n\varphi\right)}} \Bigg\{\sqrt{1+b_{\zeta}} \arcsin\left[\sqrt{1+b_{\zeta}} \sin \left(\frac{n\varphi}{2}\right)\right]\nonumber
	\\
	&\qquad\qquad\qquad\qquad\qquad\qquad\qquad\qquad\qquad\qquad\qquad\quad\;\;+\sqrt{-b_{\zeta}} \operatorname{arctanh}\left[\frac{\sqrt{2} \sqrt{-b_{\zeta}} \sin \left(\frac{n\varphi}{2}\right)}{\sqrt{(b_{\zeta}+1) \cos \left(n\varphi\right)-b_{\zeta}+1}}\right]\Bigg\},\label{scalarzetaI}
	\end{align}
where the following constants have been defined,
\begin{align*}
a_{\phi}&=-\left(1-p\right)\frac{n^{2}}{16r^{2}}+\left(p-4\right)c_{u}{}^{2},
\\
b_{\phi}&=\frac{\left(5-2p\right)c_{u}{}^{2}}{a_{\phi}},
\\
a_{\zeta}&=p\,c_{u}{}^{2}+\left(3+p\right)\frac{n^{2}}{16r^{2}},
\\
b_{\zeta}&=-\frac{\left(3+2p\right)n^{2}}{16r^{2}a_{\zeta}},
\end{align*}
through which the constraints $a_{\phi}\geq0$, $b_{\phi}\geq-1$, $a_{\zeta}\geq0$ and $b_{\zeta}\leq0$ are sufficient and necessary conditions for obtaining real scalar fields, $\phi$ and $\zeta$ (cf. Eqs.~\eqref{scalarphiI} and \eqref{scalarzetaI}).

Besides the singularities exhibited by the scalar field $\zeta^{I}$, the behavior of the variable $u$ suggests that an infinite amount of energy to achieve model $I$ configuration is required (see appendix \ref{longermodelIandII}).

To avoid such a shortcoming, the model $II$, with $p\leq 3$ and $Re(c_{u})=0$, can be introduced.
In this case, the metric can be stated as ($u_{0}=v_{0}=0$),
\begin{equation}
\boldsymbol{g}^{II}=\sqrt{\left|\cos \left(\frac{l\theta}{2}\right)\cos \left(\frac{n\varphi}{2}\right)\right|}\eta_{\mu\nu}\mathrm{d}x^{\mu}\otimes\mathrm{d}x^{\nu}+\sqrt{\frac{\left|\cos \left(\frac{n\varphi}{2}\right)\right|^{p}}{\left|\cos \left(\frac{l\theta}{2}\right)\right|^{p-1}}}\left({\rho}^{2}\mathrm{d}\theta\otimes\mathrm{d}\theta+{r}^{2}\mathrm{d}\varphi\otimes\mathrm{d}\varphi\right).\label{p<3}
\end{equation}
Notice that, for $p\geq3$, the metric \eqref{p<3} would imply into an infinite effective volume. The scalar fields and potential for such a configuration are as follows,
	\begin{align}
	\mathcal{V}^{II}&=M^{4}{\sec}^{\left(1-p\right)/2} \left(\frac{l\theta}{2}\right){\sec}^{p/2}
	\left(\frac{n\varphi}{2}\right)\left(\frac{{l}^{2}}{2{\rho}^{2}}+\frac{{n}^{2}}{2{r}^{2}}\right),\label{potentialII}
	\\
	\phi^{II}&=\pm M^{2}\cos \left(\frac{l\theta}{2}\right)\frac{4\rho\sqrt{2a_{\phi}} \sqrt{1-b_{\phi} \tan ^2\left(\frac{l\theta}{2}\right)}}{l \sqrt{1- b_{\phi}+ (1+b_{\phi}) \cos \left(l\theta\right)}} \Bigg\{\sqrt{1+b_{\phi}} \arcsin\left[\sqrt{1+b_{\phi}} \sin \left(\frac{l\theta}{2}\right)\right]\nonumber
	\\
	&\qquad\qquad\qquad\qquad\qquad\qquad\qquad\qquad\qquad\qquad\qquad\quad\;\;+\sqrt{-b_{\phi}} \operatorname{arctanh}\left[\frac{\sqrt{2} \sqrt{-b_{\phi}} \sin \left(\frac{l\theta}{2}\right)}{\sqrt{(b_{\phi}+1) \cos \left(l\theta\right)-b_{\phi}+1}}\right]\Bigg\},\label{scalarphiII}
	\\
	\zeta^{II}&=\pm M^{2}\cos \left(\frac{n\varphi}{2}\right)\frac{4r\sqrt{2a_{\zeta}} \sqrt{1-b_{\zeta} \tan ^2\left(\frac{n\varphi}{2}\right)}}{n \sqrt{1- b_{\zeta}+ (1+b_{\zeta}) \cos \left(n\varphi\right)}} \Bigg\{\sqrt{1+b_{\zeta}} \arcsin\left[\sqrt{1+b_{\zeta}} \sin \left(\frac{n\varphi}{2}\right)\right]\nonumber
	\\
	&\qquad\qquad\qquad\qquad\qquad\qquad\qquad\qquad\qquad\qquad\qquad\quad\;\;+\sqrt{-b_{\zeta}} \operatorname{arctanh}\left[\frac{\sqrt{2} \sqrt{-b_{\zeta}} \sin \left(\frac{n\varphi}{2}\right)}{\sqrt{(b_{\zeta}+1) \cos \left(n\varphi\right)-b_{\zeta}+1}}\right]\Bigg\},\label{scalarzetaII}
	\end{align}
where one identifies the following constants,
\begin{align*}
a_{\phi}&=-\left(1-p\right)\frac{n^{2}}{16r^{2}}-\left(p-4\right)\frac{l^{2}}{16\rho^{2}},
\\
b_{\phi}&=-\frac{\left(5-2p\right)l^{2}}{16\rho^{2}a_{\phi}},
\\
a_{\zeta}&=-p\frac{l^{2}}{16\rho^{2}}+\left(3+p\right)\frac{n^{2}}{16r^{2}},
\\
b_{\zeta}&=-\frac{\left(3+2p\right)n^{2}}{16r^{2}a_{\zeta}}.
\end{align*}

In this case, $a_{\phi}\geq0$, $b_{\phi}\leq0$, $a_{\zeta}\geq0$ and $b_{\zeta}\leq0$ are the sufficient and necessary conditions for assuring real scalar fields. The scalar fields exhibit several singularities, depending on the values for $n$ and $l$. These singularities explains the number of cusps in the warp factor. In order to realize physically consistent solutions, the required energy to achieve their internal structure must be finite. Even though model $II$ exhibits several singularities as depicted by the scalar fields, the total energy necessary to accomplish model $II$ is finite (see appendix \ref{longermodelIandII}). This is an evinced advantage with respect to the model $I$. Although model $II$ has finite total energy, one may still argue against its physical significance, due to its recurrent singularities, a shortcoming that must be considered in the following model issues.

\subsubsection{The $p=0$ (or $p=1$) Case (Model $III$)}

The third option of analytical models with flat branes, with two scalar fields and $p=0$, starts from setting $\hat{f}=0$ and $\tilde{h}=\tilde{A}$, which leads to the subsequent metric,
\begin{equation}
\boldsymbol{g}=e^{-2\hat{A}}e^{-2\tilde{A}}\omega_{\mu\nu}\mathrm{d}x^{\mu}\otimes\mathrm{d}x^{\nu}+e^{-2\tilde{f}}\mathrm{d}u\otimes\mathrm{d}u+e^{-2\hat{h}}e^{-2\tilde{A}}\mathrm{d}v\otimes\mathrm{d}v.\label{metricp=0}
\end{equation}
Again, from Eqs.~\eqref{einsteinsepV}-\eqref{einsteinsep2}, after separation of variables and some straightforward manipulations, one finds the following system of equations,
\begin{align}
\frac{\mathcal{V}}{8M^{4}}&= e^{2 \tilde{f}}\left(-5{\tilde{A}_{,u}}{}^{2}+\tilde{f}_{,u} \tilde{A}_{,u}+\tilde{A}_{,uu}\right)+Ce^{2 \tilde{A}},\label{generalpotential2}
\\
\frac{{\phi_{,u}}^{2}}{4M^{4}}&=4\tilde{f}_{,u}\tilde{A}_{,u}+4\tilde{A}_{,uu}-Ce^{2\tilde{A}}e^{-2 \tilde{f}},\label{generalscalar1}
\\
C&=-e^{2 \hat{h}}\left(4 {\hat{A}_{,v}}{}^{2}-\hat{A}_{,v} \hat{h}_{,v}-\hat{A}_{,vv}\right),\label{flatequation}
\\
\frac{{\zeta_{,v}}^{2}}{4M^{4}}&=\;3 \hat{h}_{,v} \hat{A}_{,v}+3\hat{A}_{,vv},\label{flatscalar2}
\end{align}
where $C\in\mathbb{R}$ is some separation constant.
Here one can interpret Eqs.~\eqref{generalpotential2} and \eqref{generalscalar1} as defining the potential and the scalar field $\phi$, respectively, and one can actually solve Eqs.~\eqref{flatequation} and \eqref{flatscalar2}. By choosing $\hat{h}=0$ straightforwardly implies into the solution
\begin{align}
\hat{A}^{III}&=\hat{A}_{0}-\frac{1}{4} \ln \left|\cos \left[2\sqrt{C} \left(v+v_{0}\right)\right]\right|,\label{solutionwarpIII}
\\
\zeta^{III}&=\pm2\sqrt{3}M^{2} \operatorname{arctanh}\left\{\sin \left[2 \sqrt{C} \left(v+v_{0}\right)\right]\right\},\label{solutionscalarIII}
\end{align}
which, from now on, shall be called model $III$ and for which, without loss of generality, one can set $\hat{A}_{0}=0$ and $v_{0}=0$.

Given the periodicity of $\hat{A}^{III}$, one departs from the choice of $v=r\varphi$, where $\varphi\in\mathbb{S}^{1}$ and $r$ is the radius of $\mathbb{S}^{1}$. Since the metric must be continuous in $\mathbb{S}^{1}$ one must have that $e^{-2\hat{A}}$ must also be continuous in $\mathbb{S}^{1}$, which means that
$$
\left|\cos \left(2\sqrt{C} r2\pi\right)\right|=\left|\cos(0)\right|=1\implies C=\frac{n^{2}}{16r^{2}}\text{, }n\in\mathbb{N}.
$$
Therefore one may write the metric, with $\hat{A}_{0}=0$ and $v_{0}=0$, as
\begin{equation}
\boldsymbol{g}^{III}= \sqrt{\left|\cos\left(\frac{n\varphi}{2}\right)\right|}e^{-2\tilde{A}}\eta_{\mu\nu}\mathrm{d}x^{\mu}\otimes\mathrm{d}x^{\nu}+e^{-2\tilde{f}}\mathrm{d}u\otimes\mathrm{d}u+r^{2}e^{-2\tilde{A}}\mathrm{d}\varphi\otimes\mathrm{d}\varphi,\label{flat2metricgeneral}
\end{equation}
which expresses a setup with a Minkowski brane $(\Lambda=0)$ with two scalar fields. Both the scalar field $\zeta^{III}$ and warp factor $\hat{A}^{III}$ are, up to some constant, equivalent in form to those ones from model $I$, as depicted in Figs.~\ref{warpI} and \ref{zetaI}. Despite of such similarities, distinctive coordinates lead to inequivalent fields so that solutions must be re-discussed and re-interpreted. Considering the possible values of $n$, only $n=1$ configuration does not require the modulus in $\sqrt{\left|\cos\left({n\varphi}/{2}\right)\right|}$, since $\cos\left({\varphi}/{2}\right)$ is strictly positive in this region. 

In this case, one may be tempted to interpret each of the cusps of the warp factor as forming different branes. However, since the unique localizing parameter in this model is the radius $r$ of $\mathbb{S}^{1}$, it is better to interpret such a configuration as a single brane with some internal structure as the same is true for models $I$ and $II$.

\begin{figure}[!htb]
	\includegraphics[scale=0.95]{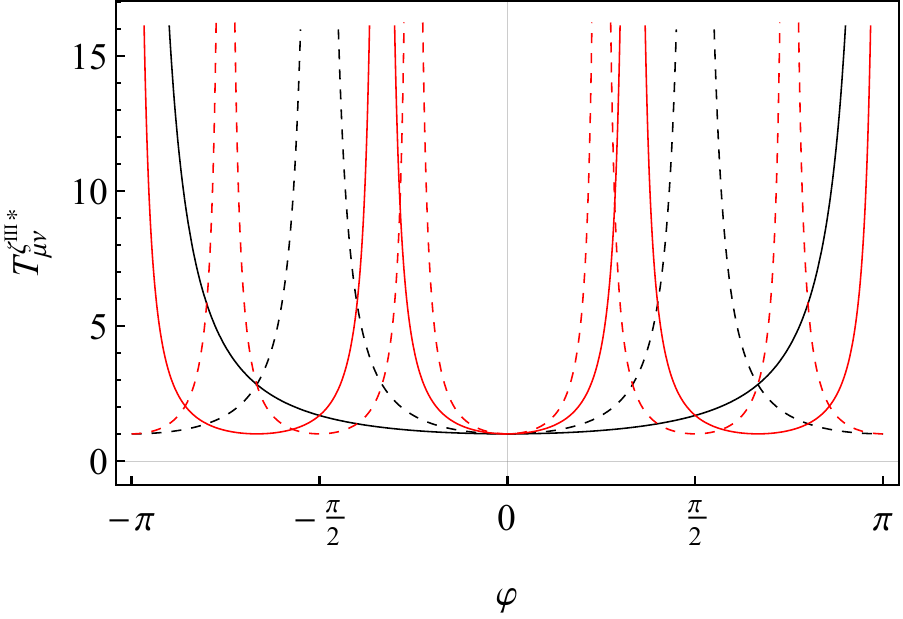}
	\caption{(Color online) Stress energy tensor $T^{\zeta^{III*}}_{\mu\nu}=-r^{2}T^{\zeta^{III}}_{\mu\nu}/3 M^4 n^2\eta_{\mu\nu}$ as a function of $\varphi$, for $n=1$ (black line), $n=2$ (black dashed line), $n=3$ (red line) and $n=4$ (red dashed line).}\label{stresstensorflatp=0}
\end{figure}

In this case, before evaluating metric $\boldsymbol{g}^{III}$, scalar field $\phi$ and potential configurations, one should turn the attention to the associated stress energy tensor. For $p=0$ models, from Eqs.~\eqref{generalpotential2} and \eqref{generalscalar1}, the total stress energy tensor can be separated as follows,
\begin{align}
T_{MN}&=T^{\zeta}_{MN}+T^{\phi}_{MN},\label{stress1}
\\
T^{\zeta}_{MN}&=\zeta_{,M}\zeta_{,N}-g_{MN}\frac{\zeta^{,K}\zeta_{,K}}{2},\label{stress2}
\\
T^{\phi}_{MN}&=\phi_{,M}\phi_{,N}-g_{MN}\left(\frac{\phi^{,K}\phi_{,K}}{2}+\mathcal{V}\right).\label{stress3}
\end{align}
which allows one to focus on the stress energy tensor driven by the scalar field $\zeta$ \eqref{stress2},
$$
T^{\zeta^{III}}_{\mu\nu}=-\frac{3 M^4 n^2\eta_{\mu\nu}}{2r^{2}}\left|\sec\left(\frac{n\varphi}{2}\right)\right|^{3/2}.
$$
which is depicted Fig.~\ref{stresstensorflatp=0}.
Of course, Fig.~\ref{stresstensorflatp=0} evinces that the $T^{\zeta^{III}}_{\mu\nu}$ singularities shall support a number of cusps in the warp factor. From the perspective of the bulk, the energy necessary to achieve such a configuration can be computed in terms of
\begin{align*}
	E^{\zeta^{III}}_{\mu\nu}&=\int_{\mathbb{E}^{6}}T^{\zeta^{III}}_{\mu\nu}\sqrt{-\mathrm{g}}\mathrm{d}^{6}x\propto\int^{\pi}_{-\pi}\left|\sec\left(\frac{n\varphi}{2}\right)\right|^{1/2}\mathrm{d}\varphi,
\end{align*}
where the last integral converges for all values of $n$. Therefore, the energy of these models, as far as $\zeta$ is concerned, is finite. Although the energy needed for this configuration is finite, one may still argue against the physical significance of this model, due to the number of singularities in the stress energy tensor.

To complete the model one now only lacks the dependence of the warp factor to the $u$ coordinate related to $\tilde{A}$, to the scalar field $\phi$, as well as to the potential $\mathcal{V}$. For model $III$ these fields must satisfy Eqs.~\eqref{generalpotential2} and \eqref{generalscalar1}. Notice here that while $\mathcal{V}$, $\phi$ and $\tilde{A}$ are still undetermined, $\tilde{f}$ is a mere choice of coordinates. Therefore one has complete freedom for choosing one of such fields, as long as further analytical integration is allowed for the other two fields. This means that a multitude of solutions can be find as to fit such a building procedure. As will be presented later a similar set of equations for $\tilde{A}$, $\phi$ and $\mathcal{V}$ will be found for different solutions of $\hat{A}$ and $\zeta$, this is to be expected since the equations are separated in the variables $u$ and $v$. Later a couple of examples will be proposed, all of which start by assuming $\tilde{A}$. This is simply to achieve an intended geometry for $\left(\mathbb{B}^{2},\boldsymbol{\sigma}\right)$, which shall lead to a common solution set for $\tilde{A}$, $\phi$ and $\mathcal{V}$ for all the models with $p=0$.

To resume, model $III$ also contains a trivial extension of well known models of $(4+1)$-dimensional brane-worlds. Looking at Eqs.~\eqref{generalpotential2} and \eqref{generalscalar1}, one should notice that, for $\tilde{f}=n=0$ (which is nothing but a choice of coordinates and $C=0$), exactly the same equations, up to some constants, are obtained from such a five dimensional case \cite{Bernardini2016,Almeida2009,Bazeia2004,Dzhunushaliev2010,Bazeia2002,DeWolfe2000,Ahmed2013,Gremm2000a,Kehagias2001,Kobayashi2002,Bronnikov2003,BAZEIA2009b,BarbosaCendejas2013,Chinaglia2016}. These models contain, which is nothing surprising, trivial extensions of the five dimensional brane-world models so deeply considered in the previously quoted works. One may call it trivial because the metric takes the form,
\begin{equation}
\boldsymbol{g}=e^{-2\tilde{A}}\left(\eta_{\mu\nu}\mathrm{d}x^{\mu}\otimes\mathrm{d}x^{\nu}+r^{2}\mathrm{d}\varphi\otimes\mathrm{d}\varphi\right)+\mathrm{d}u\otimes\mathrm{d}u,\label{trivialmetric}
\end{equation}
which is nothing but the same metric of five dimensional setup with an additional co-dimensional compactified structure as $\mathbb{S}^{1}$, and with the defect generated by the scalar field $\phi$ and potential $\mathcal{V}$ ($\zeta=0$).

\subsection{The Bent Brane Case $\left(p=0\text{, }\Lambda\neq0\right)$}

Considering the bent brane case, Eqs. \eqref{einsteinsepV}-\eqref{einsteinsep2} with $\Lambda\neq0$ and $p=0$, no preliminary assumption about the curvature of $\mathbb{M}^{4}$ (i.e. about $\Lambda \neq 0$) is required. Departing from the metric Eq.~\eqref{metricp=0}, and after some straightforward manipulations involving Eqs.~\eqref{einsteinsepV}-\eqref{einsteinsep2} (for $p=0$), they can be cast in the form of
\begin{align}
C&=\Lambda e^{2 \hat{A}}-e^{2 \hat{h}}\left(4 {\hat{A}_{,v}}{}^{2}-\hat{A}_{,v} \hat{h}_{,v}-\hat{A}_{,vv}\right),\label{bentequation}
\\
\frac{{\zeta_{,v}}^{2}}{4M^{4}}&=\;3 \hat{h}_{,v} \hat{A}_{,v}+3\hat{A}_{,vv}-\Lambda e^{2\hat{A}}e^{-2\hat{h}},\label{bentscalar2}
\end{align}
where $C\in\mathbb{R}$ is the separation constant. Again, the expressions defining the potential $\mathcal{V}$ and the scalar field $\phi$ are given by Eqs.~\eqref{generalpotential2} and \eqref{generalscalar1}, which correspond to the flat brane model with $p=0$.

To solve Eq.~\eqref{bentequation}, one can set:
$i$) either $\zeta_{,v}=0$ $(\zeta=0)$; $ii$) or $C=0$, but $\zeta_{,v}\neq 0$.

It means that only when two scalar fields are present and the brane is not flat that some additional supposition $(C=0)$ about the solution must be made, in all other cases one can generally solve these equations.
In particular, the first case, with $\zeta=0$, is the most interesting one. It corresponds to a model with a single scalar field which drives a smooth behavior with no singularities in the stress energy tensor, which are ingrained in the other configurations ($I$, $II$ and $III$).

\subsubsection{The Single Scalar Field Case (Model $IV$)}
Starting from the constraint imposed by $\zeta=0$, model $IV$ is resumed by the behavior of a single scalar field.
To solve Eqs.~\eqref{bentequation} and \eqref{bentscalar2} one can set $\hat{h}=0$ in order to obtain some simplifications.
Thus, combining Eqs.~ \eqref{bentequation} and \eqref{bentscalar2}, one can write
$$
\frac{\Lambda}{3} e^{2 \hat{A}}-{\hat{A}_{,v}}{}^{2}=\frac{C}{4}\iff \int\frac{\mathrm{d}\hat{A}}{\sqrt{\frac{\Lambda}{3} e^{2\hat{A}}-\frac{C}{4}}}=\pm\left(v+v_{0}\right),
$$
which exhibits three different solution which depends on the values of $\Lambda$ and $C$, i.e. 
\begin{align}
	&\hat{A}=\ln\left(\frac{\sqrt{3}}{\sqrt{\Lambda}\left|v+v_{0}\right|}\right)\text{, if }C=0\text{ and }\Lambda>0,\label{solution1}
	\\
	&\hat{A}=-\ln\left\{2\sqrt{\frac{\left|\Lambda\right|}{3\left|C\right|}}\cosh\left[\frac{\sqrt{\left|C\right|}}{2}\left(v+v_{0}\right)\right]\right\}\text{, if }C,\Lambda<0,\label{solution2}
	\\
	&\hat{A}^{IV}=-\ln\left\{2\sqrt{\frac{\Lambda}{3C}}\left|\cos\left[\frac{\sqrt{C}}{2}\left(v+v_{0}\right)\right]\right|\right\}\text{, if }C,\Lambda>0,\label{solution3}
\end{align}
which are all consistent with Eqs.~\eqref{bentequation} and \eqref{bentscalar2}. 

Clearly, the solutions from \eqref{solution1} and \eqref{solution2} do not depict RS-like features: gravity is not localized along the corresponding extra dimension, unless one could force $v$ to be periodic. Nevertheless, the warp factors \eqref{solution1} and \eqref{solution2} are not periodic and no thin brane can supply the required boundary conditions. Gravity can be localized only by setting $C=\Lambda=0$, which leads to constraining $\hat{A}(v)=\hat{A}_{0}$ ($\hat{A}_{0}\in \mathbb{R}$), and by supposing $v=r\varphi$, with $\varphi\in \mathbb{S}^{1}$. In this case, one has the same trivial case from Eq.~ \eqref{trivialmetric}, which corresponds to a trivial extension of five dimensional brane-worlds.

On the other hand, looking at solution \eqref{solution3}, which is periodic, i.e. with $v=r\varphi$, where $\varphi\in \mathbb{S}^{1}$, one does find more appealing localization features, which emerge from its compact characteristic. Since one expects the metric to be continuous, then the warp factor $e^{-2\hat{A}}$ shall also be periodic and continuous in $\mathbb{S}^{1}$, i.e. (for $v_{0}=0$),
$$
\cos^{2}\left(\frac{\sqrt{C}}{2}r2\pi\right)=\cos^{2}\left(0\right)=1\implies C=\frac{n^{2}}{r^{2}}\text{, }n\in\mathbb{N}^{+},
$$
where one should notice that $n\neq0$, since the warp factor is ill defined for $n=0$.

Since the peculiarities related to the solutions from Eqs.~\eqref{solution1} and \eqref{solution2} have already been discussed, one should pay more attention to the solution from Eq.~\eqref{solution3}.

\begin{figure}[!htb]
	\includegraphics[scale=0.69]{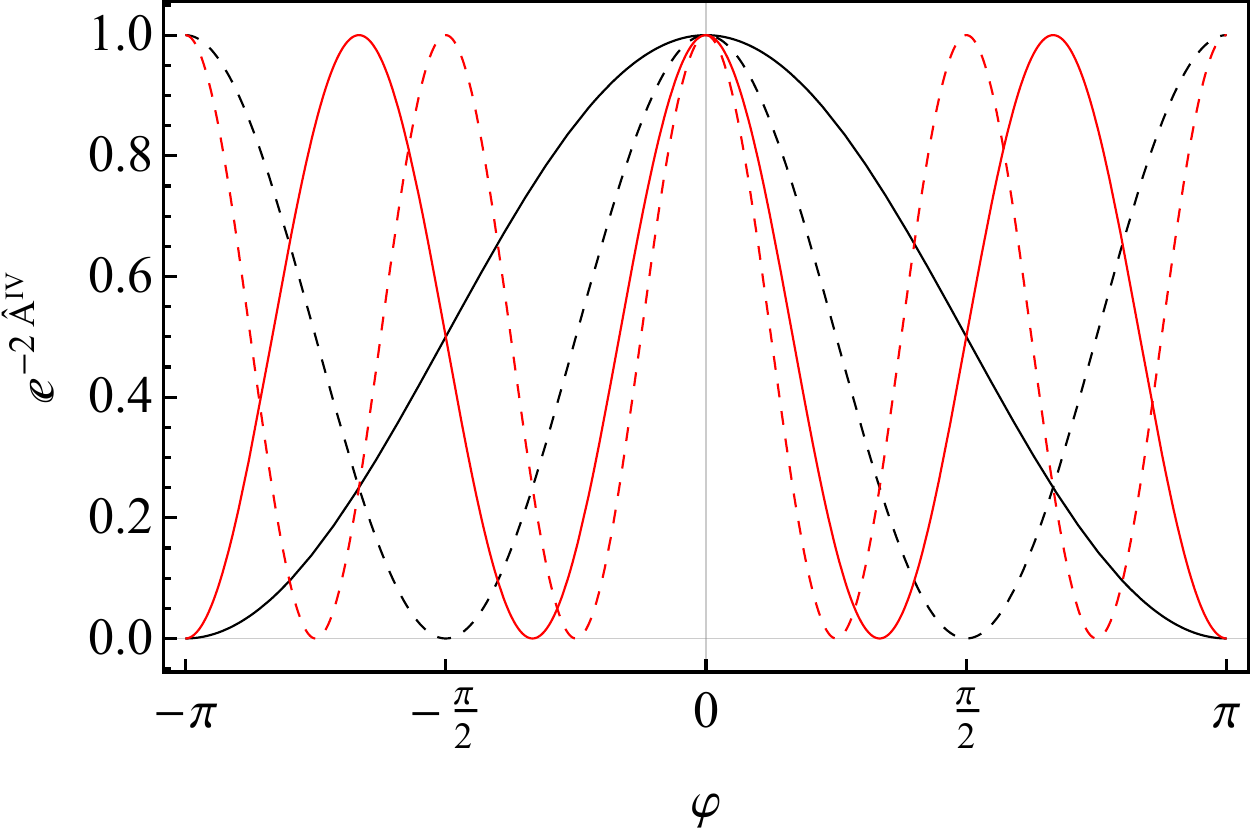}$\;\;$
	\caption{(Color online) Warp factor $e^{-2\hat{A}^{IV}}$ of model $IV$ as a function of $\varphi$, for $n=1$ (black line), $n=2$ (black dashed line), $n=3$ (red line) and $n=4$ (red dashed line).}\label{warpfactorsinglescalar}
\end{figure}

In this case, the related metric, with $v_{0}=0$, is written as
\begin{equation}
\boldsymbol{g}^{IV}= \frac{4r^{2}\Lambda}{3n^{2}}\cos^{2}\left(\frac{n\varphi}{2}\right) e^{-2\tilde{A}}\omega^{+}_{\mu\nu}\mathrm{d}x^{\mu}\otimes\mathrm{d}x^{\nu}+e^{-2\tilde{f}}\mathrm{d}u\otimes\mathrm{d}u+r^{2}e^{-2\tilde{A}}\mathrm{d}\varphi\otimes\mathrm{d}\varphi,\label{bentmetricgeneral}
\end{equation}
which corresponds to the most appealing solutions once some physical conditions are imposed.
In particular, it only works either for a de Sitter brane $(\Lambda>0)$ or, at least, for a space with positive constant curvature. Clearly, since no scalar field $\zeta$ is effective, the energy to achieve such a configuration is finite. Fig.~\ref{warpfactorsinglescalar} depicts the form of the warp factor $e^{-2\hat{A}^{IV}}$, which explains why this model should be more relevant then models $I$, $II$ and $III$: there are no cusps in the warp factor. This corresponds to a straightforward consequence of no singularities in the stress energy tensor.

Even with singularities eliminated from the stress energy tensor, this model still exhibits curvature singularities. Whenever ${\cos^{2}\left({n\varphi}/{2}\right)=0}$ the warp factor is null and the metric would have vanishing components. This could be an effect of a badly defined choice of coordinates, and represent some form of horizon.
In this context, the Kretschmann scalar (see expression \eqref{Kretschmannp=0} in Appendix \ref{Kretschmann}) for model IV reads,
$$
K= e^{4\tilde{A}}\left\{\frac{3n^{2}}{r^{2}}\sec^{2}\left(\frac{n\varphi}{2}\right)\left[\frac{n^{2}}{r^{2}}\tan^{2}\left(\frac{n\varphi}{2}\right)+4\tilde{A}_{,u}{}^{2}\right]+\frac{5n^{4}}{2r^{4}}+e^{-4\tilde{A}}\mathfrak{K}(u)+16\tilde{A}_{,uu}{}^{2}+40\tilde{A}_{,u}{}^{4}-\tilde{A}_{,u}{}^{2}\right\},
$$
which results into curvature singularities, since this scalar invariant is singular whenever ${\cos^{2}\left({n\varphi}/{2}\right)=0}$.

Another interesting property of model $IV$ is the constant ${4\Lambda}/{3C}$ that multiplies the warp factor. This constant can not be removed from the warp factor, otherwise it will not be a solution of Eqs.~\eqref{bentequation} and \eqref{bentscalar2}. Yet if one increases the value of $C$, the warp factor becomes not only more localized, but also exhibits a decreasing amplitude. In fact, one could expect the maximum value of the warp factor to be one, thus one could impose ${4\Lambda}/{3C}=1$. For $\Lambda$ assuming tiny values, one should have tiny values for $C$. Therefore, the warp factor would not be exceptionally localized. Here, no concerns to such relation between $C$ and $\Lambda$ will be made, then $C$ will be regarded as a completely independent value.

One again only lacks the dependence of the warp factor on the $u$ coordinate related to $\tilde{A}$, to the scalar field $\phi$, as well as to the potential $\mathcal{V}$. In this case, the involved fields must satisfy Eqs.~\eqref{generalpotential2} and \eqref{generalscalar1}, with $C=n^{2}/r^{2}$ for model $IV$. Therefore, the dependence of these quantities on $u$ is equivalent to that obtained for the flat brane model $III$, with metric \eqref{flat2metricgeneral}, and with the distinction being only due to the value of $C$: for model $III$, the constant $C=n^{2}/16r^{2}$, $n\in\mathbb{N}$, while for model $IV$, the constant $C=n^{2}/r^{2}$, $n\in\mathbb{N}^{+}$.  Eqs.~\eqref{generalpotential2} and \eqref{generalscalar1} will be solved in a redundant way, for models $III$, $IV$ and $V$ in section \ref{predetermined}.

\subsubsection{The C=0 Case (Model $V$)}

So far one has built models over a flat and de Sitter branes, the whole spectrum of possible values of $\Lambda$ can be filled by anti-de Sitter brane solutions.

\begin{figure}[!htb]
	\subfloat[]{\label{warpfactorbentp=0}\includegraphics[scale=0.69]{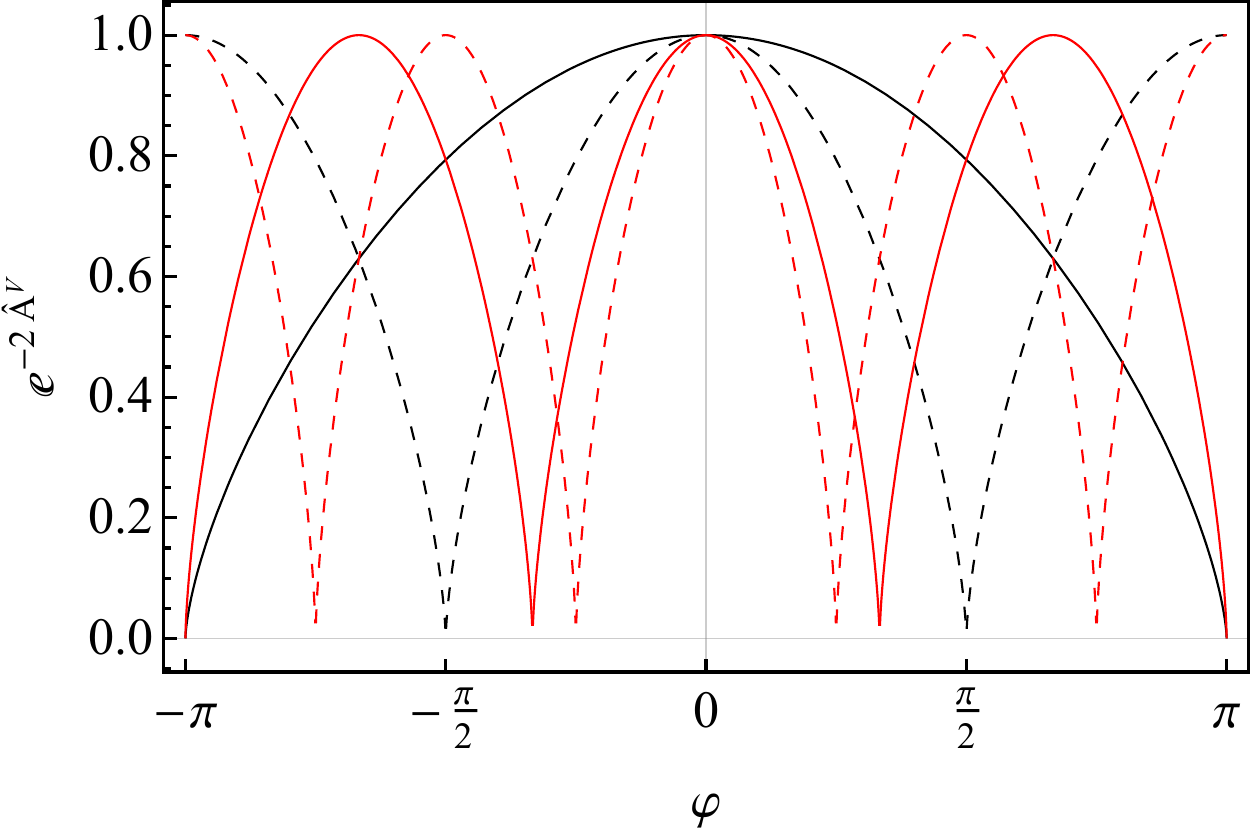}}$\;\;$
	\subfloat[]{\label{stressbentp=0}\includegraphics[scale=0.69]{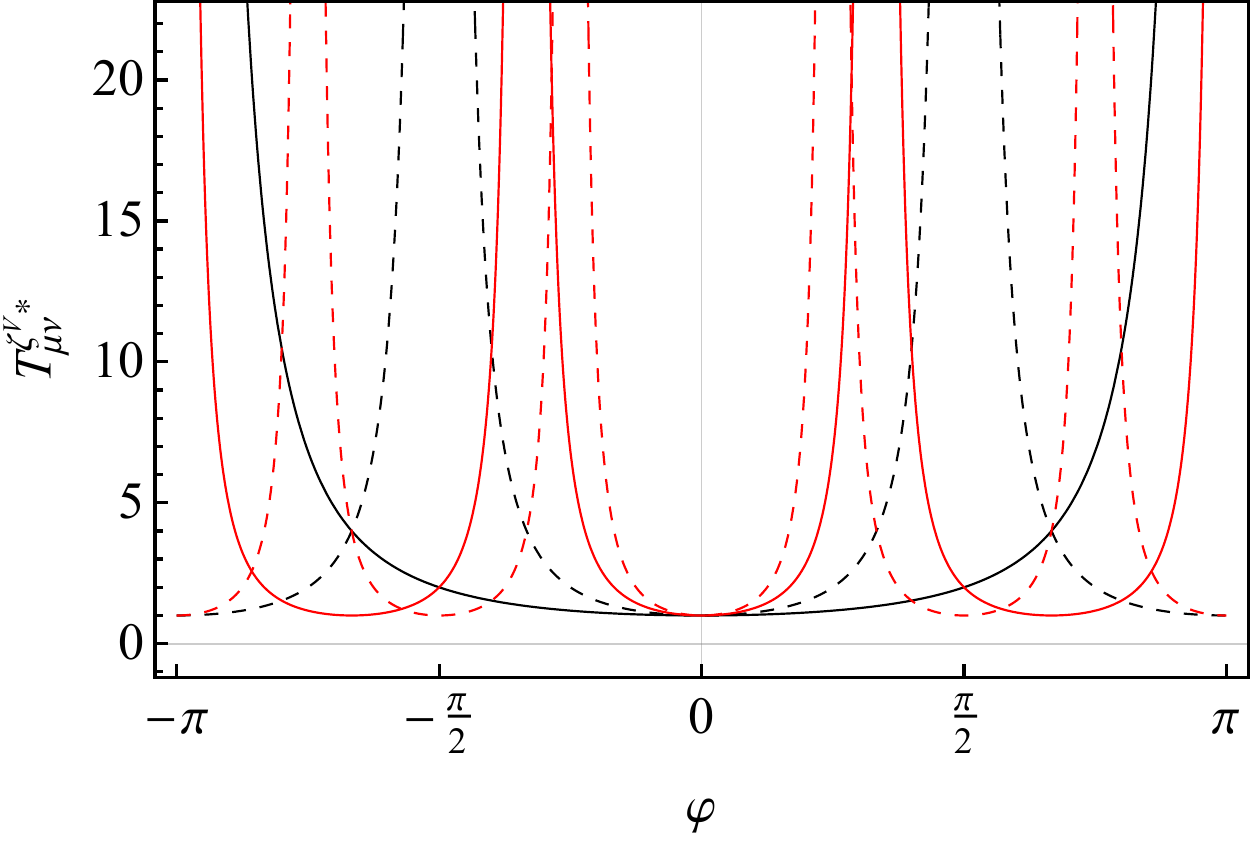}}
	\caption{(Color online) (a) The warp factor $e^{-2\hat{A}^{V}}$ of model $V_{0}$ as a function of $\varphi$. (b) The stress energy tensor $T^{\zeta^{*}}_{\mu\nu}=-T^{\zeta}_{\mu\nu}{3 r^{2}}/{4 M^4 n^2\omega^{-}_{\mu\nu}}$ of model $V$ as a function of $\varphi$. The plots are for $n=1$ (solid black line), $n=2$ (black dashed line), $n=3$ (solid red line) and $n=4$ (red dashed line).}
\end{figure}

To realize analytical solutions of Eq.~\eqref{bentequation} when two scalar fields are present and the brane is bent, i.e. the space-time curvature of $\mathbb{M}^{4}$ is non null, one must constrain $C$ to $0$. Other values of $C$ do not allow strict analytical calculations. A simplified scenario is accomplished by setting $\hat{h}=\hat{A}$ so as to reduce  Eqs.~\eqref{bentequation} and \eqref{bentscalar2} for the warp factor $\hat{A}$ and $\zeta$, respectively, to the now called model $V$, for which
\begin{align}
&\hat{A}^{V}=\hat{A}_{0}-\frac{1}{3}\ln\left|\cos\left[\sqrt{3\left|\Lambda\right|}\left(v+v_{0}\right)\right]\right|,\label{warpfactor3}
\\
&\zeta^{V}=\pm\frac{4 M^2}{\sqrt{3}}\operatorname{arctanh}\left\{\sin \left[\sqrt{3\left|\Lambda\right|} \left(v+v_0\right)\right]\right\},\label{scalar3}
\end{align}
with $\Lambda < 0$. The solution $\hat{A}$ for positive values of $\Lambda$ does not exhibit RS-like features. Due to the periodicity of $\hat{A}^{V}$, one is able to choose $v=r\varphi$, where $\varphi\in\mathbb{S}^{1}$. Since the metric must be continuous, one must also have $e^{-2\hat{A}}$ continuous in $\mathbb{S}^{1}$, therefore
$$
\left[\cos \left(\sqrt{3\left|\Lambda\right|} r2\pi\right)\right]^{2/3}=1\implies r=\frac{n}{2\sqrt{3\left|\Lambda\right|}}\text{, }n\in\mathbb{N}^{+},
$$
where simplified expressions are yielded from choosing $\hat{A}_{0}=0$ and $v_{0}=0$.
For such a completely contrasting result, obviously there is no relation between $C$ and the radius $r$ of $\mathbb{S}^{1}$, as well as $\Lambda$ is a free parameter. In fact, the radius $r$ of $\mathbb{S}^{1}$ is constrained by the value of the cosmological constant $\Lambda$ one chooses for the space-time $\mathbb{M}^{4}$, and the metric is written as
\begin{equation}
\boldsymbol{g}^{V}=\cos^{2/3}\left(\frac{n\varphi}{2}\right)e^{-2\tilde{A}}\omega^{-}_{\mu\nu}\mathrm{d}x^{\mu}\otimes\mathrm{d}x^{\nu}+e^{-2\tilde{f}}\mathrm{d}u\otimes\mathrm{d}u+r^{2} \cos^{2/3}\left(\frac{n\varphi}{2}\right)e^{-2\tilde{A}}\mathrm{d}\varphi\otimes\mathrm{d}\varphi,\label{bent2metricgeneral}
\end{equation}
which expresses a compactified setup for an anti-de Sitter brane $(\Lambda < 0)$ scenario at $\mathbb{M}^{4}$, with constant negative curvature and two scalar fields. In Fig.~\ref{warpfactorbentp=0} the form of the warp factor is exhibited for different values of $n$. The form of the scalar field is exactly the same as the one depicted in Fig.~\ref{zetaI}. But the scalar field now depends on the $\varphi$ coordinate, which is different from that one considered in model $I$.

Fig.~\eqref{stressbentp=0} depicts the stress energy tensor, ${T^{\zeta}}_{\mu\nu}$, for the scalar field $\zeta$,
$$
T^{\zeta^{V}}_{\mu\nu}=-\frac{4 M^4 n^2\omega^{-}_{\mu\nu}}{3 r^{2}} \sec^{2}\left(\frac{n \varphi }{2}\right),
$$
which evidently exhibit singularities correlated to the number of cusps exhibited by the warp factor. 
Again, from the perspective of the bulk, one has the finite formation energy given by
\begin{align*}
E^{\zeta^{V}}_{\mu\nu}&=\int_{\mathbb{E}^{6}}T^{\zeta^{V}}_{\mu\nu}\sqrt{-\mathrm{g}}\mathrm{d}^{6}x\propto\int^{\pi}_{-\pi}\sec^{1/3}\left(\frac{n \varphi }{2}\right)\mathrm{d}\varphi.
\end{align*}

To complete the model, one notices that the fields must satisfy Eqs.~\eqref{generalpotential2} and \eqref{generalscalar1} with $C=0$, such that
\begin{align*}
\frac{\mathcal{V}}{8M^{4}}&= e^{2 \tilde{f}}\left(-5{\tilde{A}_{,u}}{}^{2}+\tilde{f}_{,u} \tilde{A}_{,u}+\tilde{A}_{,uu}\right),
\\
\frac{{\phi_{,u}}^{2}}{4M^{4}}&=4\tilde{f}_{,u}\tilde{A}_{,u}+4\tilde{A}_{,uu}.
\end{align*}

By choosing coordinates such that $\tilde{f}=0$, one recovers the same equations, up to some constants, as in the five dimensional thick brane-worlds with a single scalar field. It means that, one more time one has a non-trivial extension of the usual five dimensional brane-world models, which can be ratified by setting $\tilde{f}=n=0$ into Eqs.~ \eqref{bent2metricgeneral}, \eqref{warpfactor3} and \eqref{scalar3}.

\section{Setups From Predetermined Internal Spaces}\label{predetermined}

In the previous sections, a first subset of models $I$ and $II$ for intersecting thick branes was obtained and discussed in terms of the model degenerate dependence on a single co-dimensional coordinate $v \leftrightarrow u$. A second subset, for models $III$, $IV$ and $V$, which include a splitted dependence between $v$ and $u$ and admit some additional freedom in the choice of the field parameters $\tilde{A}$, $\phi$, and $\mathcal{V}$, has also been evaluated.
In this section, the hypothesis of constraining such additional degree of freedom by imposing a geometry for $\left(\mathbb{B}^{2},\sigma\right)$ shall be considered.

As previously argued, Eqs.~\eqref{generalpotential2} and \eqref{generalscalar1} form a common set of equations for all the $p=0$ models. These two equations involve three field parameters $\tilde{A}$, $\phi$, and $\mathcal{V}$. Due to the remnant degree of freedom, Eqs.~\eqref{generalpotential2} and \eqref{generalscalar1} can be recast in to a first order configuration (see \cite{Afonso2006}) to be solved.
Given that $p=0$, one finds that the metric of the internal space $\mathbb{B}^{2}$ takes the form of
\begin{equation}
\boldsymbol{\sigma}=e^{-2\tilde{f}}\mathrm{d}u\otimes\mathrm{d}u+e^{-2\tilde{A}}e^{-2\hat{h}}\mathrm{d}v\otimes\mathrm{d}v.\label{metricpredetermined}
\end{equation}
Thus the choice of $\tilde{A}$ and $\tilde{f}$ fixes the geometry of $\mathbb{B}^{2}$, since $\hat{h}$ is nothing but a choice of coordinates which has been previously specified for each model. That makes choosing the field $\tilde{A}$ a better option than fixing either $\phi$ or $\mathcal{V}$, in manner that one can achieve an intended geometry. For this reason, these spaces have a predetermined geometry, since one does not determine it from the field equations, but chooses $\tilde{A}$ and $\tilde{f}$ such that an expected geometry is achieved. As long as one is able to cast the metric of the internal space as \eqref{metricpredetermined}, the geometrical interpretation that follows is straightforward. The tricky thing here is finding a combination of $\tilde{A}$ and $\tilde{f}$ that allows for the integration at Eq.~\eqref{generalscalar1}. In the following subsections, solutions to these equations will be provided by a choice of the metric of the internal space $\mathbb{B}^{2}$ that allows for the respective analytical integration of Eq.~\eqref{generalscalar1}.

In particular, when $\hat{h}=0$ and the coordinate $v$ is compactified as $\mathbb{S}^{1}$ (models $III$ and $IV$) one is able to cast the metric \eqref{metricpredetermined} in a particular fashion so that the internal space could be a sphere or Spheroid (subsecs.~\ref{modelsphere} and \ref{modelspheroid}). Since models $III$ and $IV$ have the same common geometry and topology for the internal space, the solutions that follow are common to both of them. Model $V$ can also share these specific solutions for $\tilde{A}$, $\phi$ and $\mathcal{V}$, but the applied geometrical interpretation shall not be valid in the latter cases.

\subsection{Solving Eqs.~\eqref{generalpotential2} and \eqref{generalscalar1}}

When one chooses coordinates such that $\tilde{f}=0$, Eqs.~\eqref{generalpotential2} and \eqref{generalscalar1} are similar in structure to the equations that define five dimensional bent braneworlds \cite{Gremm2000,Afonso2006,Sasakura2002}, reminded that there are some constraints imposed by the separation constants, $\Lambda$ and $C$. Therefore, a departure solution as, for instance, due to Ref.~\cite{Gremm2000},
\begin{equation}
\tilde{A}=-\ln\Big|\cos\big[a\left(u+u_{0}\right)\big]\Big|,\label{generalwarp}
\end{equation}
can be considered.
Notice that one is able to choose $a\left(u+u_{0}\right)=\theta$, with $\theta \in \left[-\pi/2,\pi/2\right]$, as long as one allows for the singularities at $\pm\pi/2$ for the scalar field. Thus one is able to consider $u$ to be compactified as $\mathbb{S}^{1}$, just imposing $a=1/2r$, where $r$ is the radius of $\mathbb{S}^{1}$. It allows one to run $\theta$ from $-\pi$ to $\pi$. As an example, for the sphere models that follow, one regards $a=1/r$, $u_{0}=-r\pi/2$, and thus $u\in\left[0,r \pi\right]$.

\begin{figure}[!htb]
	\includegraphics[scale=0.65]{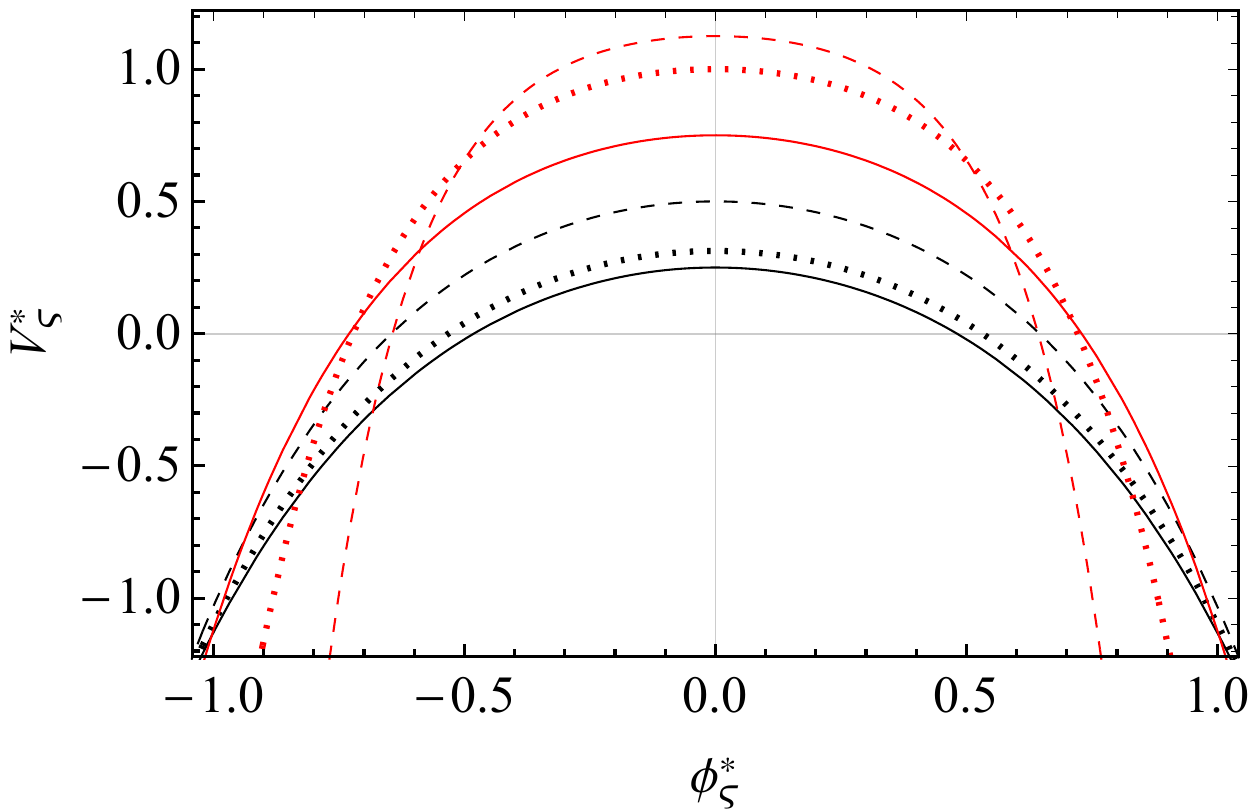}
	\caption{(Color online) Potential $\mathcal{V}_{\varsigma}^{*}=\mathcal{V}_{\varsigma}/32M^{4}a^{2}$ as a function of the scalar field $\phi_{\varsigma}^{*}={\phi_{\varsigma}}/{4M^{2}}$. The plots are for $C=0$ (solid black line), $C=a^{2}/4$ (dotted black line), $C=a^{2}$ (dashed black line), $C=2a^{2}$ (solid red line), $C=3a^{2}$ (dotted red line), $C=7a^{2}/2$ (dashed red line).}\label{potential(phi)}
\end{figure}

From Eqs.~\eqref{generalpotential2} and \eqref{generalscalar1}, scalar field, $\phi$, and potential, $\mathcal{V}$, are cast as
\begin{align}
\phi_{\varsigma}&=\pm 2 M^{2}\sqrt{4-\frac{C}{a^{2}}}\operatorname{arctanh}\Big\{\sin\big[a\left(u+u_{0}\right)\big]\Big\},
\\
\mathcal{V}_{\varsigma}&=8M^{4}a^{2}\Bigg\{5-\left(4-\frac{C}{a^{2}}\right)\sec^{2}\big[a\left(u+u_{0}\right)\big]\Bigg\}.
\end{align}
which, in this case, allows for an explicit correspondence given by
$$
\mathcal{V}_{\varsigma}=8M^{4}a^{2}\Bigg\{5-\left(4-\frac{C}{a^{2}}\right)\cosh^{2}\bigg[\frac{a\phi}{2M^{2}\sqrt{4a^{2}-C}}\bigg]\Bigg\}.
$$
When $C=4a^{2}$ the scalar field $\phi_{\varsigma}$ is null and the potential $\mathcal{V}_{\varsigma}$ is a constant, thus one either has either a single scalar field $\zeta$, as for models $III$ and $V$, or no scalar field, as for model $IV$. For model $IV$, since no scalar field is present, $\zeta=\phi=0$, the potential is a constant. Once returning to Einstein equations, one then finds $G_{MN}=-{5C}g_{MN}/{2}$, and thus $\mathbb{E}^{6}$ is nothing but a de Sitter space of six dimensions $\left(\mathbbm{d}\mathbb{S}^{6}\right)$ written in some unusual system of coordinates. The form of the scalar field $\phi_{\varsigma}$ can be seen in Fig.~\ref{zetaI}, while the potential $\mathcal{V}_{\varsigma}$ is depicted in Fig.~\ref{potential(phi)}.

The corresponding metric of such configuration is given by
\begin{equation}
\boldsymbol{g}^{J}_{\varsigma}=\cos^{2}\big[a\left(u+u_{0}\right)\big]e^{-2\hat{A}^{J}}\omega_{\mu\nu}\mathrm{d}x^{\mu}\otimes\mathrm{d}x^{\nu}+\cos^{2}\big[a\left(u+u_{0}\right)\big]e^{-2\hat{h}^{J}}\mathrm{d}v\otimes\mathrm{d}v+\mathrm{d}u\otimes\mathrm{d}u.\label{metricpredeterminedgeneral}
\end{equation}
where the index $J$ in $\boldsymbol{g}^{J}_{\varsigma}$, $\hat{A}^{J}$ and $\hat{h}^{J}$ refers to one of the models $III$, $IV$ or $V$ (i.e. $J=III$, $\hat{A}^{J}=\hat{A}^{III}$, refers to the warp factor of model $III$). See that the warp factor from Eq.~\eqref{metricpredeterminedgeneral} exhibit the same pattern as for model $IV$ \eqref{bentmetricgeneral} cf. Fig.~\ref{warpfactorsinglescalar}.

Other configurations can also be achieved by choosing $\tilde{f}$ to be non-null, thus even if the warp factor given by Eq.~\eqref{generalwarp}, the configuration would be different. As an example, one may consider the following choice of $\tilde{f}$,
$$
\tilde{f}=-\frac{1}{2}\ln\bigg\{1-\kappa\cos^{2}\big[a\left(u+u_{0}\right)\big]\bigg\},
$$
where $\kappa$ is a constant such that $\kappa\in\left(0,1\right)$. For $\kappa=0$, one recovers the metric from \eqref{metricpredeterminedgeneral}. As it shall be clarified in the following subsection, 
this choice corresponds to a reduction of the spheroid model, for which
\begin{equation}
\boldsymbol{g}^{J}_{\epsilon}=\cos^{2}\left[a\left(u+u_{0}\right)\right]\left(e^{-2\hat{A}^{J}}\omega_{\mu\nu}\mathrm{d}x^{\mu}\otimes\mathrm{d}x^{\nu}+e^{-2\hat{h}^{J}}\mathrm{d}v\otimes\mathrm{d}v\right)+\left\{1-\kappa\cos^{2}\big[a\left(u+u_{0}\right)\big]\right\}\mathrm{d}u\otimes\mathrm{d}u,\label{generalspheroid}
\end{equation}
where $\hat{A}^{J}$ and $\hat{h}^{J}$ could be any of the functions determined in models $III$, $IV$ or $V$.

Analytical solution for Eq.~\eqref{generalscalar1} are constrained by the choice of $C=4a^{2}$, i.e. with $J=III$ and $IV$ at \eqref{generalspheroid}. In this cases, upon an integration of Eq.~\eqref{generalscalar1}, one has
\begin{equation}
\frac{\phi_{\epsilon}}{4M^{2}}=\mp\sqrt{1-\kappa} \operatorname{arctanh} \left(\frac{\sqrt{\kappa } \sin \left[\frac{\sqrt{C} \left(u+u_{0}\right)}{2}\right]}{\sqrt{1-\kappa \cos^{2} \left[\frac{\sqrt{C} \left(u+u_{0}\right)}{2}\right]}}\right),\label{scalarspheroidgeneral}
\end{equation}
\begin{equation}
\frac{\mathcal{V}_{\epsilon}}{2CM^{4}}=\frac{5\left(1-\kappa\right)-4\kappa\left(1-\kappa\right)\cos^{2}\left[\frac{\sqrt{C}\left(u+u_{0}\right)}{2}\right]}{\left\{1-\kappa\cos^{2}\left[\frac{\sqrt{C}\left(u+u_{0}\right)}{2}\right]\right\}^{2}},
\end{equation}
and no longer does the scalar field shall exhibit singularities at $\pm\pi/2$.
\begin{figure}[!htb]
	\subfloat[]{\includegraphics[scale=0.69]{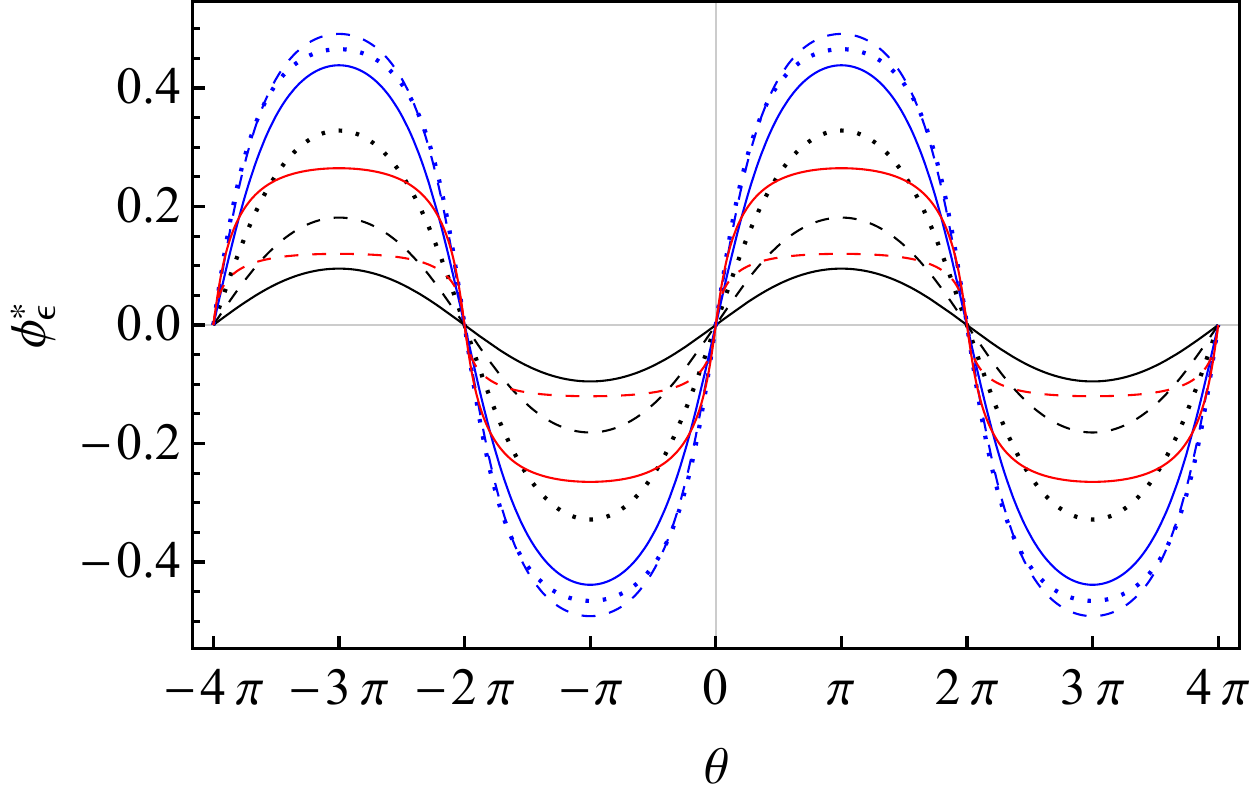}}$\;\;$
	\subfloat[]{\label{potential(scalar)}\includegraphics[scale=0.69]{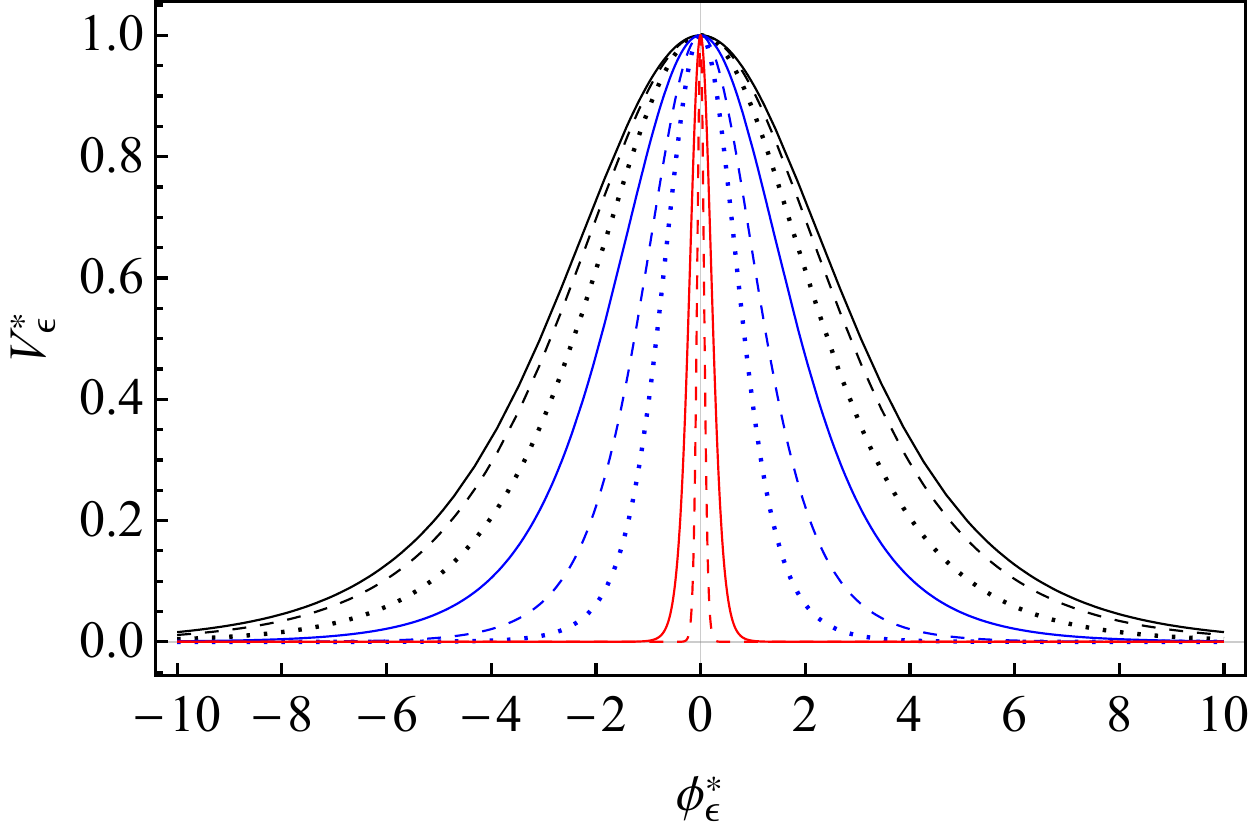}}
	\caption{(Color online) (a) Scalar field $\phi^{*}_{\epsilon}=\phi_{\epsilon}/4M^{2}$ as a function of $\theta=a\left(u+u_{0}\right)$. (b) Potential $\mathcal{V}^{*}_{\epsilon}=\left(1-\kappa\right)\mathcal{V}_{\epsilon}/{2CM^{4}}\left(5-4\kappa\right)$ as a function of $\phi^{*}_{\epsilon}=\phi_{\epsilon}/4M^{2}\sqrt{1-\kappa}$. The plots are for $\kappa=0.1$ (solid black line), $\kappa=0.2$ (dashed black line), $\kappa=0.4$ (dotted black line), $\kappa=0.6$ (solid blue line), $\kappa=0.8$ (dashed blue line), $\kappa=0.9$ (dotted blue line), $\kappa=0.99$ (solid red line) and $\kappa=0.999$ (dashed red line).}\label{scalargeneral}
\end{figure}
It is straightforward to invert the expression for $\phi$ so as to write the potential $\mathcal{V}$ as a function of $\phi$. After some forthright manipulations one finds,
\begin{equation}
\mathcal{V}_{\epsilon}=2CM^{4}\left[\frac{1+4\left(1-\kappa\right)\cosh^{2}\left(\frac{\phi_{\epsilon}}{4M^{2}\sqrt{1-\kappa} }\right)}{\left(1-\kappa\right)\cosh^{4}\left(\frac{ \phi_{\epsilon}}{4M^{2}\sqrt{1-\kappa} }\right)}\right],\label{spheroidpotentialphi}
\end{equation}
from which scalar field and potential forms are depicted in Fig.~\ref{scalargeneral}.

\begin{figure}[!htb]
	\subfloat[]{\includegraphics[scale=0.69]{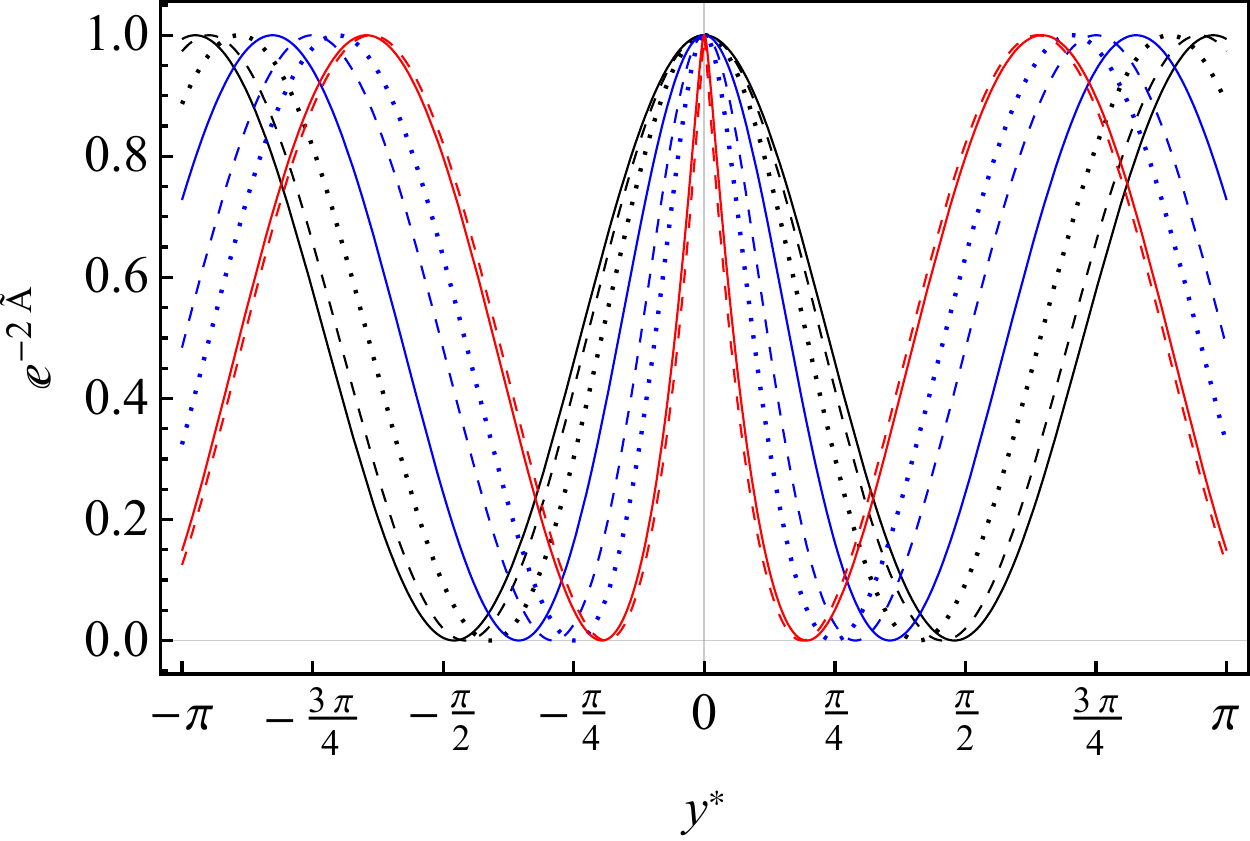}}$\;\;$
	\subfloat[]{\includegraphics[scale=0.69]{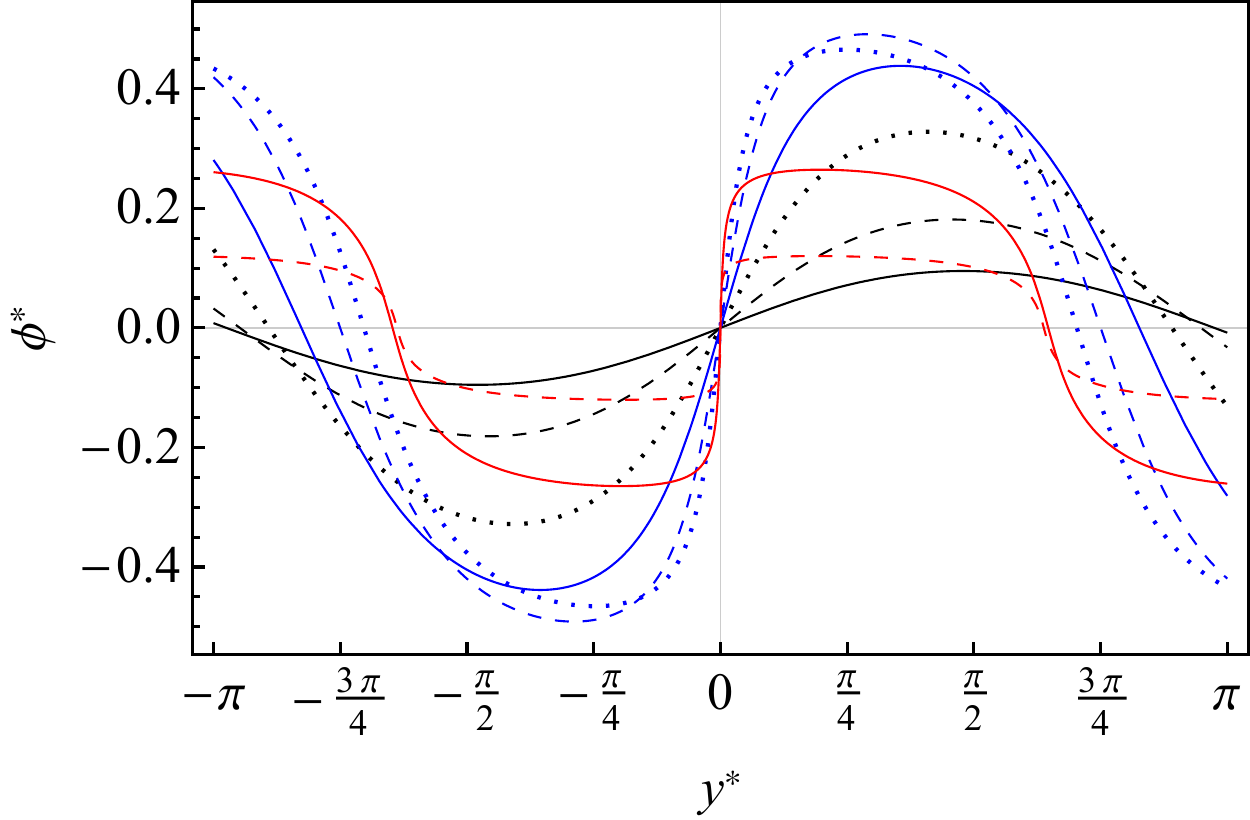}}
	\caption{(Color online) (a) Warp factor $e^{-2\tilde{A}}$ as a function of $y^{*}=\sqrt{C}y/2$. (b) Scalar field $\phi^{*}_{\epsilon}=\phi_{\epsilon}/4M^{2}$ as a function of $y^{*}=\sqrt{C}y/2$. The plots are for $\kappa=0.1$ (solid black line), $\kappa=0.2$ (dashed black line), $\kappa=0.4$ (dotted black line), $\kappa=0.6$ (solid blue line), $\kappa=0.8$ (dashed blue line), $\kappa=0.9$ (dotted blue line), $\kappa=0.99$ (solid red line) and $\kappa=0.999$ (dashed red line).}\label{kappawarpscalar}
\end{figure}

The warp factor for the metric \eqref{generalspheroid} is exactly the same as in \eqref{metricpredeterminedgeneral}, but due to the contribution from $g_{uu}$, the $u$ coordinate has a different meaning. Thus it would be interesting to change coordinates to be able to better compare how the metric \eqref{generalspheroid} fares against the one from \eqref{metricpredeterminedgeneral}. To this end, one chooses a new coordinate $y$, with such luck that
$$
\mathrm{d}y=\sqrt{1-\kappa\cos^{2}\left[\frac{\sqrt{C}}{2}\left(u+u_{0}\right)\right]}\mathrm{d}u,
$$
upon which, after an integration, one finds
\begin{equation}
y=\frac{2 \sqrt{1-\kappa}}{\sqrt{C} }\,E\left(\frac{\sqrt{C}}{2} (u+u_{0})\left|\frac{\kappa }{\kappa -1}\right.\right),\label{variblechangespheroid}
\end{equation}
where $E\left(x\left|m\right.\right)$ is the elliptic integral of second kind. 
The inverted expression results into 
$$
\frac{\sqrt{C}}{2} (u+u_{0})=E^{-1}\left(\frac{\sqrt{C} y}{2 \sqrt{1-\kappa}}\left|\frac{\kappa }{\kappa -1}\right.\right),
$$
where $E^{-1}$ is the inverse function of the elliptic integral of second kind. Then one may write the metric \eqref{generalspheroid} in the term of the coordinate $y$ as
\begin{equation}
\boldsymbol{g}^{J}_{\epsilon}=\cos^{2}\left[E^{-1}\left(\frac{\sqrt{C} y}{2 \sqrt{1-\kappa}}\left|\frac{\kappa }{\kappa -1}\right.\right)\right]\left(e^{-2\hat{A}^{J}}\omega_{\mu\nu}\mathrm{d}x^{\mu}\otimes\mathrm{d}x^{\nu}+e^{-2\hat{h}^{J}}\mathrm{d}v\otimes\mathrm{d}v\right)+\mathrm{d}y\otimes\mathrm{d}y.\label{generalspheroid(y)}
\end{equation}
Finally, the warp factor $e^{-2\tilde{A}}$ and the scalar field $\phi$ as functions of $y$ can be seen in Fig.~\ref{kappawarpscalar}. Clearly, from Fig.~\ref{kappawarpscalar}, as $\kappa$ gets closer to $1$, the warp factor becomes more localized, and in the limit of $\kappa$ going to $1$, a thin brane is recovered. Hence $\kappa$ is the localizing parameter in this model: as it gets closer to $1$, the brane should be closer to a thin brane and the matter distribution in this model should look more like a cusped function, which can only be realized by looking at metric \eqref{generalspheroid(y)}. Otherwise, one generally prefers to work with \eqref{generalspheroid} since a straightforward  geometrical interpretation is achieved when one applies this geometry to $\mathbb{S}^{2}$.

Note that the setup from Eq.~\eqref{generalspheroid} could also be considered in the five dimensional context, since the equations are, up to some constant, equivalent. From the previous choice of $\tilde{f}$, one can thus construct some novel models of bent branes in five dimensions, since the metric is just given by
\begin{equation}
\boldsymbol{g}=\cos^{2}\left[\frac{\sqrt{\left|\Lambda\right|}}{2}\left(u+u_{0}\right)\right]\omega_{\mu\nu}\mathrm{d}x^{\mu}\mathrm{d}x^{\nu}+\left\{1-\kappa\cos^{2}\left[\frac{\sqrt{\left|\Lambda\right|}}{2}\left(u+u_{0}\right)\right]\right\}\mathrm{d}u^{2},\label{spheroidbased4+1}
\end{equation}
where $\Lambda$ is the curvature of space-time $\left(\mathbb{M}^{4},\boldsymbol{\omega}\right)$.

\subsection{The Sphere Models}\label{modelsphere}

An interesting application of the models constructed in previous sections is concerned with the possibility of constructing brane-worlds over $\mathbb{S}^{2}$.

The sphere models, for instance, starts with the assumption that the internal space $\left(\mathbb{B}^{2},\boldsymbol{\sigma}\right)$ is a sphere, or in other words, $\left(\mathbb{B}^{2},\boldsymbol{\sigma}\right)\equiv\left(\mathbb{S}^{2},\boldsymbol{\varsigma}\right)$, where (cf. Eq.~\eqref{sphere})
\begin{equation}
\boldsymbol{\varsigma}=r^{2}\mathrm{d}\theta\otimes\mathrm{d}\theta+r^{2}\sin^{2}\left(\theta\right)\,\mathrm{d}\varphi\otimes\mathrm{d}\varphi.\label{sphere2}
\end{equation}
In this case, one has chosen $u\equiv r\theta$, $\tilde{f}\equiv0$ and $\tilde{A}\equiv-\ln\left[\sin\left(\theta\right)\right]$, as well as $\varphi\in\left[-\pi,\pi\right]$ and $\theta\in\left[0,\pi\right]$. 
This choice corresponds to exactly the same as the one from Eq.~\eqref{generalwarp}, where now one choses $u_{0}=\pi/2r$, $a=1/r$ and $u$ only takes values at the subinterval $\left[0\,,\,r\pi\right]$. See that this choice for $\tilde{f}$ and $\tilde{A}$ is also allowed for model $V$, which however does not have the internal space metric as from Eq.~\eqref{sphere2}. For this reason, model $V$ will be disregarded in this section.

Turning to the point from Eq.~\eqref{sphere2}, Eqs.~\eqref{generalpotential2} and \eqref{generalscalar1} are easily solved so as to return the quantities
\begin{align*}
& \mathcal{V}=\frac{8M^{4}}{r^{2}}\left[5-4\left(1-\frac{Cr^{2}}{4}\right)\csc^{2}\theta\right],
\\
&\phi=\pm 4M^{2}\sqrt{1-\frac{Cr^{2}}{4}}\ln\left[\tan\left(\frac{\theta}{2}\right)\right],
\end{align*}
such that the potential as a function of $\phi$ is given by
$$
\mathcal{V}=\frac{8M^{4}}{r^{2}}\left\{5-\left(4-Cr^{2}\right)\cosh^{2}\left[\frac{\phi}{4M^{2}\sqrt{1-\frac{Cr^{2}}{4}}}\right]\right\}.
$$
For $\phi$ read as a real scalar field, one has
$$
1-\frac{Cr^{2}}{4}\geq0\iff C\leq \frac{4}{r^{2}},
$$
from which, for models $III$ and $IV$, the constraints over $C$ restrict the number of possible models to its dependence on the value of $n$,
\begin{enumerate}
	\item $C^{III}={n^{2}}/{16r^{2}}\implies n\in\left\{0,1,2,3,4,5,6,7,8\right\}$;
	\item $C^{IV}={n^{2}}/{r^{2}}\implies n\in\left\{1,2\right\}$.
\end{enumerate}

Thus one can have, for model $III$, nine different configurations for the scalar field and potential, each for different values of $n$. Meanwhile, for model $IV$, there are only two different configurations.

\begin{figure}[!htb]
	\subfloat[]{\label{figure1}\includegraphics[scale=0.69]{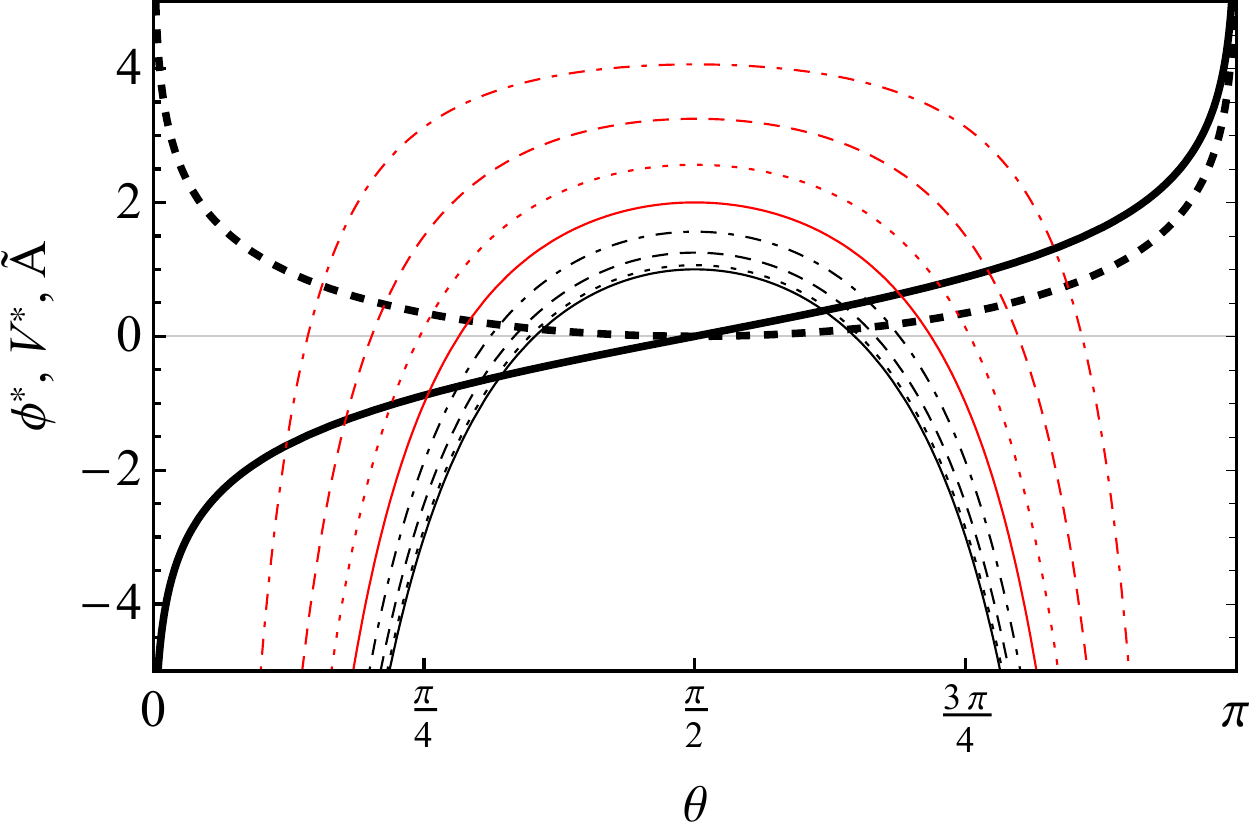}}$\;\;$
	\subfloat[]{\label{stresssphere}\includegraphics[scale=0.69]{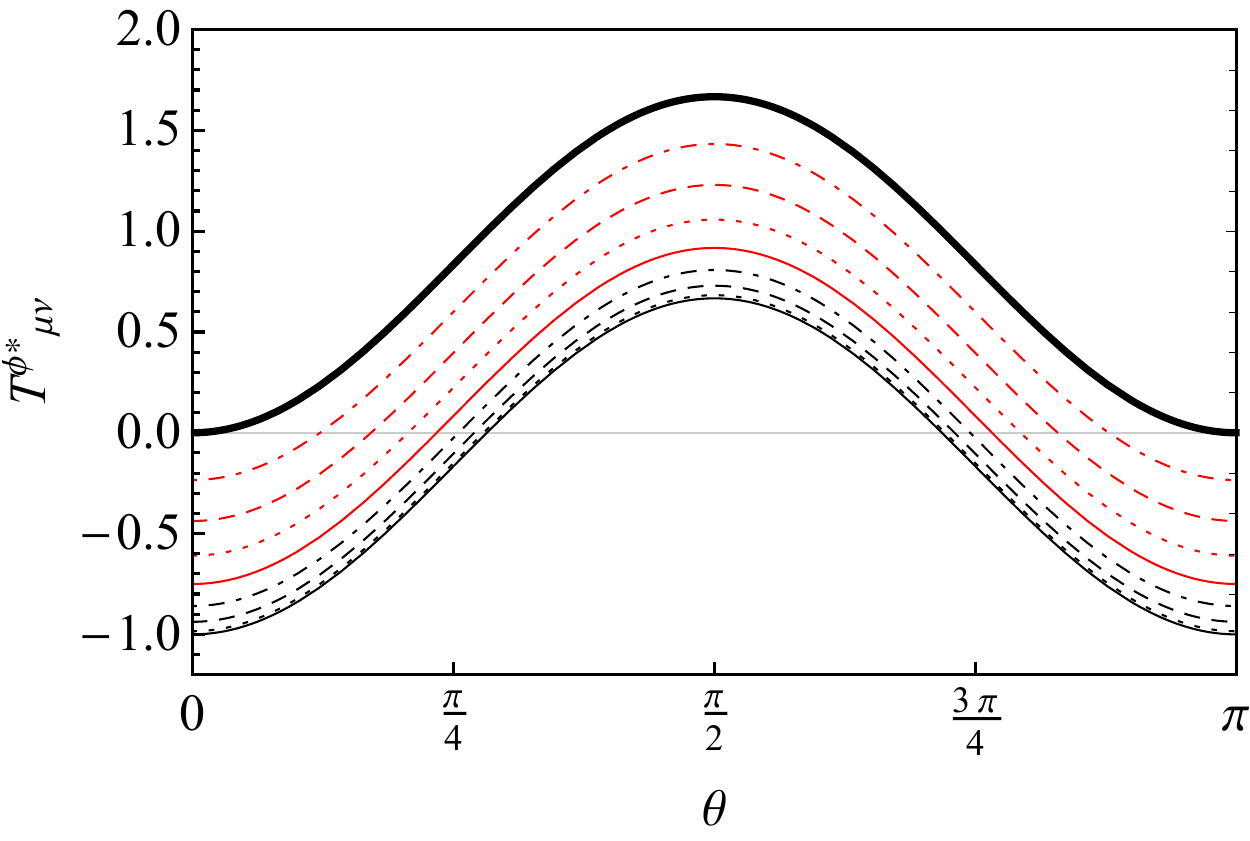}}
	\caption{(Color online) (a) Scalar field $\phi^{*}=\phi/4\sqrt{1-{C}/{4}}$ (thick black line), warp factor $\tilde{A}$ (thick black dashed line) and potential $\mathcal{V}^{*}=\mathcal{V}/4$. (b) The $\theta$ dependence of the stress energy tensor ${T^{\phi}}_{\mu\nu}$, ${T^{{\phi}^{*}}}_{\mu\nu}=-\sin^{2}\theta\;{{T^{\phi}}^{\mu}}_{\nu}/8$. The plots are for $C=0$ (thin black line), $C=1/16$ (thin black dotted line), $C=1/4$ (thin black dashed line), $C=9/16$ (thin black dot-dashed line), $C=1$ (thin red line), $C=25/16$ (thin red dotted line), $C=9/4$ (thin red dashed line), $C=49/16$ (thin red dot-dashed line) and $C=4$ (thick black solid line), with $M=r=1$.}
\end{figure}

\begin{figure}[!htb]
	\includegraphics[scale=0.27]{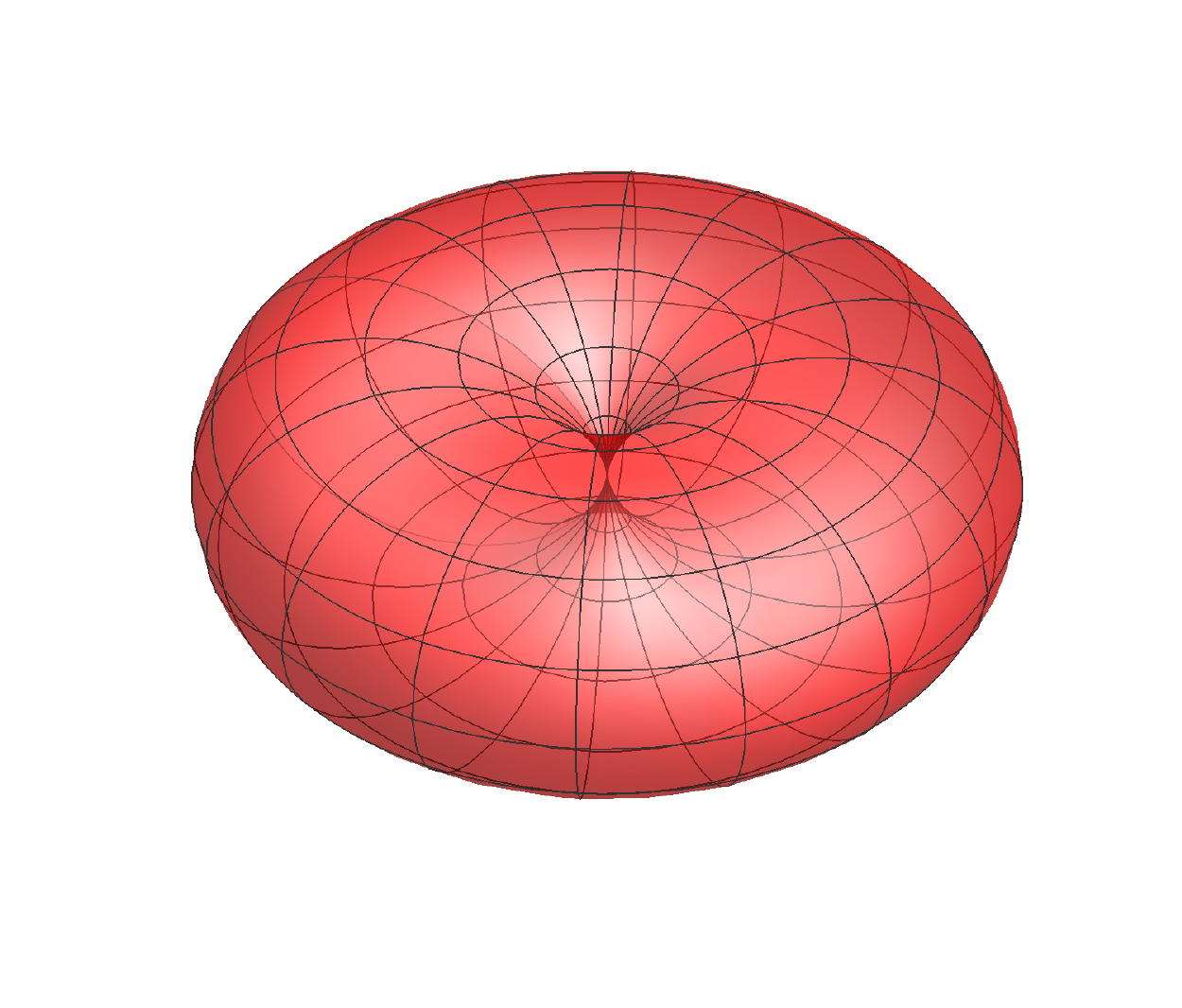}
	\includegraphics[scale=0.27]{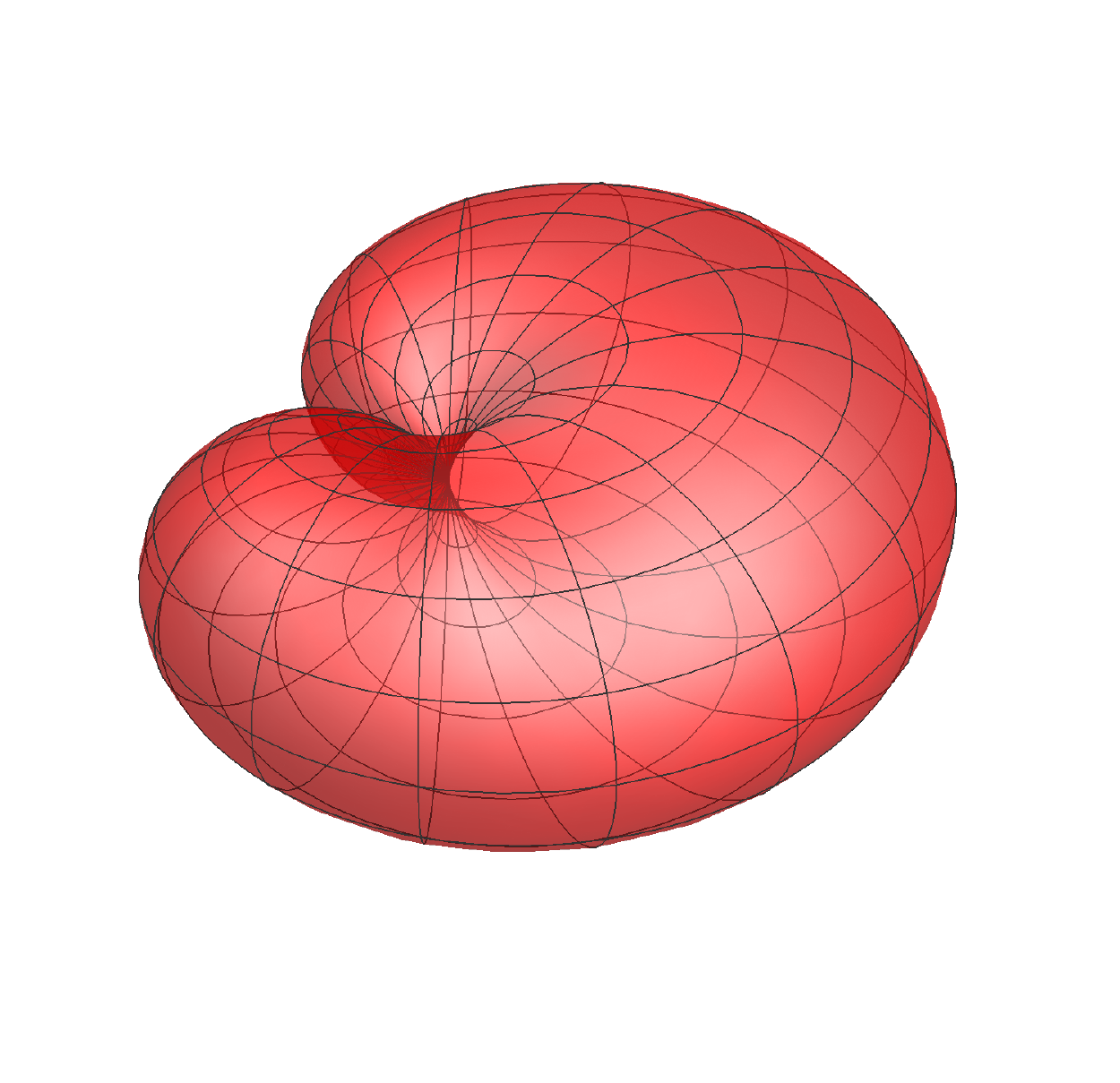}
	\includegraphics[scale=0.27]{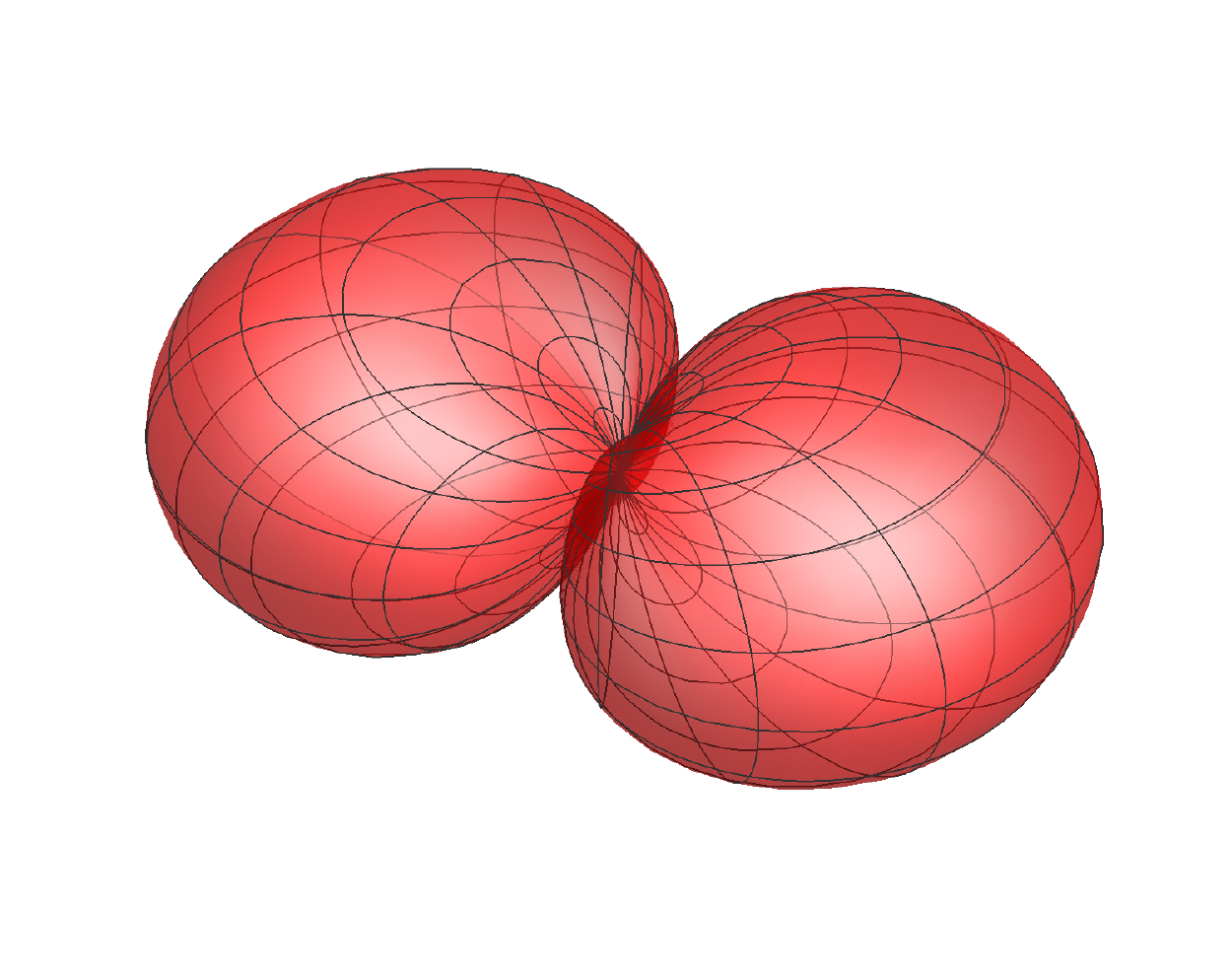}
	\includegraphics[scale=0.27]{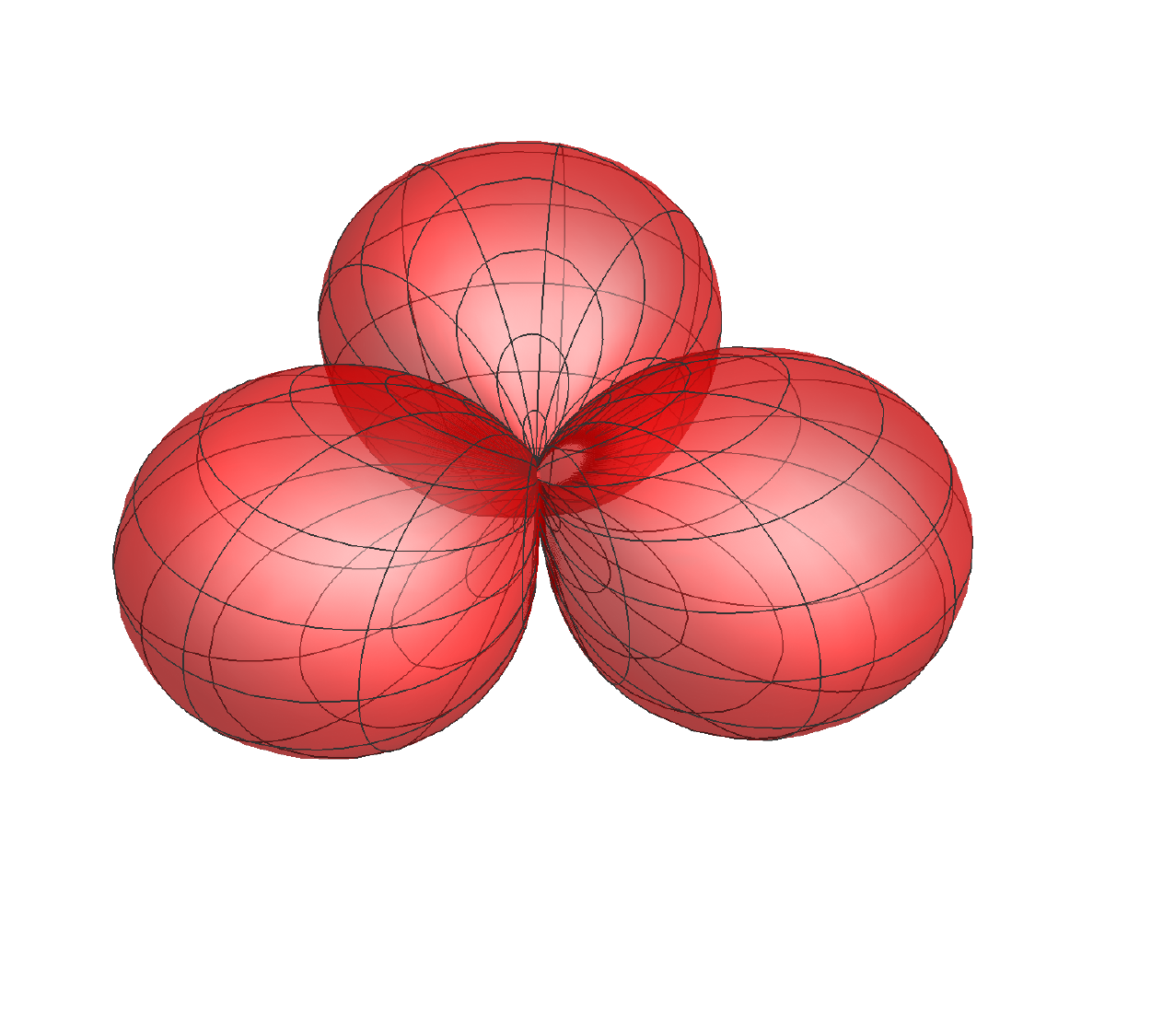}
	\includegraphics[scale=0.27]{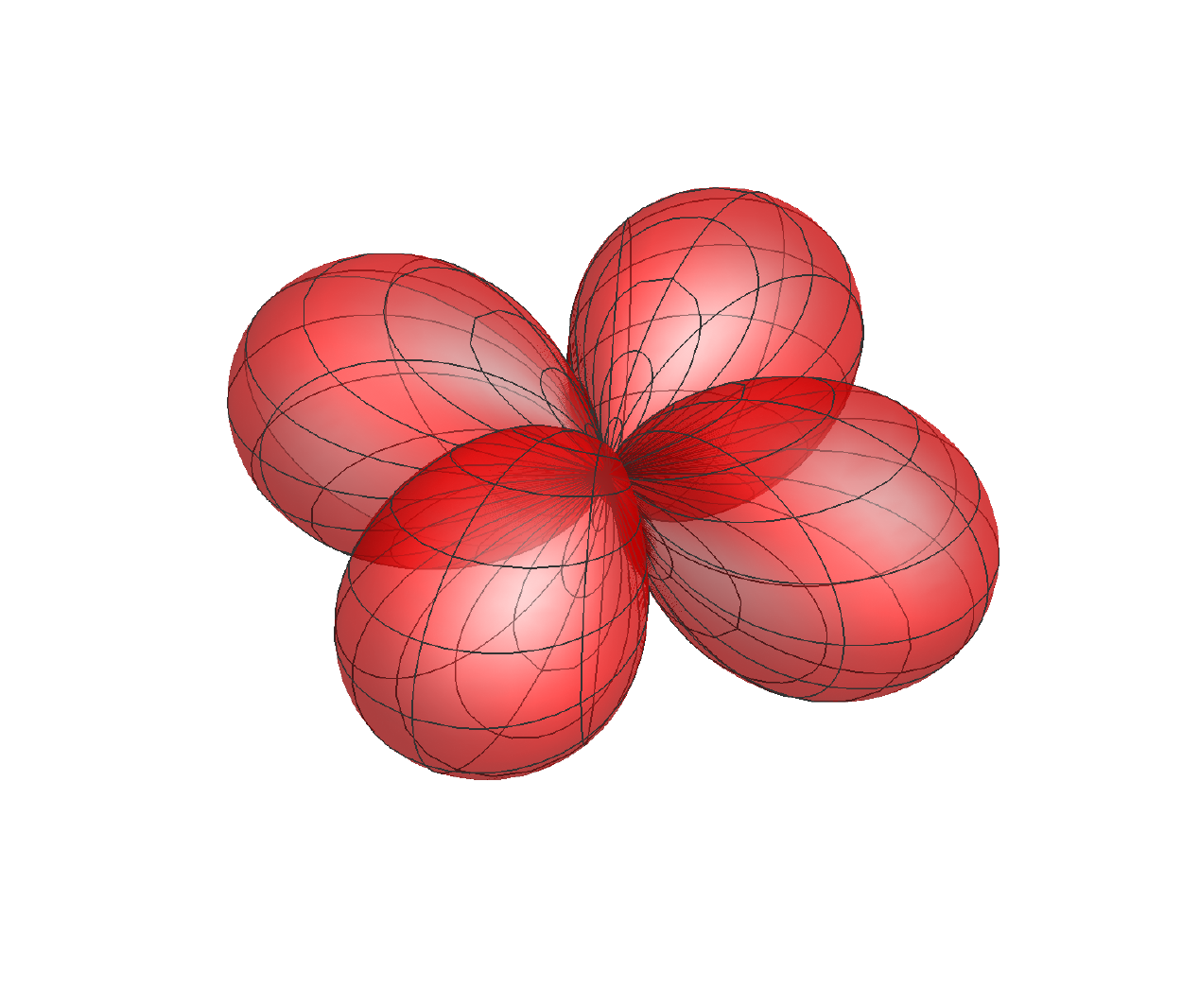}
	
	\includegraphics[scale=0.335]{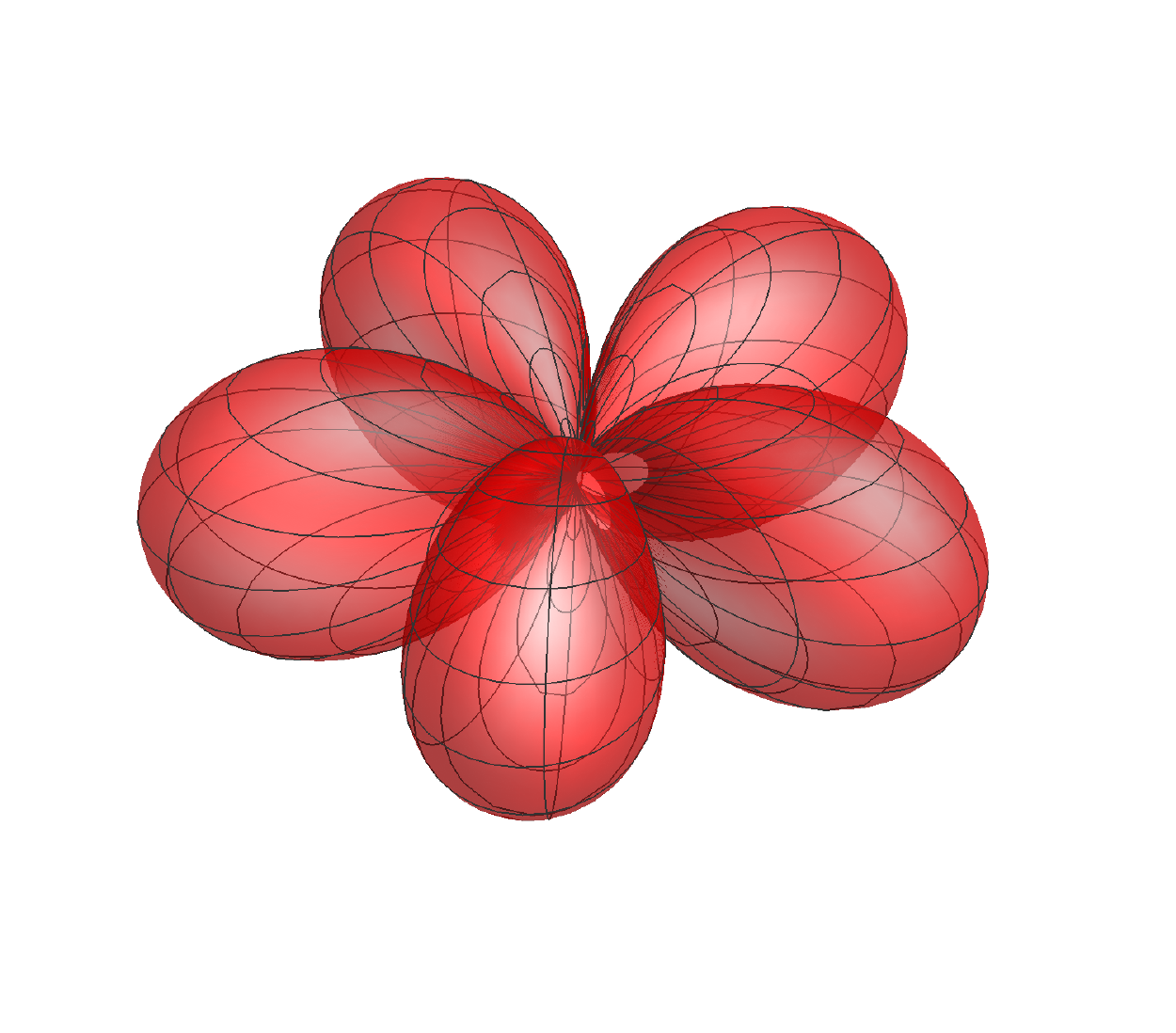}
	\includegraphics[scale=0.335]{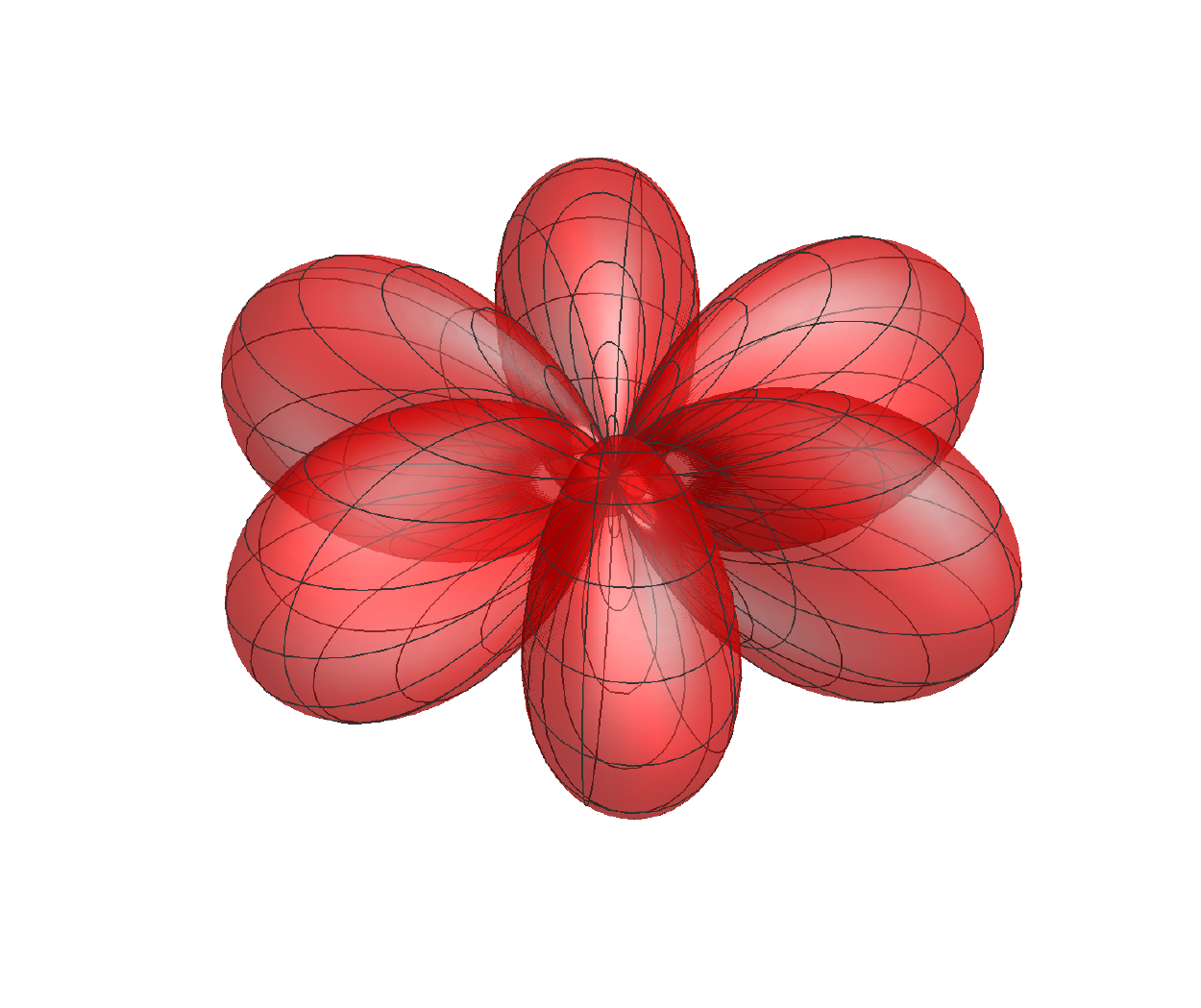}
	\includegraphics[scale=0.335]{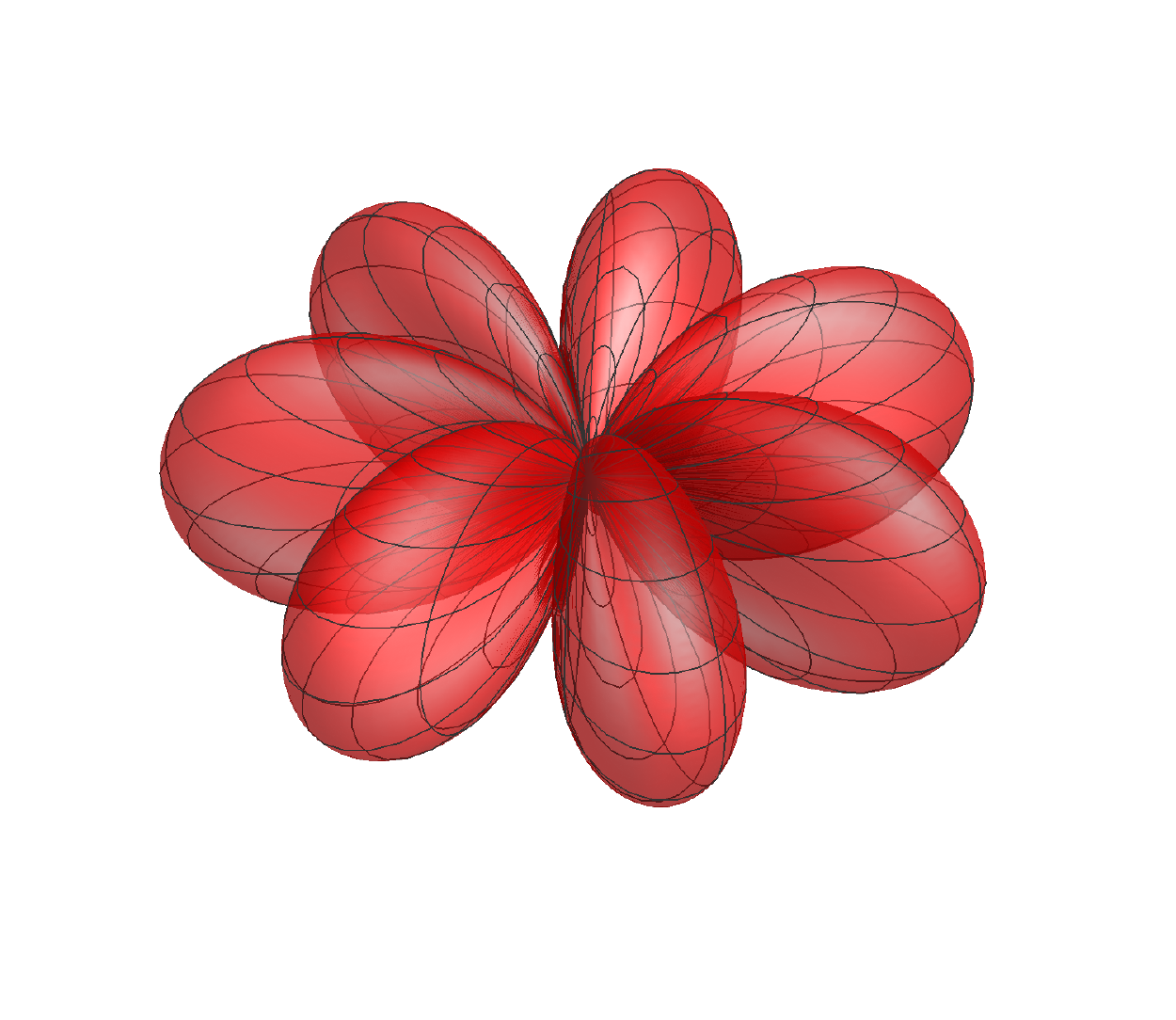}
	\includegraphics[scale=0.335]{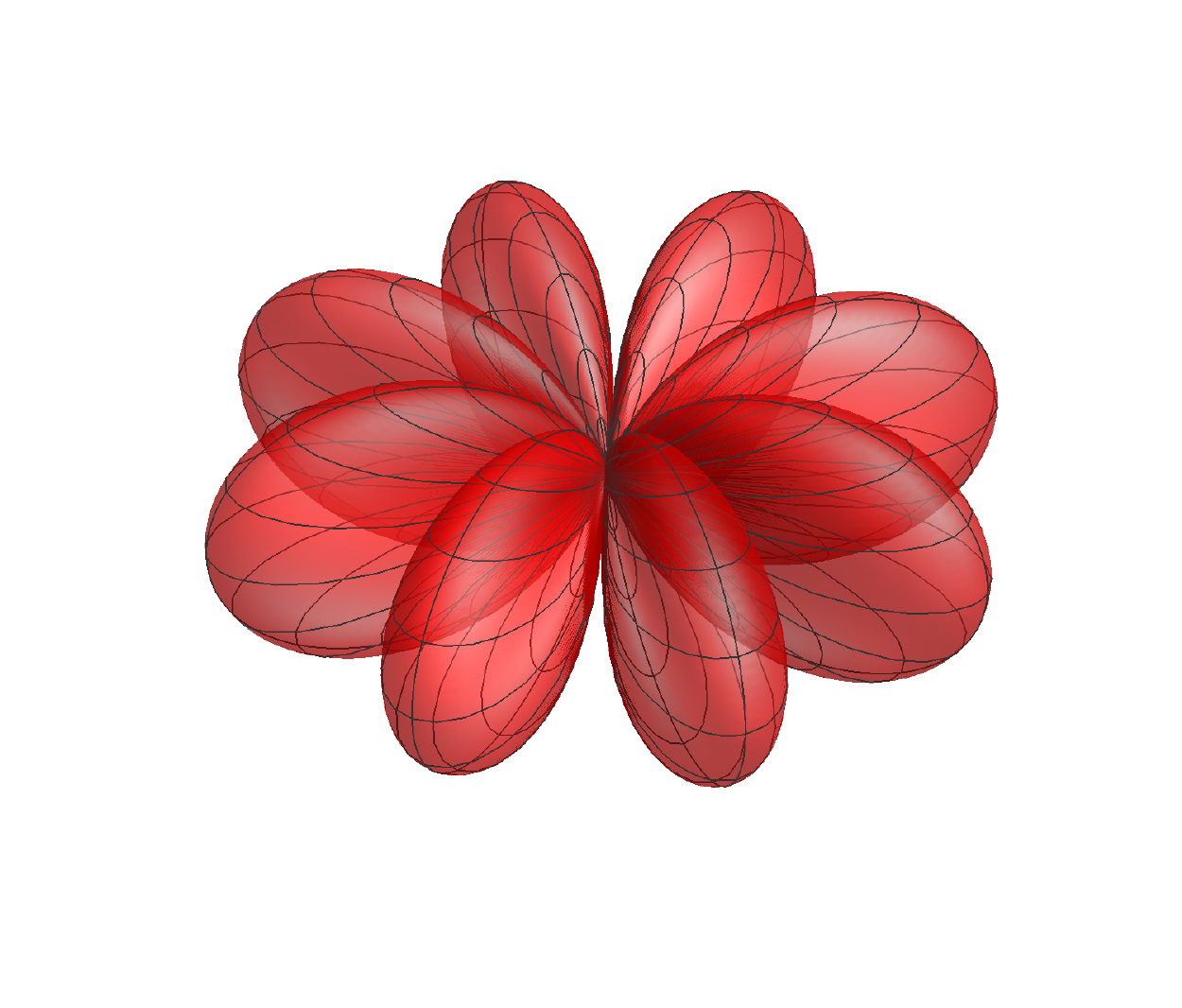}
	
	\caption{(Color online) Warp factor $e^{-2A}$ of model $III$ in a spherical plot, with $r=\Lambda=1$. The top figures are for $n=0$, $1$, $2$, $3$ and $4$ (from left to right) and the bottom ones are for $n=5$, $6$, $7$ and $8$ (from left to right).}\label{warpfactor1}
\end{figure}

\begin{figure}[!htb]
	\includegraphics[scale=0.325]{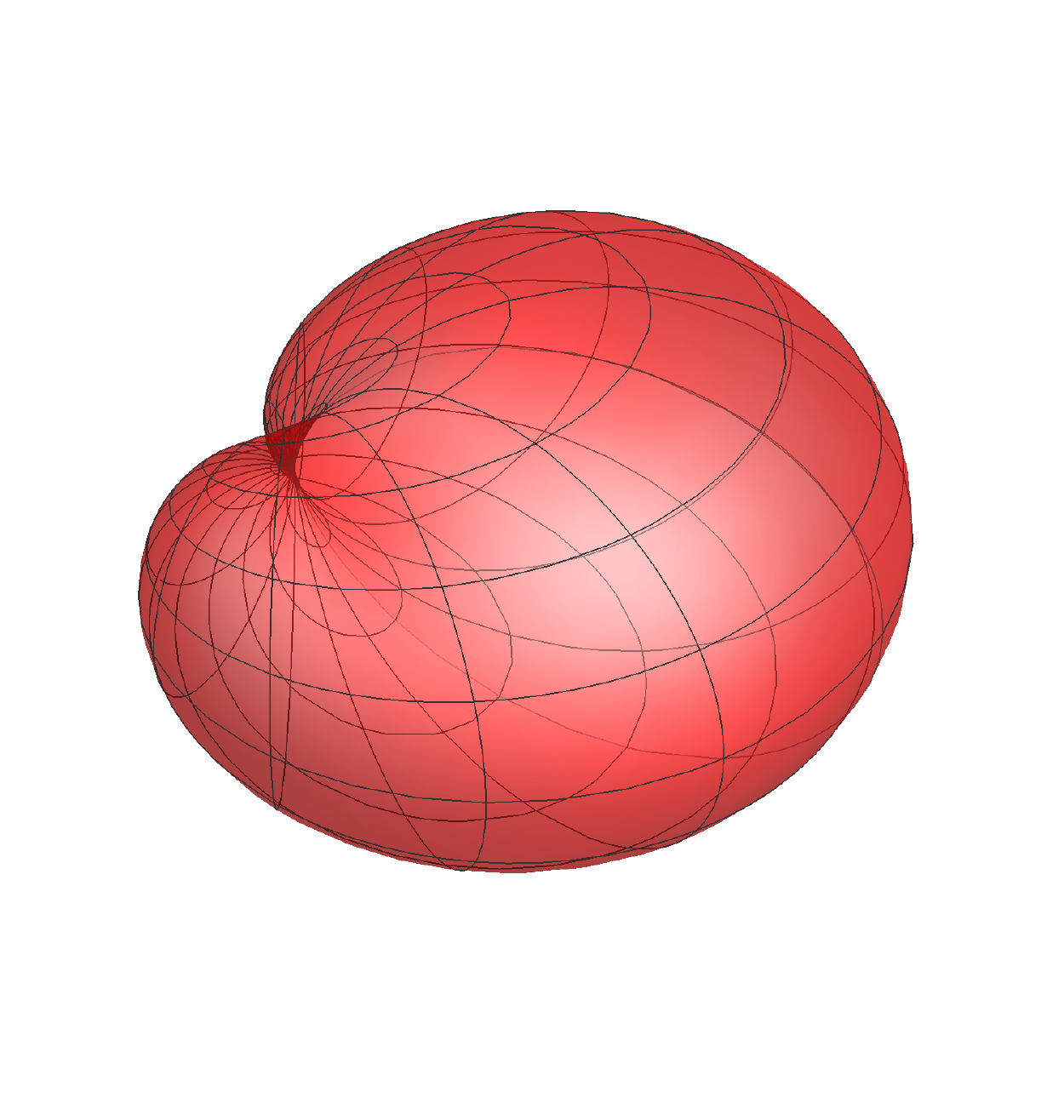}
	\includegraphics[scale=0.325]{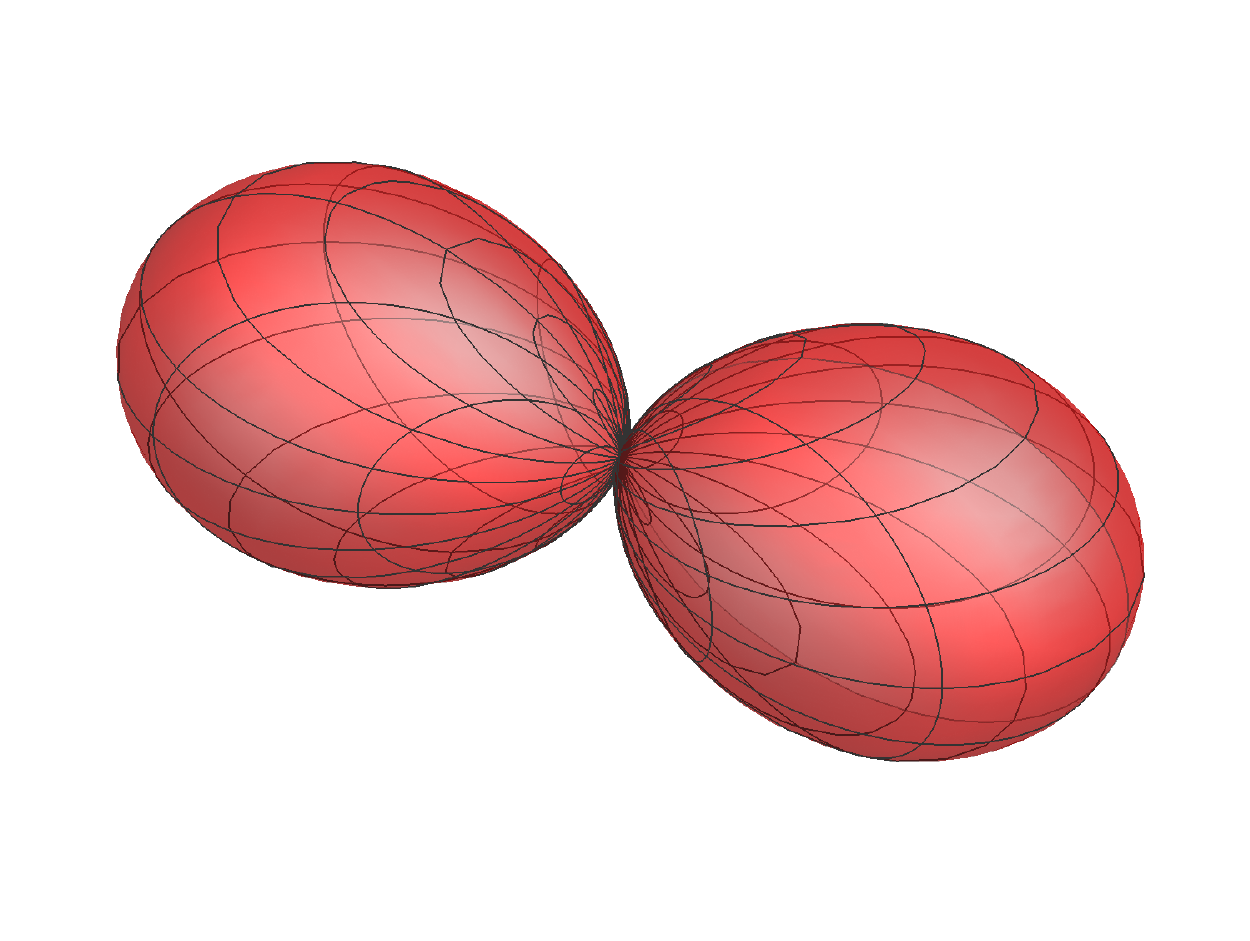}
	\caption{(Color online) Warp factor $e^{-2A}$ of model $IV$ for $n=1$ (left figure) and $n=2$ (right figure) in a spherical plot, with $r=\Lambda=1$.}\label{warpfactor2}
\end{figure}

When $C=4/r^{2}$ ($n=8$ for model $III$ or $n=2$ for model $IV$) one finds a vacuum: the scalar field $\phi$ is null and the potential $\mathcal{V}$ is a constant. For model $IV$, this configuration turns out to be $\mathbbm{d}\mathbb{S}^{6}$. In Fig.~\ref{figure1} the scalar, warp factor and potential, for different values of $C$, are presented. The potential as a function of $\phi$ can be seen in Fig.~\ref{potential(scalar)}.
For models $III$ and $IV$ the complete metric can be written in the form,
\begin{align}
\boldsymbol{g}^{III}=& \sqrt{\left|\cos\left(\frac{n\varphi}{2}\right)\right|}\sin^{2}\theta\;\eta_{\mu\nu}\;\mathrm{d}x^{\mu}\otimes\mathrm{d}x^{\nu}+r^{2}\;\mathrm{d}\theta\otimes\mathrm{d}\theta+r^{2}\sin^{2}\left(\theta\right)\,\mathrm{d}\varphi\otimes\mathrm{d}\varphi,
\\
\boldsymbol{g}^{IV}=& \frac{4r^{2}\Lambda}{3n^{2}}\cos^{2}\left(\frac{n\varphi}{2}\right) \sin^{2}\theta\;\omega^{+}_{\mu\nu}\;\mathrm{d}x^{\mu}\otimes\mathrm{d}x^{\nu}+r^{2}\;\mathrm{d}\theta\otimes\mathrm{d}\theta+r^{2}\sin^{2}\left(\theta\right)\,\mathrm{d}\varphi\otimes\mathrm{d}\varphi.
\end{align}
Figs.~\ref{warpfactor1} and \ref{warpfactor2} depict the warp factor $e^{-2A}$ of models $III$ and $IV$ for various values of $n$.

Finally, using the definition from Eq.~\eqref{stress3}, the stress energy tensor of the scalar field $\phi$ can be obtained for different models. Redundantly, the explicit form of ${{T^{\phi}}^{\mu}}_{\nu}$ is common to all models ($III$ and $IV$) and is only a function of $\theta$,
$$
{{T^{\phi}}^{\mu}}_{\nu}=\frac{8M^{4}}{r^{2}}{\delta^{\mu}}_{\nu}\left[3\left(1-\frac{Cr^{2}}{4}\right)\csc^{2}\left(\theta\right)-5\right].
$$
Despite exhibiting some singularities, ${T^{\phi}}_{\mu\nu}$ is localized and non-singular, since once it is multiplied by the warp factor it becomes well behaved. In Fig.~\ref{stresssphere}, the $\theta$ dependence of ${T^{\phi}}_{\mu\nu}$ is depicted for several values of $C$. Clearly the total energy in these models, as far as $\phi$ is concerned, is finite, given that the stress energy tensor is localized. In fact, for all these scenarios, the total stress energy tensor is given by
\begin{align}
T^{III}_{\mu\nu}&=-\frac{24M^4\eta_{\mu\nu}}{r^{2}}\sqrt{\left|\cos\left(\frac{n\varphi}{2}\right)\right|}\left[\frac{n^2}{8}\sec^{2}\left(\frac{n\varphi}{2}\right)+\left(\frac{5}{3}\sin^{2}\theta-1+\frac{n^{2}}{64}\right)\right],
\\
T^{IV}_{\mu\nu}&=\frac{32M^{4}\Lambda}{n^{2}}\omega^{+}_{\mu\nu}\cos^{2}\left(\frac{n\varphi}{2}\right) \left[1-\frac{n^{2}}{4}-\frac{5}{3}\sin^{2}\left(\theta\right)\right].
\end{align}

So far these models have been presented in spherical coordinates, the introduction of stereographic coordinates, i.e.
\begin{align*}
u=& r\cot \left(\frac{\theta}{2}\right) \cos\left(\varphi\right),
\\
v=& r\cot \left(\frac{\theta}{2}\right) \sin\left(\varphi\right),
\end{align*}
allows one to rewrite the metrics of models $III$ and $IV$ as
\begin{align}
&\boldsymbol{g}^{III}=\frac{4r^{2}\left(u^{2}+v^{2}\right)}{\left(r^{2}+u^{2}+v^{2}\right)^{2}}\Bigg\{\sqrt{\left|\cos\left[\frac{n}{2}\arccos\left(\frac{u}{\sqrt{u^{2}+v^{2}}}\right)\right]\right|}\eta_{\mu\nu}\mathrm{d}x^{\mu}\otimes\mathrm{d}x^{\nu}+\mathrm{d}u\otimes\mathrm{d}u+\mathrm{d}v\otimes\mathrm{d}v\Bigg\},\label{model1stereo}
\\
&\boldsymbol{g}^{IV}=\frac{4r^{2}\left(u^{2}+v^{2}\right)}{\left(r^{2}+u^{2}+v^{2}\right)^{2}}\Bigg\{\frac{4r^{2}\Lambda}{3n^{2}}\cos^{2}\left[\frac{n}{2}\arccos\left(\frac{u}{\sqrt{u^{2}+v^{2}}}\right)\right]\omega^{+}_{\mu\nu}\mathrm{d}x^{\mu}\otimes\mathrm{d}x^{\nu}+\mathrm{d}u\otimes\mathrm{d}u+\mathrm{d}v\otimes\mathrm{d}v\Bigg\}.\label{model2stereo}
\end{align}

\begin{figure}[!htb]
	\includegraphics[scale=0.26]{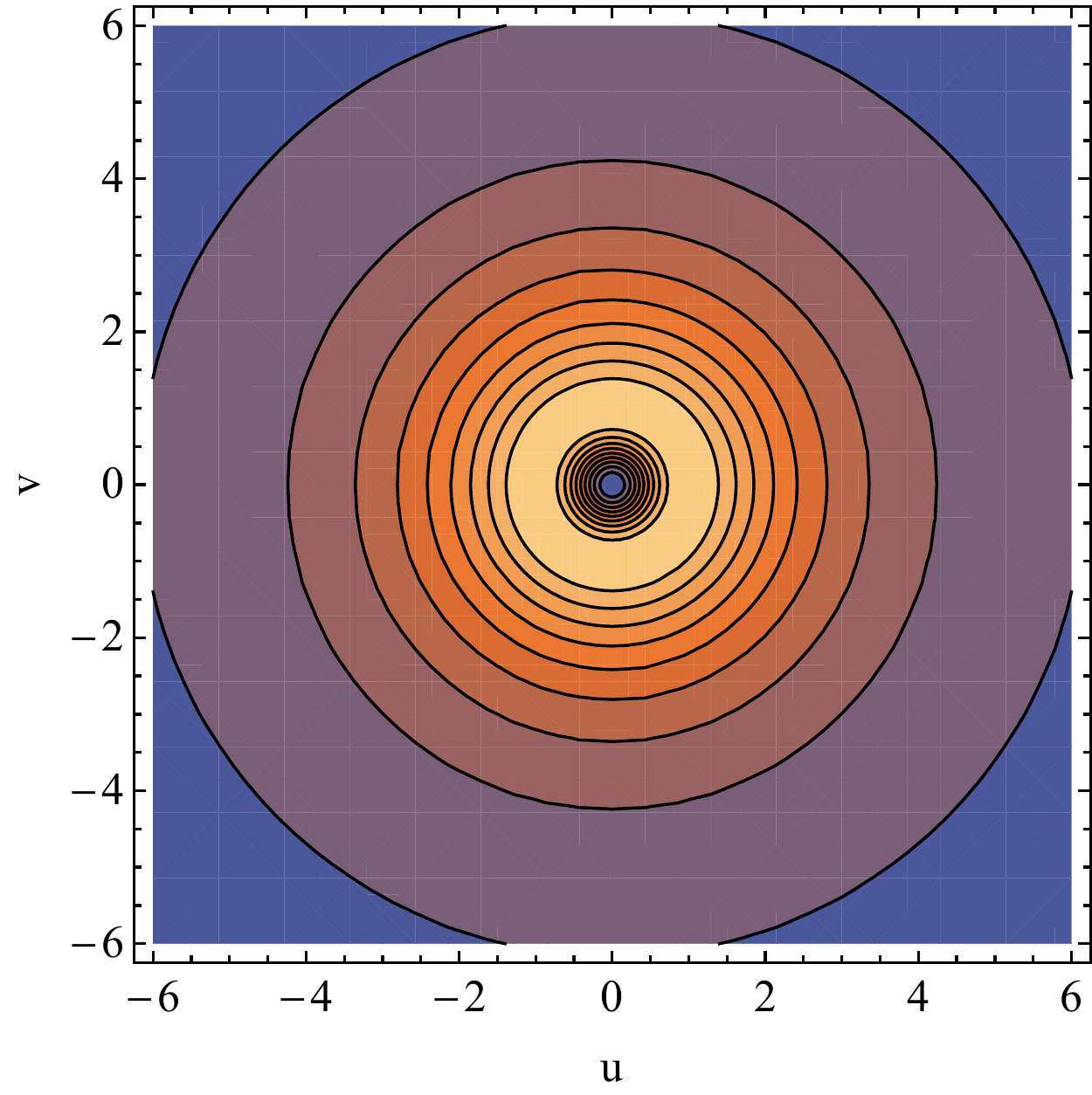}
	\includegraphics[scale=0.26]{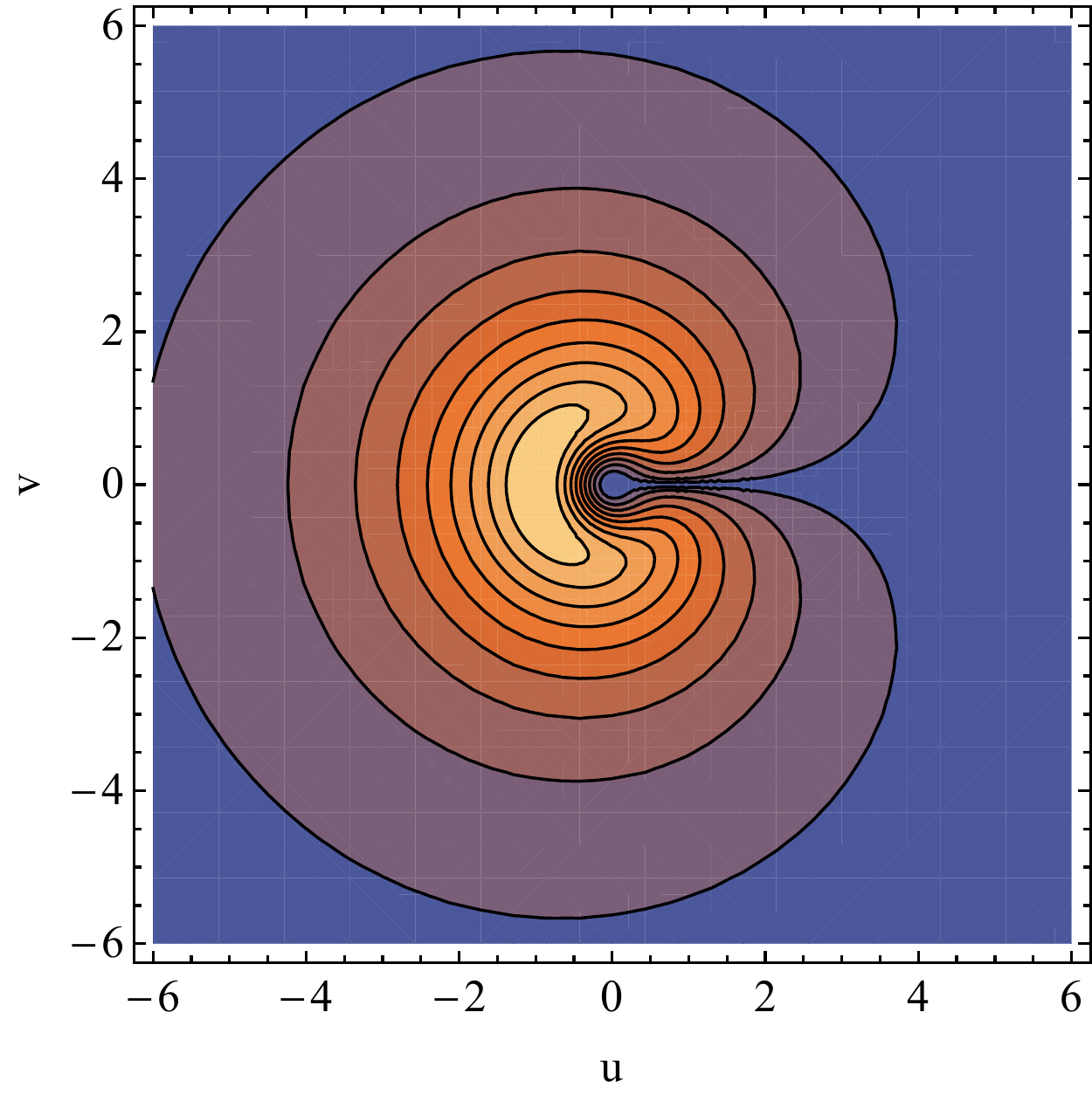}
	\includegraphics[scale=0.26]{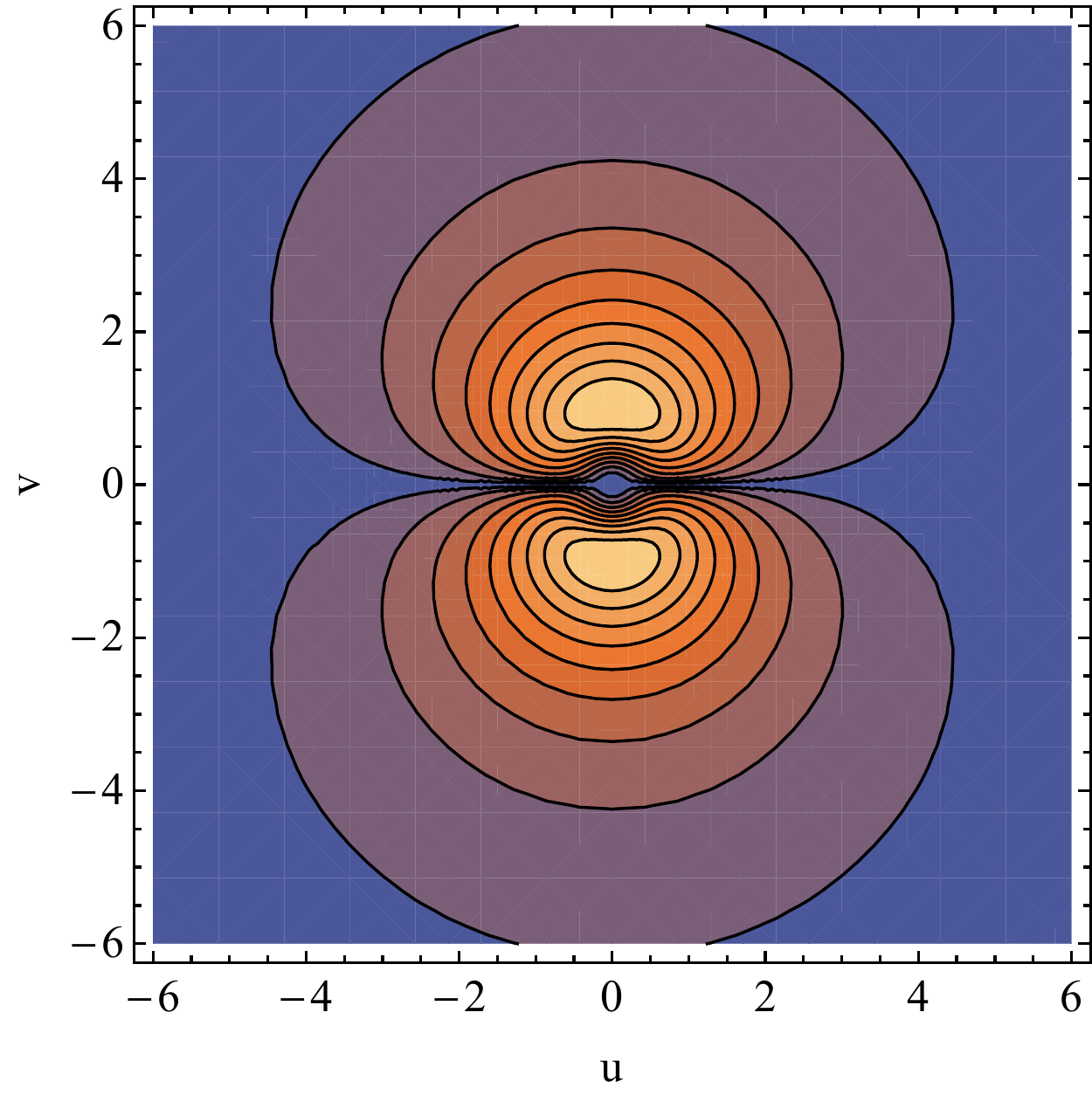}
	\includegraphics[scale=0.26]{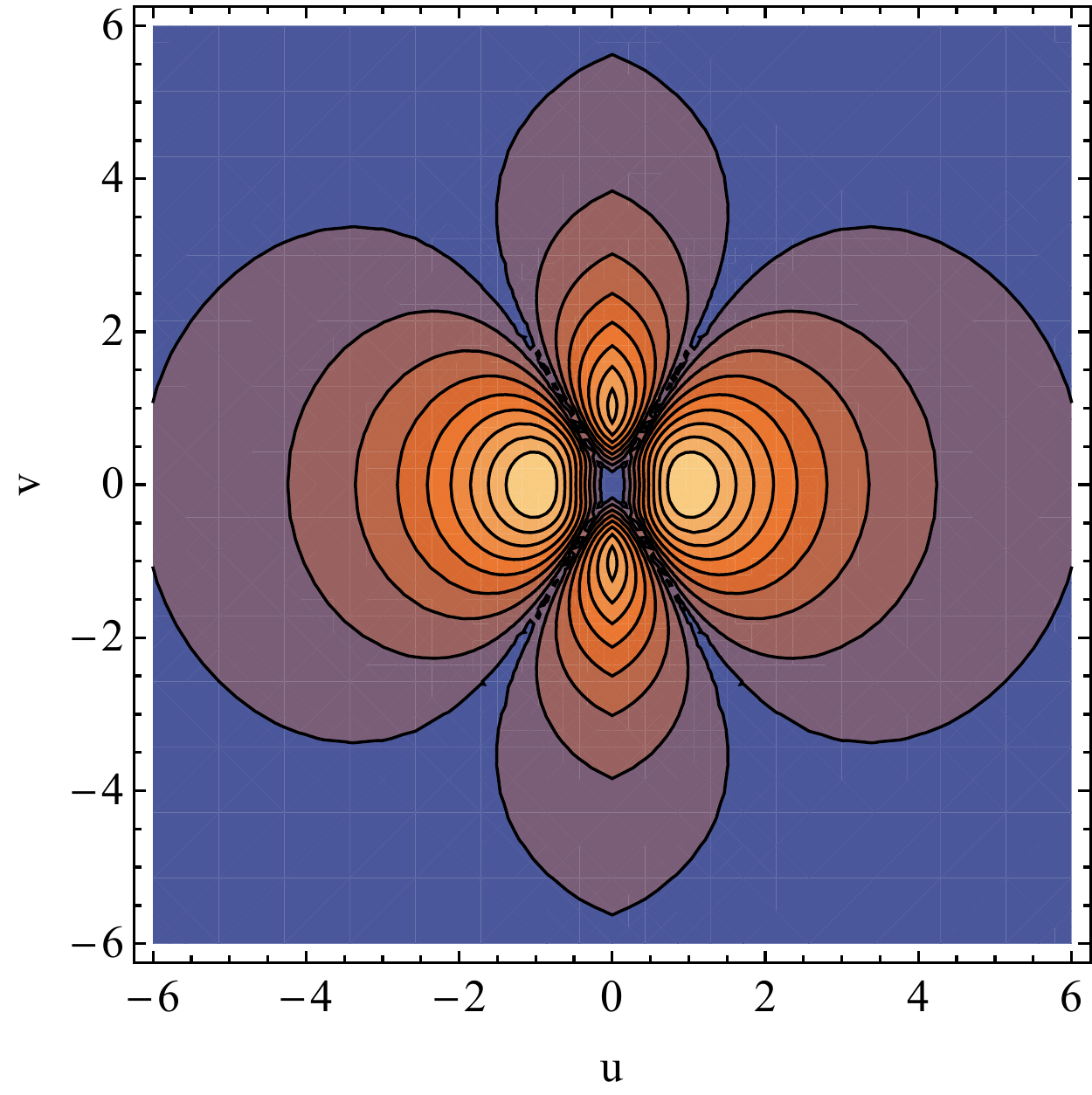}
	\includegraphics[scale=0.26]{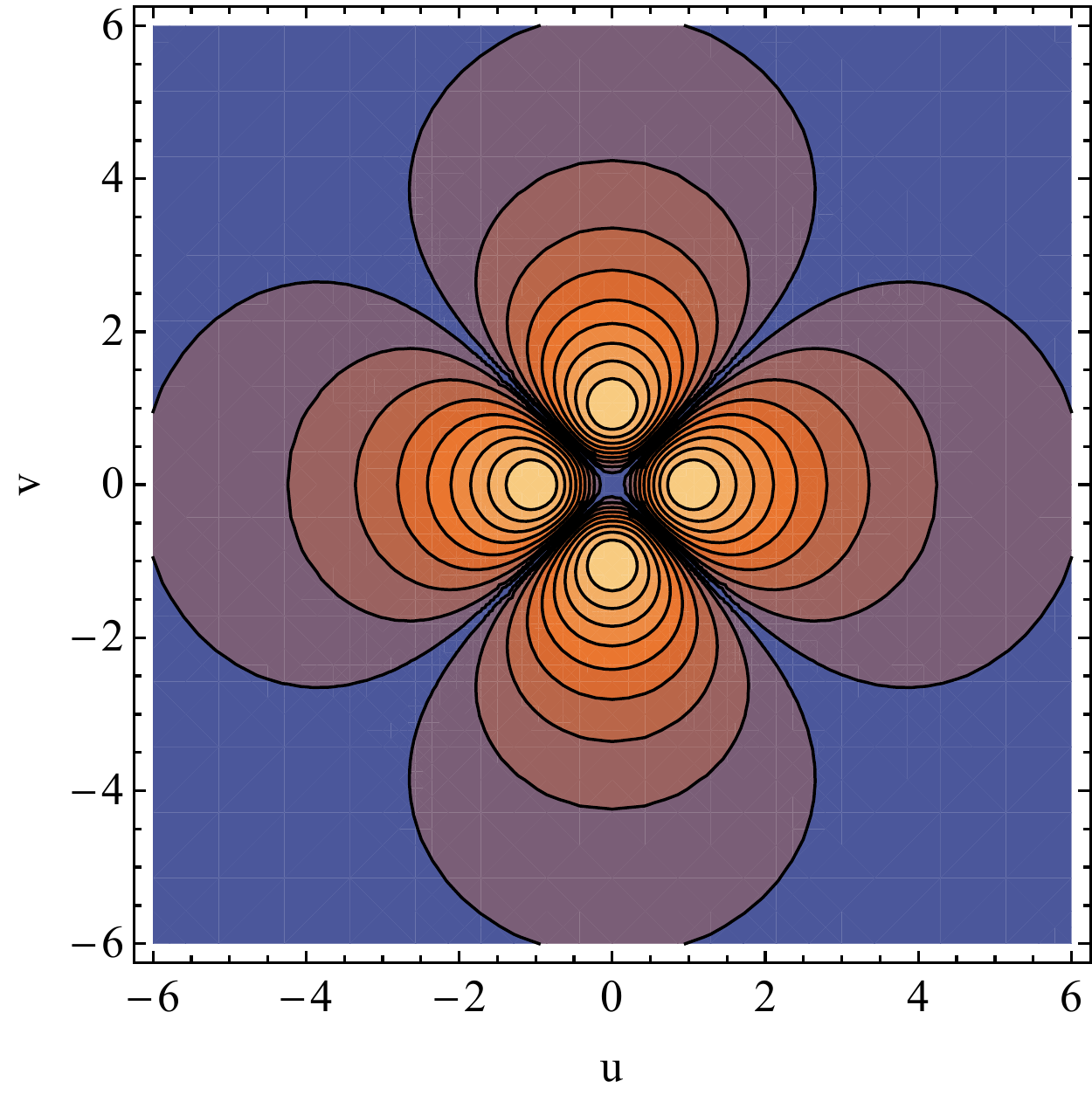}
	\includegraphics[scale=0.26]{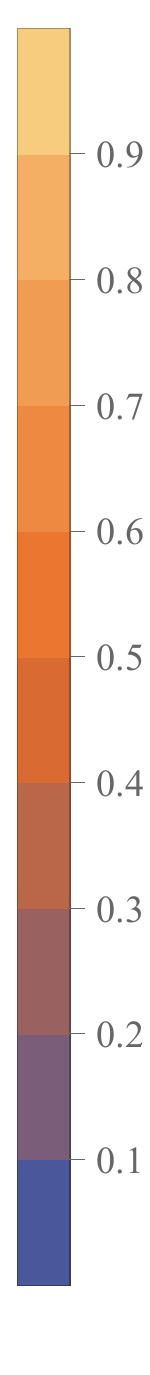}

	\includegraphics[scale=0.325]{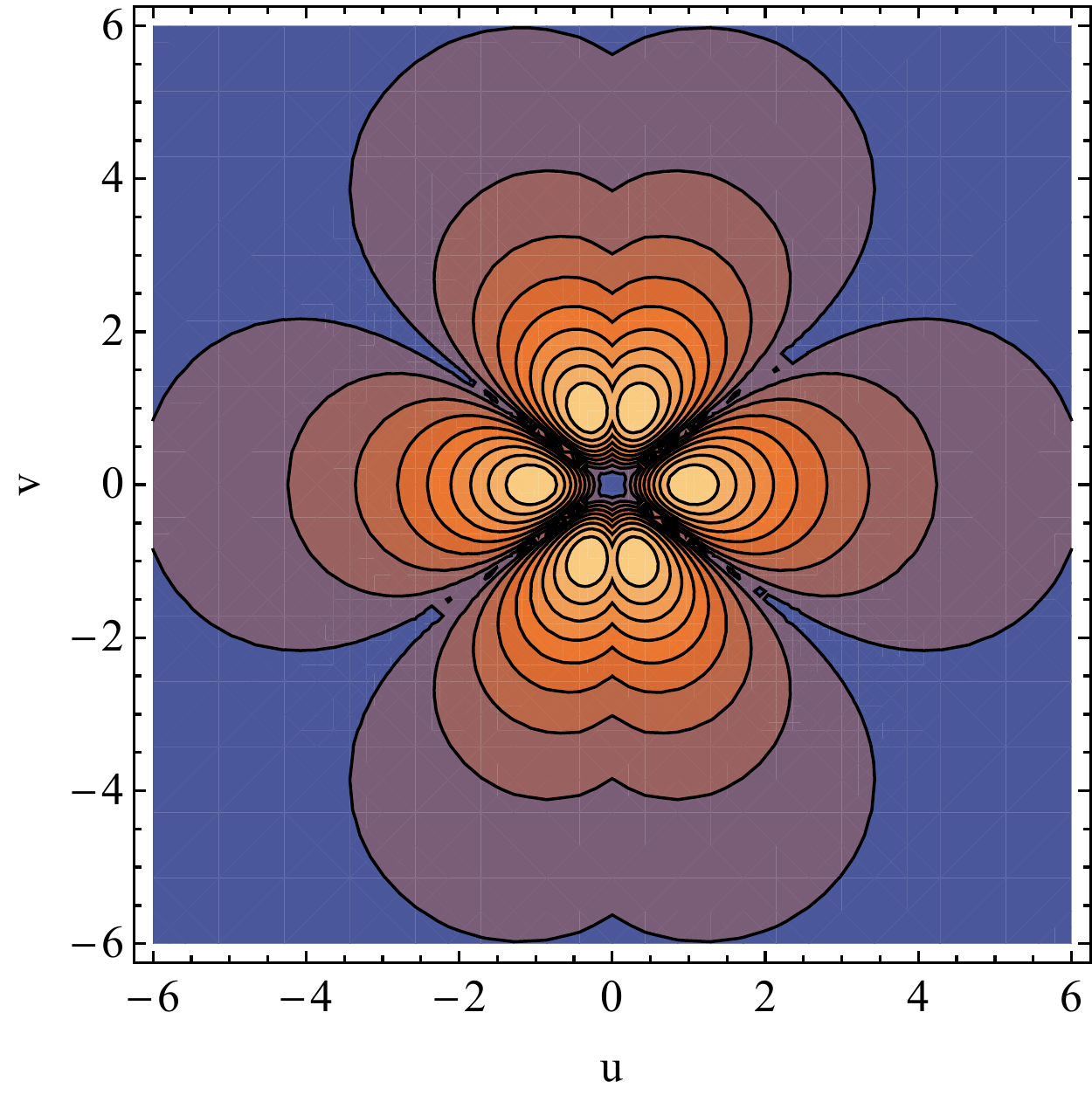}
	\includegraphics[scale=0.325]{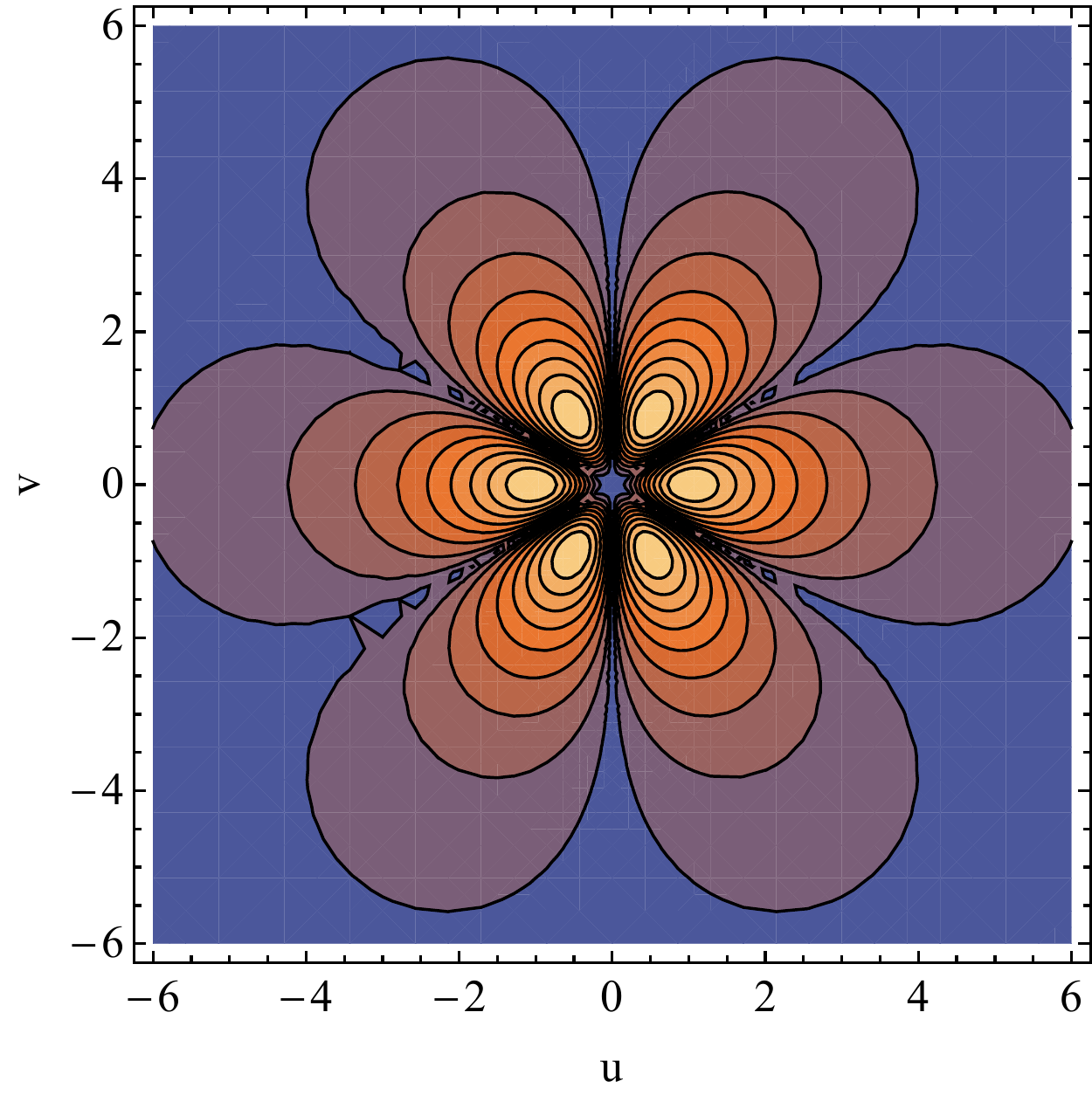}
	\includegraphics[scale=0.325]{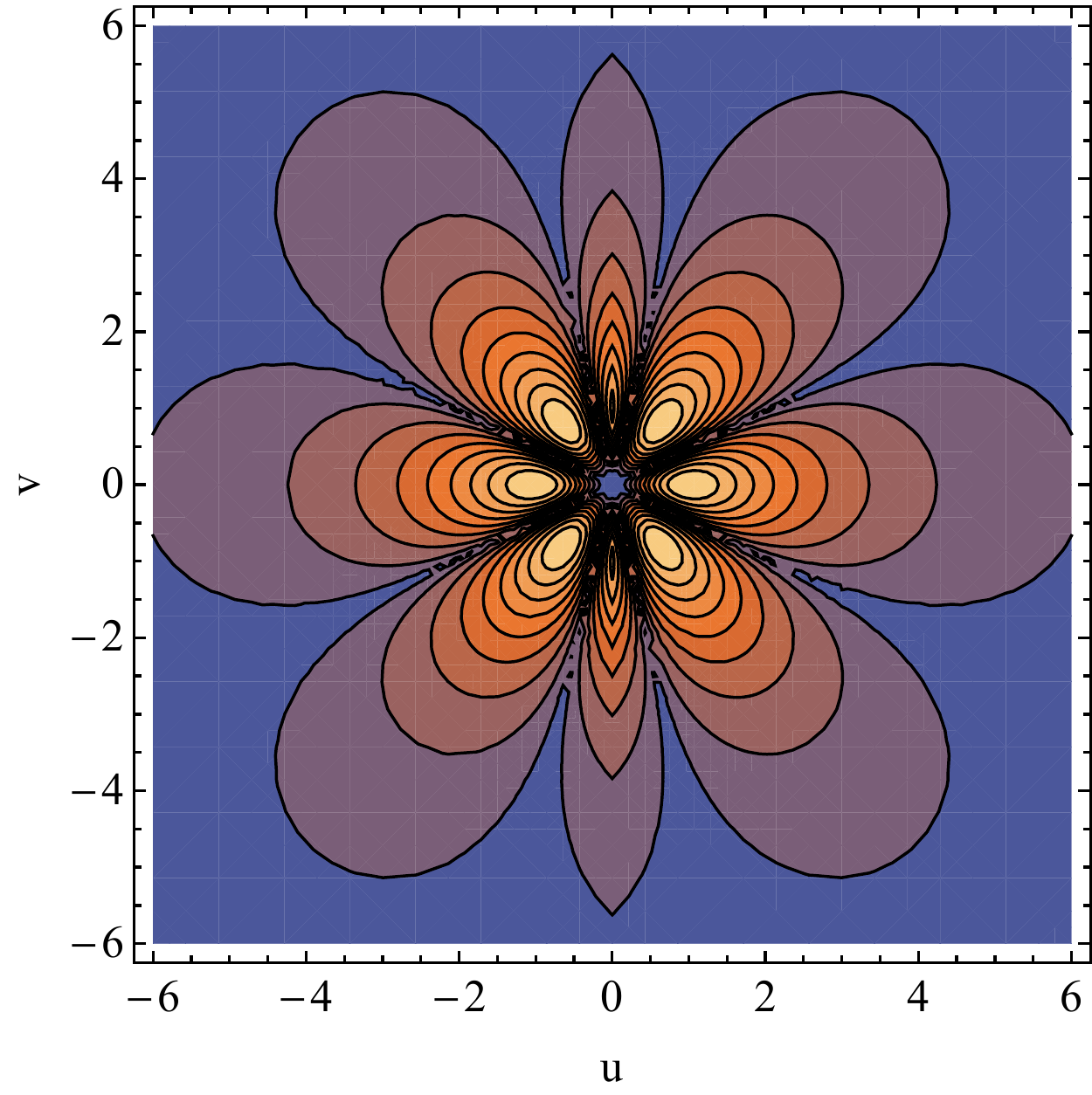}
	\includegraphics[scale=0.325]{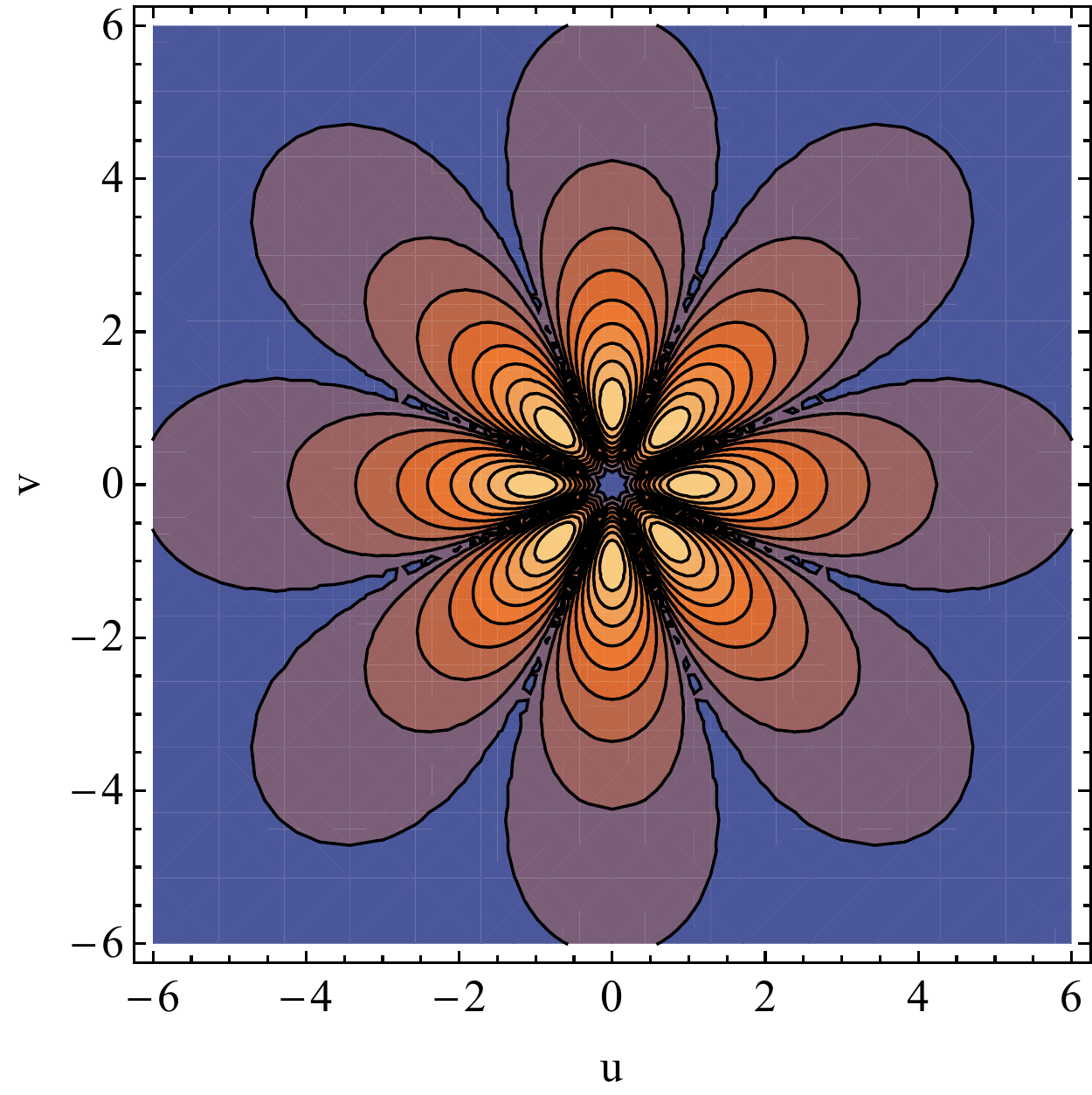}
	\includegraphics[scale=0.3]{legend_I}
	\caption{(Color online) Warp factor $e^{-2A}$ of model $III$ in stereographic coordinates. The top figures are for $n=0$, $1$, $2$, $3$ and $4$ (from left to right) and the bottom ones are for $n=5$, $6$, $7$ and $8$ (from left to right).}\label{warpIII}
\end{figure}

\begin{figure}[!htb]
	\includegraphics[scale=0.31]{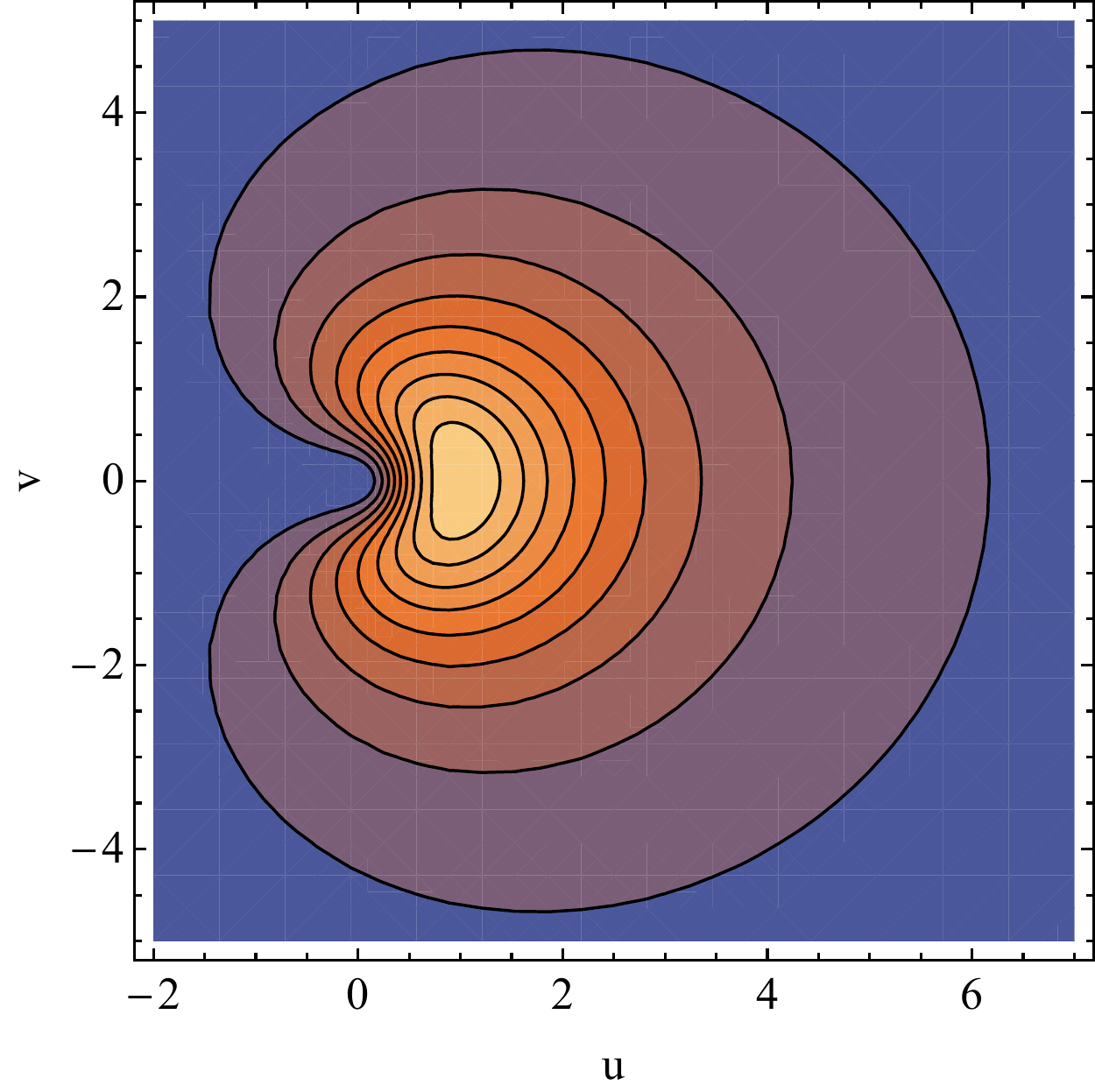}
	\includegraphics[scale=0.31]{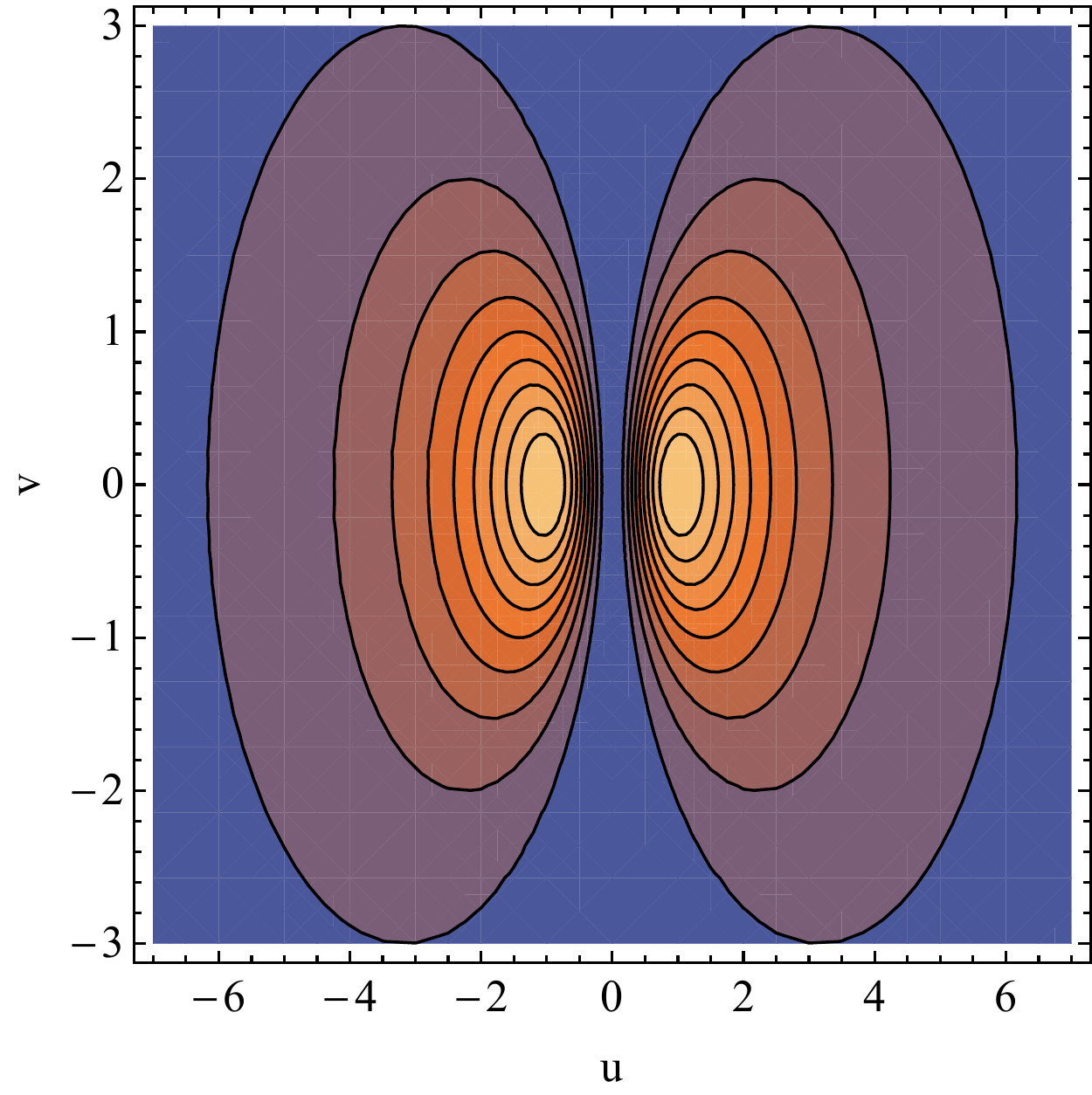}
	\includegraphics[scale=0.3]{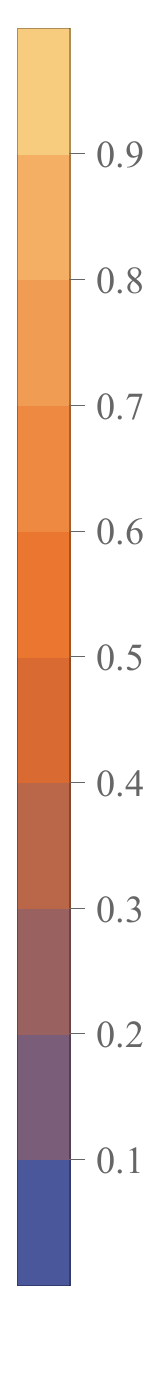}
	\caption{(Color online) Warp factor $e^{-2A}$ of model $IV$ for $n=1$ (left figure) and $n=2$ (right figure), in stereographic coordinates.}\label{warpIV}
\end{figure}

One can thus notice the advantage of choosing the initial metric of $\mathbb{B}^{2}$ as from Eq.~\eqref{assumptionmetric} if, on the other hand, one had started with a conformally flat form. As can be seen from expressions \eqref{model1stereo} and \eqref{model2stereo}, it would not be straightforward finding these solutions, since the warp factor is, most notably, not separable in $u$ and $v$. Moreover, one could express all the warp factors without the use of $\arccos$ and so on. 
In this case, the warp factor of models $III$ and $IV$ for all the allowed values of $n$ can be depicted as they appear in Figs.~\ref{warpIII} and \ref{warpIV}.

From Figs.~\ref{warpIII} and \ref{warpIV} it is clear the localization of the warp factor, even when the space goes to infinity. Therefore these models give rise to thick branes over the sphere where the only adjustable localization parameter is the radius $r$ of the sphere. This corresponds to a detriment to the model since it would be interesting to have a parameter to make the brane thiner while maintaining the radius of the sphere fixed.

\subsection{The Spheroid Models}\label{modelspheroid}

Departing from the model over the sphere, one may consider that the ground space $\left(\mathbb{B}^{2},\boldsymbol{\sigma}\right)$ is a spheroid. In some other words, $\left(\mathbb{B}^{2},\boldsymbol{\sigma}\right)\equiv\left(\mathbb{S}^{2},\boldsymbol{\epsilon}\right)$, with (cf. Eq.~\eqref{spheroid})
\begin{equation*}
\boldsymbol{\epsilon}=r^{2}\left[1+\left(\frac{\rho^{2}}{r^{2}}-1\right)\sin^{2}\theta\right]\mathrm{d}\theta\otimes\mathrm{d}\theta+r^{2}\sin^{2}\left(\theta\right)\,\mathrm{d}\varphi\otimes\mathrm{d}\varphi,
\end{equation*}
the difference between sphere and spheroid models is simply due to the geometry represented by the metric $\boldsymbol{\epsilon}$. The spheroid built here is a di-axial ellipsoid, with radii $r$ and $\rho$. In this case, the setup variables are $$u= r\theta,$$
$$
\tilde{f}=-\frac{1}{2}\ln\left\{\left[1-\kappa\sin^{2}\left(\theta\right)\right]\right\},
$$
and
$$\tilde{A}\equiv-\ln\left[\sin\left(\theta\right)\right],$$
where $\kappa=1-\rho^{2}/r^{2}$. Setting $\kappa=0$, one recovers the model over the sphere. See that this is mapped by the metric from Eq.~\eqref{generalspheroid} where one just imposes $u_{0}=\pi/2r$, $a=1/r$. Thus, through Eqs.~\eqref{generalpotential2} and \eqref{generalscalar1}, one can determine the potential and scalar field as
\begin{align}
	\frac{\mathcal{V}}{4M^{4}}&=\frac{2}{r^{2}\left[1-\kappa\sin^{2}\left(\theta\right)\right]}\left[\frac{1-\kappa}{1-\kappa\sin^{2}\left(\theta\right)}-4 \cot^{2}\left(\theta\right)\right]+2C\csc^{2}\left(\theta\right),\label{sepspheroid1}
	\\
	\frac{{\phi_{,\theta}}^{2}}{4M^{4}}&=\frac{4\left(1-\kappa\right)}{1-\kappa\sin^{2}\left(\theta\right)}+\left(4-Cr^{2}\right)\cot^{2}\left(\theta\right)-\left(1-\kappa\right) Cr^{2},\label{sepspheroid2}
\end{align}
In this case, if $C>4/r^{2}$ then the left side of equation \eqref{sepspheroid2} is not necessarily positive for all $\theta$ values. Notice that as $\theta$ approaches $\pi/2$ the term with $\cot(\theta)$ goes to infinity, while the other terms remain finite. This means that $\phi$ would necessarily be imaginary for some value of $\theta$, which is not allowed. Therefore, one has $C\leq 4/r^{2}$. 
In fact, one can not solve \eqref{sepspheroid2} in general. It can only be solved analytically when $C= 4/r^{2}$, which is translated into choosing $n=8$ for model $III$ or $n=2$ for model $IV$. Henceforward up to the end only this cases will be considered.

For $C= 4/r^{2}$, the scalar field Eq.~\eqref{sepspheroid2} is easily integrated,
\begin{equation*}
\phi=\mp4M^{2}\sqrt{1-\kappa} \operatorname{arctanh} \left[\frac{\sqrt{\kappa } \sin \left(\theta\right)}{\sqrt{1-\kappa \cos^{2} \left(\theta\right)}}\right],
\end{equation*}
and, if $\kappa=1$ or $\kappa=0$, one finds a vacuum solution. 
\begin{figure}[!htb]
	\subfloat[]{\label{figure5}\includegraphics[scale=0.69]{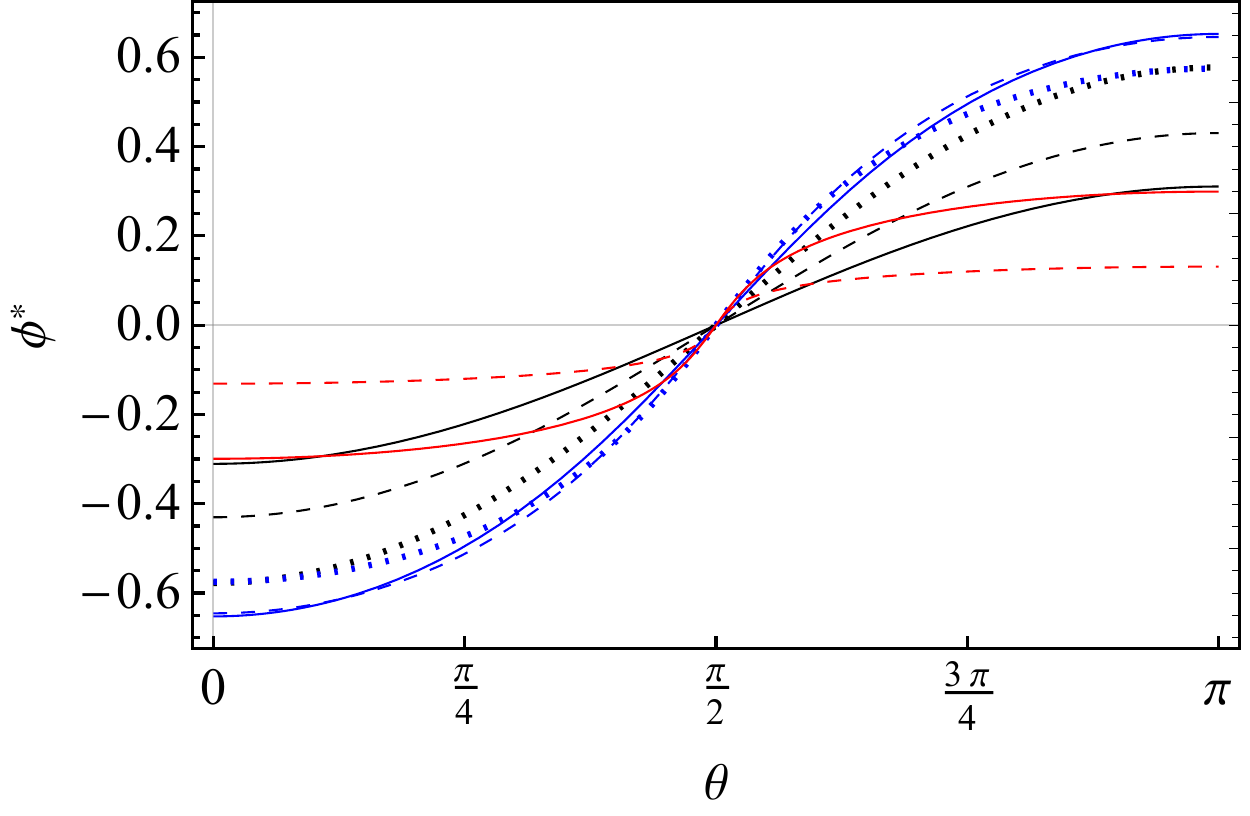}}$\;\;$
	\subfloat[]{\label{figure7}\includegraphics[scale=0.69]{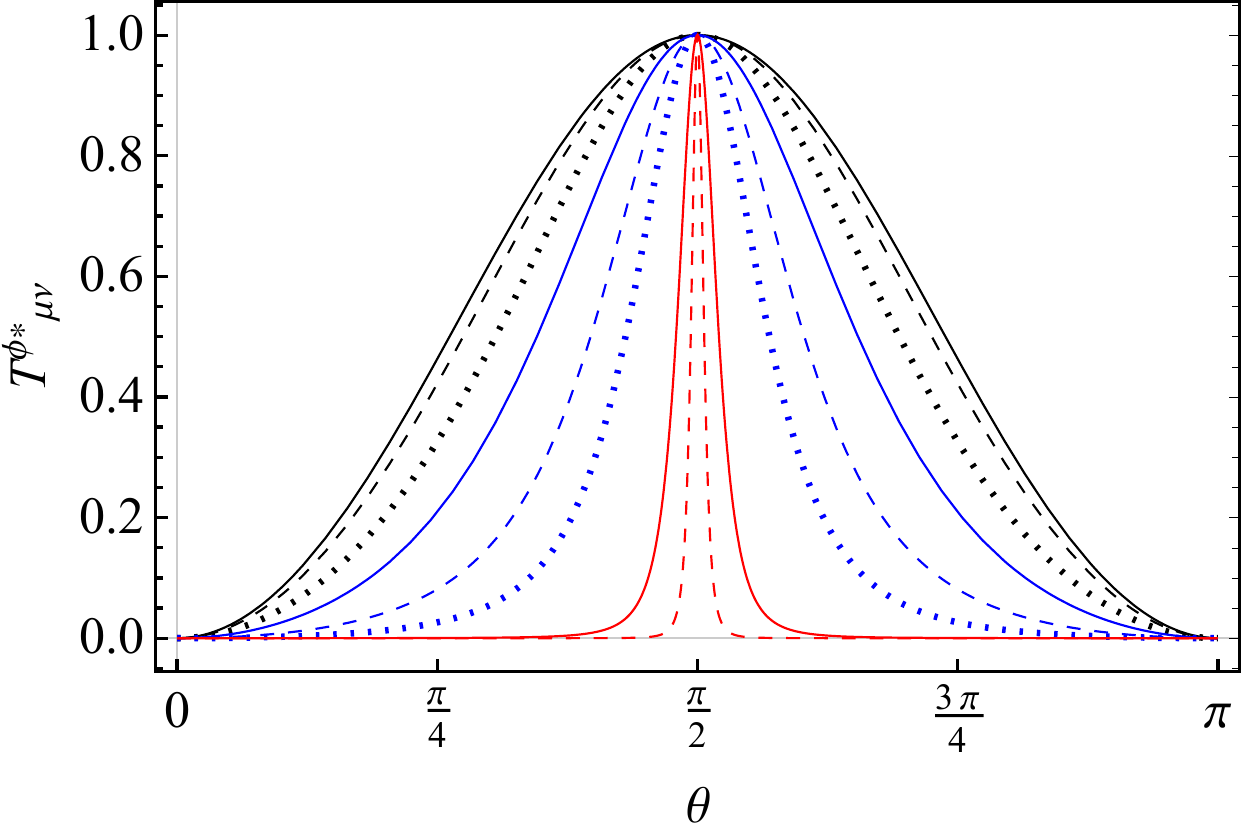}}
	\caption{(Color online) (a) Scalar field $\phi^{*}=\phi/4M^{2}$ as a function of $\theta$. (b) The $\theta$ dependence of the stress energy tensor ${T^{\phi}}_{\mu\nu}$, ${T^{{\phi}^{*}}}_{\mu\nu}=-\sin^{2}\theta\;{{T^{\phi}}^{\mu}}_{\nu}(1-\kappa)/8(5-3\kappa)$.
	The plots are for $\kappa=0.1$ (solid black line), $\kappa=0.2$ (dashed black line), $\kappa=0.4$ (dotted black line), $\kappa=0.6$ (solid blue line), $\kappa=0.8$ (dashed blue line), $\kappa=0.9$ (dotted blue line), $\kappa=0.99$ (solid red line) and $\kappa=0.999$ (dashed red line).}
\end{figure}
From Fig.~\ref{figure5} the profile of the scalar field can be read as a topological or kink-like defect. The potential $\mathcal{V}$ in terms of $\phi$ is exactly as given by Eq.~\eqref{spheroidpotentialphi} (cf. Fig.~\ref{potential(scalar)}), and the stress energy tensor of $\phi$, which is common to all models, is given by
\begin{equation}
{{T^{\phi}}^{\mu}}_{\nu}=-{\delta^{\mu}}_{\nu}\frac{8M^{4}\left(1-\kappa\right)\left[5-3\kappa\sin^{2}\left(\theta\right)\right]}{r^{2}\left[1-\kappa^{2}\sin^{2}\left(\theta\right)\right]^{2}}.
\end{equation}

Differently from the sphere models, these models possess another localizing parameter other than the radius $r$. As $\kappa$ approaches $1$ the stress energy tensor becomes more and more localized, from Fig.~\ref{figure7} 
one can notice such a behavior. Consequently, these models give rise to thick branes that are even more interesting than the spherical ones, as one chooses $\kappa$ closer to $1$ the thinner the distribution of matter becomes.

\begin{figure}[!htb]
	\subfloat[]{\includegraphics[scale=0.69]{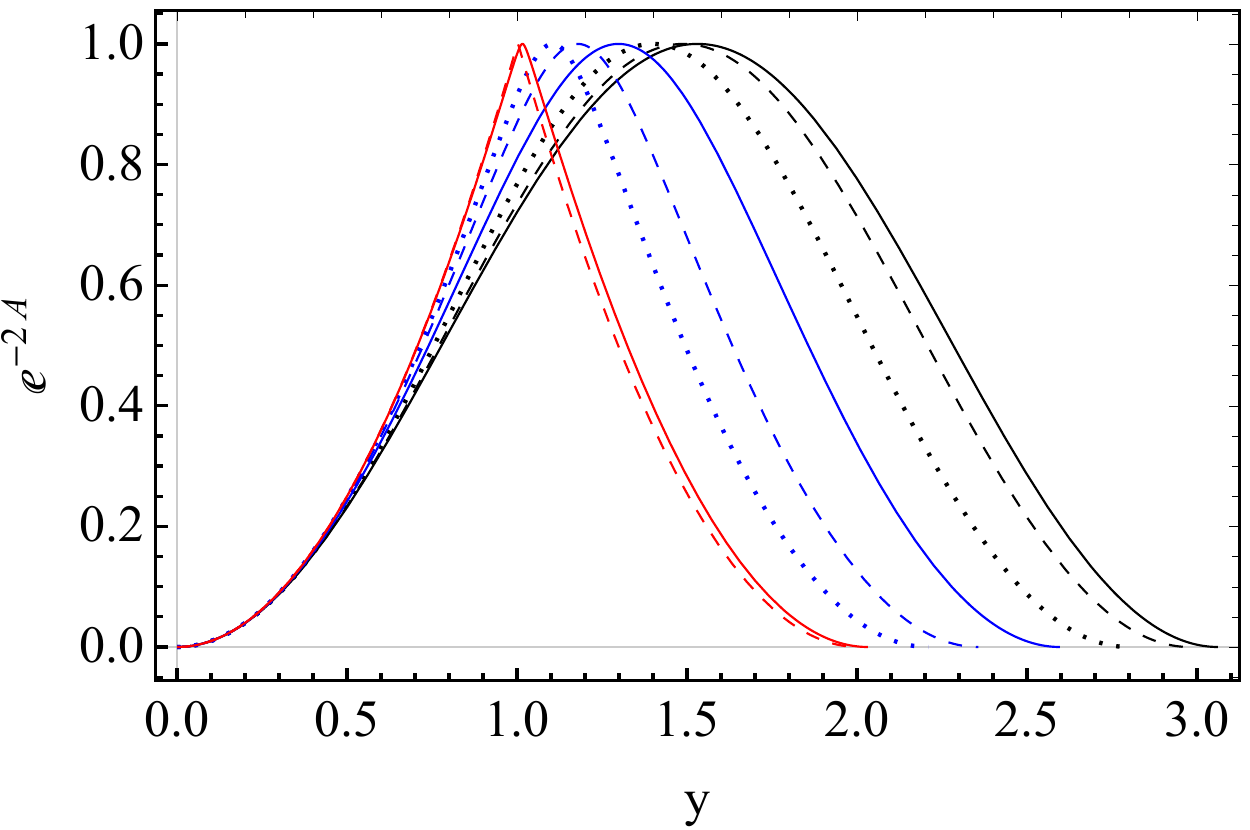}}$\;\;$
	\subfloat[]{\label{scalar(y)}\includegraphics[scale=0.69]{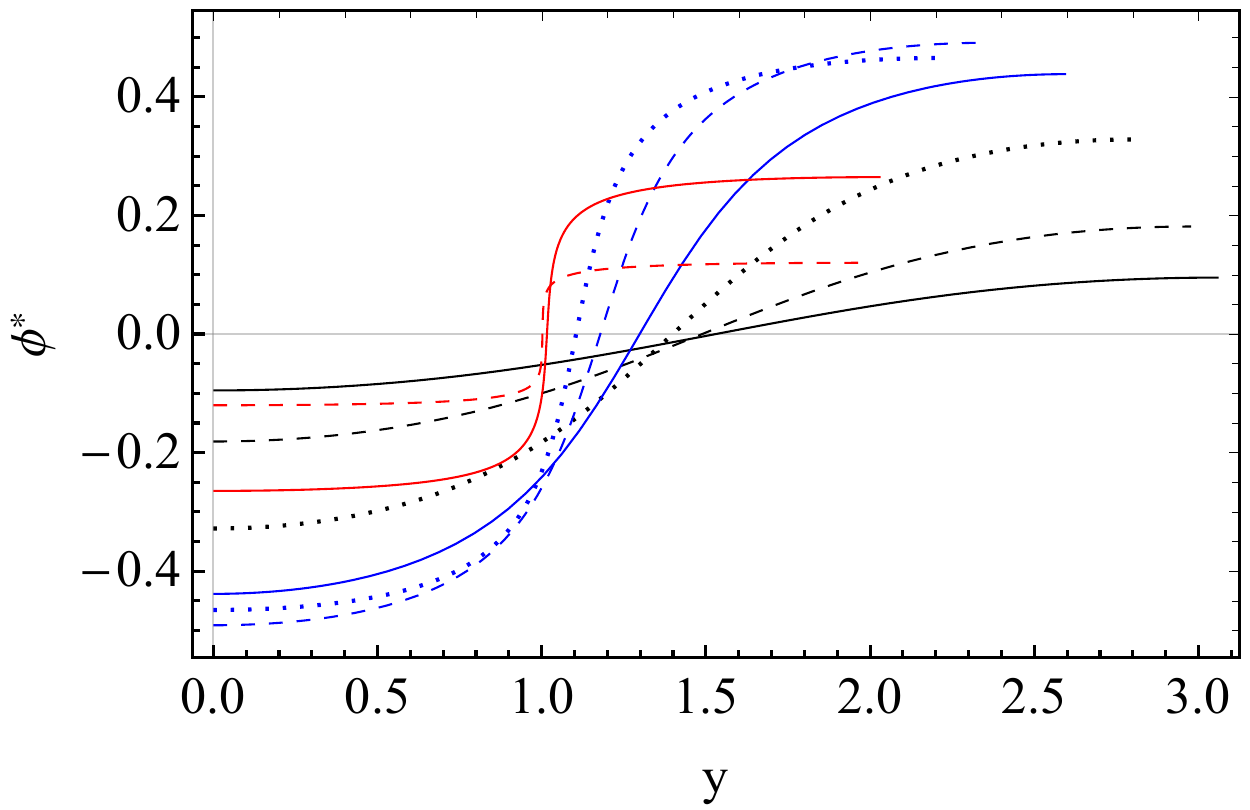}}
	\caption{(Color online) (a) Scalar field $\phi^{*}=\phi/4M^{2}$ as a function of $\theta$. (b) The $\theta$ dependence of the stress energy tensor ${T^{\phi}}_{\mu\nu}$, ${T^{{\phi}^{*}}}_{\mu\nu}=-\sin^{2}\theta\;{{T^{\phi}}^{\mu}}_{\nu}(1-\kappa)/8(5-3\kappa)$.
		The plots are for $\kappa=0.1$ (solid black line), $\kappa=0.2$ (dashed black line), $\kappa=0.4$ (dotted black line), $\kappa=0.6$ (solid blue line), $\kappa=0.8$ (dashed blue line), $\kappa=0.9$ (dotted blue line), $\kappa=0.99$ (solid red line) and $\kappa=0.999$ (dashed red line).}\label{warpscalar(y)}
\end{figure}

To more appropriately present the localizing effect that $\kappa$ has on the model, it is convenient to show how it can affect the warp factor. To do this, the change of coordinates as given by \eqref{variblechangespheroid} is preeminent. 
By writing 
$$
\mathrm{d}y=\sqrt{1-\kappa\sin^{2}\left(\theta\right)}\,\mathrm{d}\theta,
$$
thus one must choose $y=E\left(\theta \left|\kappa\right.\right)$, where $E\left(\theta \left|\kappa\right.\right)$ is the elliptic integral of second kind. The inverse is simply expressed abstractly by $\theta=E^{-1}\left(y\left|\kappa\right.\right)$, where $E^{-1}\left(y\left|\kappa\right.\right)$ is the inverse function of the elliptic integral of second kind. Then one is able to express the metric and the scalar field in terms of the coordinate $y$ by
\begin{equation*}
\boldsymbol{g}^{J}= e^{-2\hat{A}^{J}}\;\sin^{2}\left[E^{-1}\left(y\left|\kappa\right.\right)\right]\;\omega_{\mu\nu}\;\mathrm{d}x^{\mu}\otimes\mathrm{d}x^{\nu}+r^{2}\sin^{2}\left[E^{-1}\left(y\left|\kappa\right.\right)\right]\,\mathrm{d}\varphi\otimes\mathrm{d}\varphi+r^{2}\mathrm{d}y\otimes\mathrm{d}y,
\end{equation*}
$$
\phi=\mp4M^{2}\sqrt{1-\kappa} \operatorname{arctanh} \left\{\frac{\sqrt{\kappa } \sin \left[E^{-1}\left(y\left|\kappa\right.\right)\right]}{\sqrt{1-\kappa \cos^{2} \left[E^{-1}\left(y\left|\kappa\right.\right)\right]}}\right\},
$$
where both quantities are valued in the domain $\left[E\left(0\left|\kappa\right.\right)=0,E\left(\pi \left|\kappa\right.\right)\right]$, i.e. $y\in\left[0,E\left(\pi \left|\kappa\right.\right)\right]$. The warp factor $e^{-2\tilde{A}}$ and the scalar field $\phi$, in terms of $y$, are depicted in Fig.~\eqref{warpscalar(y)}. From Fig.~\eqref{warpscalar(y)} one can notice that the closer $\kappa$ gets to $1$ the more the thick brane looks like a thin brane, thus the more localized is the model. One had already depured it from the stress energy tensor pattern, but the above analysis paints a better picture of the localization of the model. Of course, express the same quantities in terms of $\theta$ instead of $y$, turns back the expected  analytical form.

Finally one can express the metric for models $III$ and $IV$ by
\begin{equation*}
\boldsymbol{g}^{III}= \sqrt{\left|\cos\left(4\varphi\right)\right|}\,\sin^{2}\left(\theta\right)\,\omega_{\mu\nu}\;\mathrm{d}x^{\mu}\otimes\mathrm{d}x^{\nu}+r^{2}\left[1-\kappa\sin^{2}\left(\theta\right)\right]\,\mathrm{d}\theta\otimes\mathrm{d}\theta+r^{2}\sin^{2}\left(\theta\right)\,\mathrm{d}\varphi\otimes\mathrm{d}\varphi,
\end{equation*}
\begin{equation*}
\boldsymbol{g}^{IV}= \frac{r^{2}\Lambda}{3}\cos^{2}\left(\varphi\right)\,\sin^{2}\left(\theta\right)\,\omega_{\mu\nu}\;\mathrm{d}x^{\mu}\otimes\mathrm{d}x^{\nu}+r^{2}\left[1-\kappa\sin^{2}\left(\theta\right)\right]\,\mathrm{d}\theta\otimes\mathrm{d}\theta+r^{2}\sin^{2}\left(\theta\right)\,\mathrm{d}\varphi\otimes\mathrm{d}\varphi.
\end{equation*}
with the warp factor being essentially the same as in the $C=4/r^{2}$ spherical model, as depicted in Figs.~\ref{warpfactor1} and \ref{warpfactor2}.
In this fashion one can realize that the spherical models with $C=4/r^{2}$ represent a brane-world model in a vacuum for $\phi$, while the spheroid models represent a topological defect that alternate between two vacuums like depicted in Fig.~\ref{figure5} (or \ref{scalar(y)}).

Unfortunately, it is not so simple to find stereographic coordinates for the spheroids as it was for the sphere. It is feasible to analytical calculations, but the expressions are too complicated for a meaningful analysis. Here no appeal to a different set of coordinates will be made to discuss the properties of these models.

\section{Conclusion}\label{Concl}

Some novel solutions of brane-world models in $(5+1)$-dimensions were classified and explored. 
As a preliminary proposal, brane-worlds generated by two scalar fields were obtained as solutions depending solely on a single coordinate of the co-dimensions, therefore constituting an intersection of two thick branes, where the adopted procedure involved constraining the metric components to be separable functions of the co-dimensions.
Brane-worlds on top of two different geometries of $\mathbb{S}^{2}$, the sphere and spheroid, were also constructed, and trivial and non-trivial extensions of the well known $(4+1)$-dimensional brane-world models were identified.
All the results implied into five different models, where two of them were strictly defined (models $I$ and $II$) up to some constant $p$, and the other three (models $III$, $IV$ and $V$) have maintained some degree freedom not specified by the field equations.

In the first subset, models $I$ and $II$ constituted strictly defined models, determined from a flat brane where the separation constant $p$ was set different from $1$ (or $0$) so as to strictly determine all the involved quantities from the Einstein field equations. The intrinsic difference between such models emerges from the choice of a constant parameter $c_{u}$: for real $c_{u}$ one finds model $I$, and imaginary $c_{u}$ one finds model $II$. For model $I$ one identifies a metric with a non-RS-like warp factor, but still noticing that the effective finite volume of the bulk allows for localizing fields in the brane. 
The biggest complication of model $I$ is its requirement of an infinite amount of energy to achieve the localized gravity configuration, which induces one to regard it as unphysical. Model $II$ is significantly more interesting in the sense that where several singularities may be identified, its total defect formation energy is finite. Nevertheless, due to the singularities in the stress energy tensor, one may regard model $II$ also as an unphysical configuration.

In the second subset, models $III$, $IV$ and $V$ consisted in brane-world configurations with some degree of freedom not strictly specified by Einstein field equations.
They were constructed by assuming that the auxiliary constant is set $p=0$ such that one is able to obtaining solutions for the whole range of possible values of the cosmological constant $\Lambda$ ($=0$, $>0$ or $<0$).
All the solutions contain extensions of some well-known five dimensional brane-worlds when the separation constant introduced for solving the coupled Einstein equations is set as $C=0$, being it either a trivial or non-trivial extension. More relevantly, some solutions for the sphere and spheroid geometries, where model $IV$ seems to be of particular interest to physics, have been scrutinized.

In particular, model $III$ was constructed upon a flat brane model with two scalar fields, for which the solution implies into some singularities where the warp factor exhibits cusped profile. Overall, the fields that constitute model $III$ have similar behavior to the ones in model $II$.
The above-mentioned model $IV$ seems to be the most relevant solution here depicted. This model consists in a de Sitter brane with a single scalar field with a consistently and smoothly well behaved warp factor. No cusps are found in the warp factor and the so ingrained singularities in the stress energy tensor, which emerge with other models, are avoided.  Regardless, after evaluating the Kretschmann scalar, for model $IV$, curvature singularities are still encountered. Model $IV$ also eventually discards the role of the scalar field $\phi$, since one could assume any generic form for the stress energy tensor as long as $T^{u}{}_{u}(u)$, $T^{\mu}{}_{\nu}(u)=\delta^{\mu}_{\nu}T^{v}{}_{v}(u)$ and $T_{uv}=0$. Thus model $IV$ may be found into other applications other than those for thick branes generated by scalar fields.  For completeness, considering the whole spectrum of possible values for the cosmological-like constant $\Lambda$, model $V$ was considered upon an anti-de Sitter brane with two scalar fields, where the separation constant has been set as $C=0$. It resulted into model $V$ possessing similar features to model $III$, which essentially exhibits the same singularities and cusps of the latter.

It is also worth to mention that, for models $III$, $IV$ and $V$, from the second subset, the Einstein equations do not define all fields. The scalar field $\phi$, the warp factor $\tilde{A}$ and the potential $\mathcal{V}$ of are not strictly defined from field equations and one thus still has some freedom in choosing such quantities. This opened the possibility for considering predetermined geometries for the internal space. By choosing $\tilde{A}$ and $\tilde{f}$  with predetermined geometry, one is able to achieve two setups that allow for the integration of the corresponding metric Eq.~ \eqref{generalscalar1}. From such choices of $\tilde{A}$ and $\tilde{f}$, one is able to accomplish a solution over the sphere and spheroid for models $III$ and $IV$. 
For the sphere models, two solutions for model $IV$ and nine solutions for model $III$ were achieved, one for each possible value of a discrete degree of freedom $n$. In particular, for model $IV$, for $n=1$, one has found a solution for which the stress energy tensor is smoothly well-behaved, even if the scalar $\phi$ exhibit some singularities. On the other hand, the $n=2$ model depicts a $\mathbbm{d}\mathbb{S}^{6}$ space, since there are no scalar fields, only the vacuum. Likewise, the spheroid seems to enjoy the most interesting features of the $\mathbb{S}^{2}$ models. Constructed for model $IV$, with $n=2$, one is able to achieve several interesting configurations for the warp factor and the scalar field, which guarantees the localization of gravity. Also, the spheroid solution pointed to a new solution for $(4+1)$-dimensional models, represented by metric \eqref{spheroidbased4+1}, where the actual localization parameter are given in terms of an arbitrary constant, $\kappa$.

To end up the most interesting results are due to an intersection of a scalar field with the vacuum (i.e. a model with, {\em de facto}, a single scalar). These results include model $III$ with $C=0$, which consist into the trivial extensions of some five dimensional brane-world models, and model $IV$ for non-vanishing values of $C$. In addition, model $III$, when $C$ is non-null, and $V$, in general, have also interesting warp factors. However, their corresponding stress energy tensor shows some unphysical singularities. In particular, the $\mathbb{S}^{2}$ configurations with $C=4/r^{2}$ seems to share the most interesting features. When the internal space is a sphere, a six dimensional de Sitter space is enclosed, since there no scalar fields, while when it is a spheroid, the only scalar field features a topological (kink-like) defect.

Of course, different one or two scalar fields in six dimensional setups could be additionally covered by this paper. For instance, the case where $\Lambda\neq0$, $C\neq0$ and both scalar fields are non-null is defined by an enhanced non-linear equation, which did not fit the scope of our work. Furthermore, while mostly considered models leads to finite energy configurations, all ended up having the deficiency of several singularities in between the different branes that form the manifold.
The difficulty {\em per se} is not upon finding analytical solutions to the Einstein equations, but indeed upon determining the ones that enclose the RS-like features, i.e. that can localize gravity and coupled fields. Even if surprising simple solutions have emerged when one assumed the internal space $\mathbb{B}^{2}$ to be $\mathbb{S}^{2}$, other setups may not be so treatable as for finding solutions that confine gravity.

Considering such an extended analysis, our next steps should include the models discussed here in the study of the localization of gravitational and matter fields, so as to realistically identify if their concerned configurations are physically appealing.

\acknowledgments
HMG is grateful for the financial support provided by
CNPq (Grant No. 141924/2019-5). The work of AEB is supported by the Brazilian Agencies FAPESP (Grant No. 2018/03960-9) and CNPq (Grant No. 301000/2019-0).

\appendix

\section{Details of models $I$ and $II$}\label{longermodelIandII}

From solutions \eqref{warpfactorv} and \eqref{warpfactoru}, one is able to write the metric in the form
\begin{align}\label{intersectingflatbranemetric}
	\boldsymbol{g}=\sqrt{\cosh \Big[2c_{u}\big(u+u_{0}\big)\Big]\cosh \Big[2c_{v}\big(v+v_{0}\big)\Big]}\eta_{\mu\nu}\mathrm{d}x^{\mu}\otimes\mathrm{d}x^{\nu}+\sqrt{\frac{\cosh^{p} \Big[2c_{v}\big(v+v_{0}\big)\Big]}{\cosh^{p-1} \Big[2c_{u}\big(u+u_{0}\big)\Big]}}\left(\mathrm{d}u\otimes\mathrm{d}u+\mathrm{d}v\otimes\mathrm{d}v\right),
\end{align}
from which it is indeed not so clear whether gravity is localized in the brane. One should notice that if $Re(c_{v})\neq0$, since $p\geq1/2$, there would be no way of ``localizing'' fields in the ``direction of $v$'', since both conformal factors which multiply $\eta_{\mu\nu}\mathrm{d}x^{\mu}\otimes\mathrm{d}x^{\nu}$ and $\mathrm{d}v\otimes\mathrm{d}v$ ``increase with $v$''. Thus one is constrained to assume $Re(c_{v})=0$ to achieve an acceptable physical solutions, which means that the $v$ coordinate can be compactified into a circle $\mathbb{S}^{1}$, or in some other words, $v=r\varphi$, where $r$ is the radius of $\mathbb{S}^{1}$ and $\varphi\in\mathbb{S}^{1}$. Since the metric must be continuous in $\mathbb{S}^{1}$, the $e^{-2\hat{A}}$ factor must be continuous in $\mathbb{S}^{1}$, i.e. 
$$
\left|\cos \left(2\left|c_{v}\right| r\,2\pi\right)\right|=\left|\cos (0)\right|=1\implies \left|c_{v}\right|=\frac{n}{4r}\text{, }n\in\mathbb{N}.
$$
Likewise, one still needs to verify the localization along the ``direction of $u$'', which can be achieved in two different ways. 

For instance, when $Im(c_{u})=0$, if one sets $p$ large enough (i.e. at least $p\geq3$), the effective volume associated with $\mathbb{B}^{2}$ becomes finite. Even though the warp factor does not have a RS-like profile, because the volume of $\mathbb{B}^{2}$ is finite, one can still possibly ``localize'' gravity and other fields. Otherwise, when $Re(c_{u})=0$, in a similar fashion to the content discussed for coordinate $v$, the space coordinate $u$ can be compactified as a circle $\mathbb{S}^{1}$. In this case, one must impose $1/2\leq p\leq3$, otherwise the effective volume is not finite (this is clearer when observing the metric from \eqref{intersectingflatbranemetric} for $p<3$). Therefore, it is imperative to choose either $Im(c_{u})=0$ for models with $p\geq3$ or $Re(c_{u})=0$ for models with $p\leq3$ in order to obtain consistent solutions with localized gravity. 

Even if one is able to localize fields in the brane, the configuration may still not be physical. If the total energy associated with the configuration is infinite, then one can argue that the solutions are not physical ones. Therefore, to realize the total energy of the system one thus writes the stress energy tensor as
\begin{equation*}
	T_{\mu\nu}=-e^{-2A}\eta_{\mu\nu}\left[g^{uu}\frac{\left(\phi_{,u}\right)^{2}}{2}+g^{\varphi\varphi}\frac{\left(\zeta_{,\varphi}\right)^{2}}{2}+\mathcal{V}\right]=-\eta_{\mu\nu}\sqrt{\frac{\cosh^{p} \left(2c_{u}u\right)}{\left|\cos \left(\frac{n\varphi}{2}\right)\right|^{p-1} }}\left[\frac{\left(\phi_{,u}\right)^{2}}{2}+\frac{\left(\zeta_{,\varphi}\right)^{2}}{2r^{2}}+\mathcal{V}\right].
\end{equation*}
As claimed above, the energy density which is computed from $\sqrt{-\mathrm{g}}T_{\mu\nu}$ must be finite, otherwise the total energy in this configuration will not be finite. One thus has
\begin{equation}
	T_{\mu\nu}\sqrt{-\mathrm{g}}=-\frac{r}{2}\eta_{\mu\nu}\left[\left(\phi_{,u}\right)^{2}+\frac{\left(\zeta_{,\varphi}\right)^{2}}{r^{2}}+2\mathcal{V}\right]\left|\cos \left(\frac{n\varphi}{2}\right)\right|^{3/2}\cosh^{3/2} \left(2c_{u}u\right)\label{density}
\end{equation}
from which it can be noticed that, if $Im(c_{u})=0$, since $\zeta=\zeta(v)$, the integration of Eq.~\eqref{density} throughout space will necessarily be infinite. Therefore one may claim that the $Im(c_{u})=0$ configurations requires an unphysical infinite amount of energy to be realized. Following a similar analysis, from Eq.~\eqref{density}, no conclusive assertion about the choice of $Re(c_{u})=0$ instead of $Im(c_{u})=0$ can be performed.
However, as a matter of completeness, the calculations will be carried out for both configurations.
\subsection{Model I}
\begin{figure}[!htb]
	\subfloat[]{\label{warpI}\includegraphics[scale=0.69]{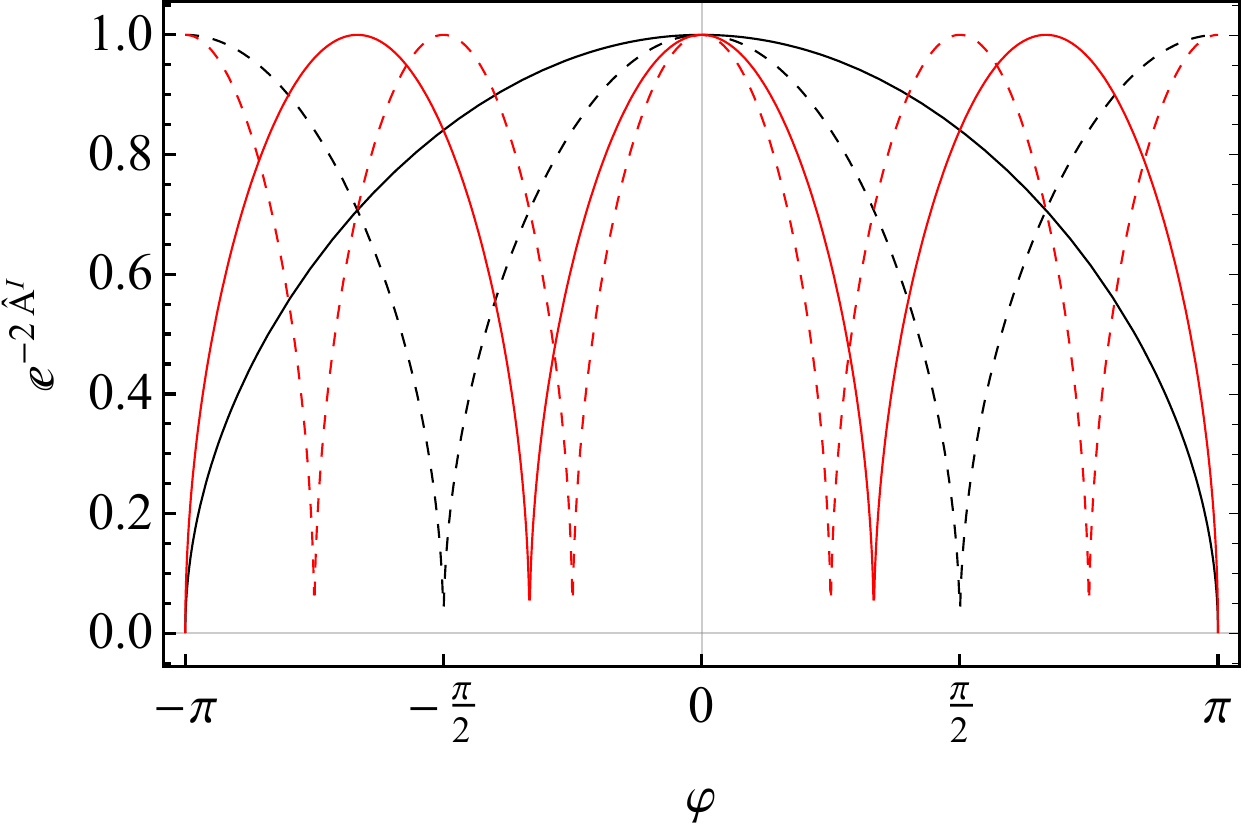}}$\;\;$
	\subfloat[]{\label{conformalI}\includegraphics[scale=0.69]{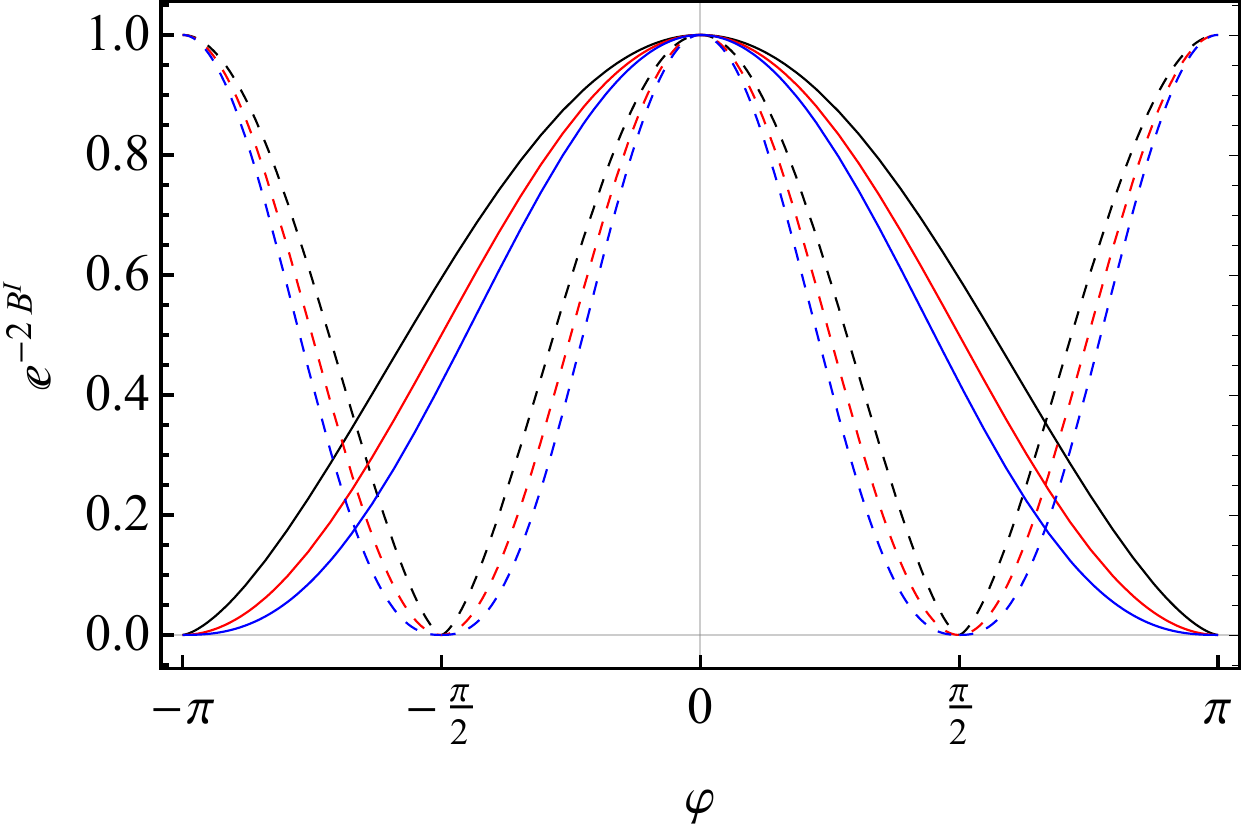}}
	\caption{(Color online) (a) Warp factor $e^{-2\hat{A}}$ of model $I$ as a function of $\varphi$, for $n=1$ (solid black line), $n=2$ (dashed black line), $n=3$ (solid red line) and $n=4$ (dashed red line). (b) Conformal factor $e^{-2B^{I}}$ of model $I$ as a function of $\varphi$, for $p=3$ (black), $p=4$ (red) and $p=5$ (blue), the solid and dashed lines correspond to $n=1$ and $n=2$, respectively.}
\end{figure}

Model $I$ is resumed by expressions \eqref{metricI}, \eqref{potentialI}, \eqref{scalarphiI} and \eqref{scalarzetaI} and the dependence on $\varphi$ for the warp and conformal factors are depicted in Figs.~\ref{warpI} and \ref{conformalI}. 

\begin{figure}[!htb]
	\subfloat[]{\includegraphics[scale=0.69]{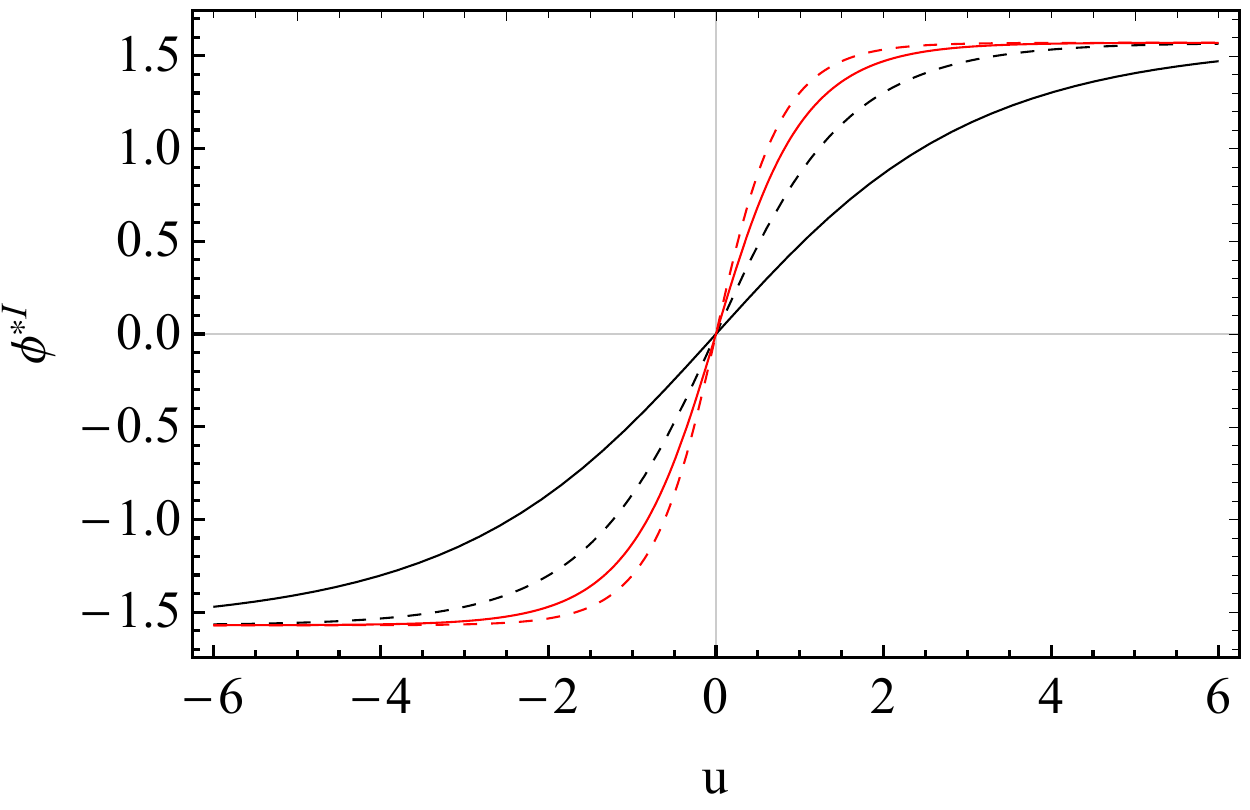}}$\;\;$
	\subfloat[]{\label{zetaI}\includegraphics[scale=0.69]{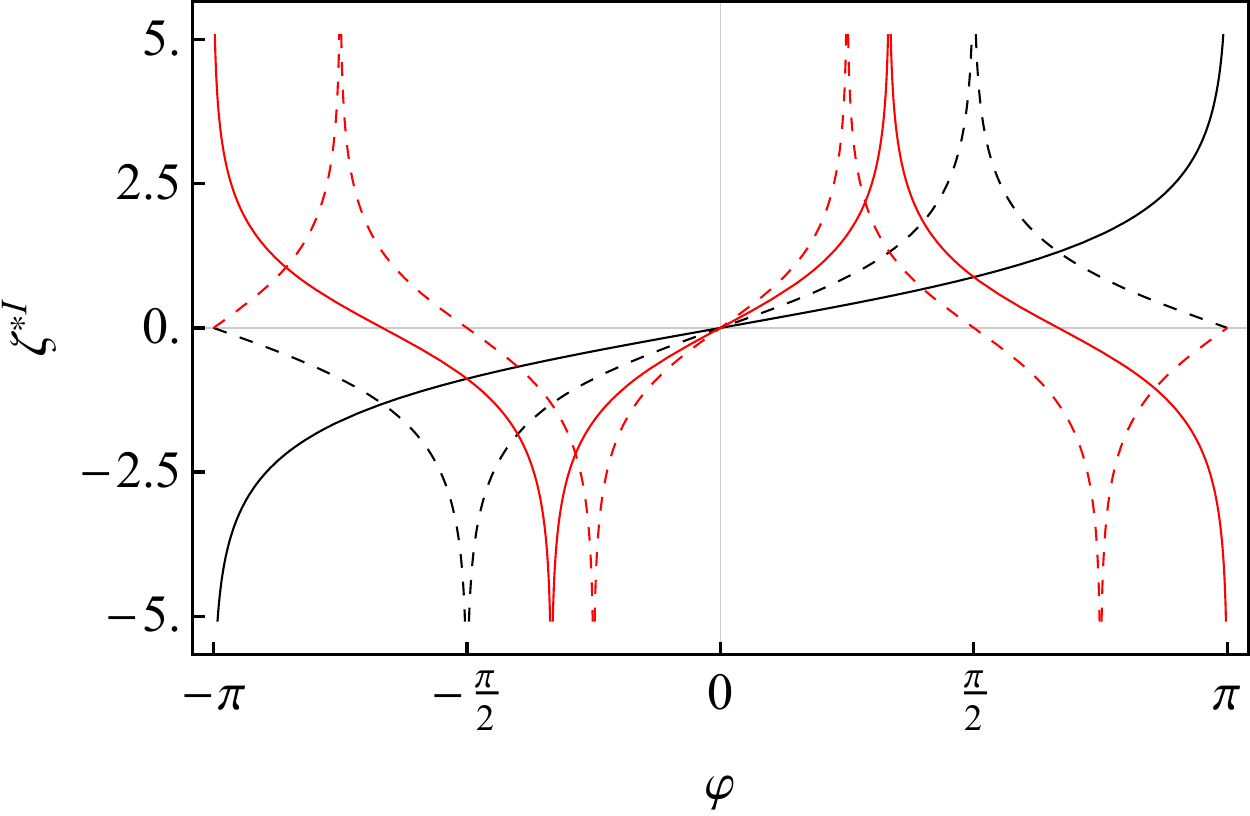}}
	\caption{(Color online) The scalar fields of model $I$: (a) $\phi^{*I}=\phi^{I}/\sqrt{2p-5}\, M^{2}$; (b) $\zeta^{*I}=\zeta^{I}/\sqrt{3+2p}\,M^{2}$. The plots are for $n=1$ (solid black line), $n=2$ (dashed black line), $n=3$ (solid red line) and $n=4$ (dashed red line).}\label{scalarfieldsI}
\end{figure}

In particular, for $b_{\phi}=b_{\zeta}=-1$, one has $c_{u}=\left|c_{v}\right|={n}/{4r}$. In this case, since $a_{\phi}\geq0$ one realizes that $p\geq5/2$, which is a tautology since $p\geq3$. For this choice, one finds the scalar fields and potential in the following form,
\begin{align}
	\mathcal{V}^{I}&=0,\nonumber
	\\
	\phi^{I}&=\pm\sqrt{2p-5}\, M^{2} \arcsin\left[\tanh\left(\frac{nu}{2r}\right)\right],\label{scalarphiI2}
	\\
	\zeta^{I}&=\pm \sqrt{3+2p}\,M^{2}\operatorname{arctanh}\left[\sin \left(\frac{n\varphi}{2}\right)\right].\label{scalarzetaI2}
\end{align}
The scalar fields are depicted in Fig.~\ref{scalarfieldsI}. The structure of the scalar field $\zeta^{I}$, as given by \eqref{scalarzetaI2}, shall recurrently appear as a driver for $(5+1)$-dimensional thick brane-worlds. As it shall be noticed in the following models, the scalar field dependence on the angular-like variables in much sense reproduce the behavior depicted in Fig.~\ref{scalarfieldsI}. Interestingly, the scalar field $\phi^{I}$ is zero when $p=5/2$, implying into a singular configuration with a single scalar field $\zeta^{I}$, with $\mathcal{V}^{I}=0$.

Finally, the stress energy tensor for model $I$ is simply written as
\begin{equation*}
	T_{\mu\nu}=\frac{M^4 n^2\eta_{\mu\nu}}{8r^{2}}\frac{ (5-2 p) \text{sech}^2\left(\frac{n u}{2 r}\right)-(2 p+3) \sec ^2\left(\frac{n \varphi }{2}\right)}{\sqrt{\operatorname{sech}^{p} \left(\frac{n u}{2 r}\right)\left|\cos \left(\frac{n\varphi}{2}\right)\right|^{p-1} }},
\end{equation*}
where its several singularities are consistent with the numerous cusps exhibited by the warp factor. Also, an infinite amount of energy is necessary to achieve such configuration, as one can check after integrating the previous expression throughout space coordinates.

\subsection{Model II}
\begin{figure}[!htb]
	\subfloat[]{\includegraphics[scale=0.69]{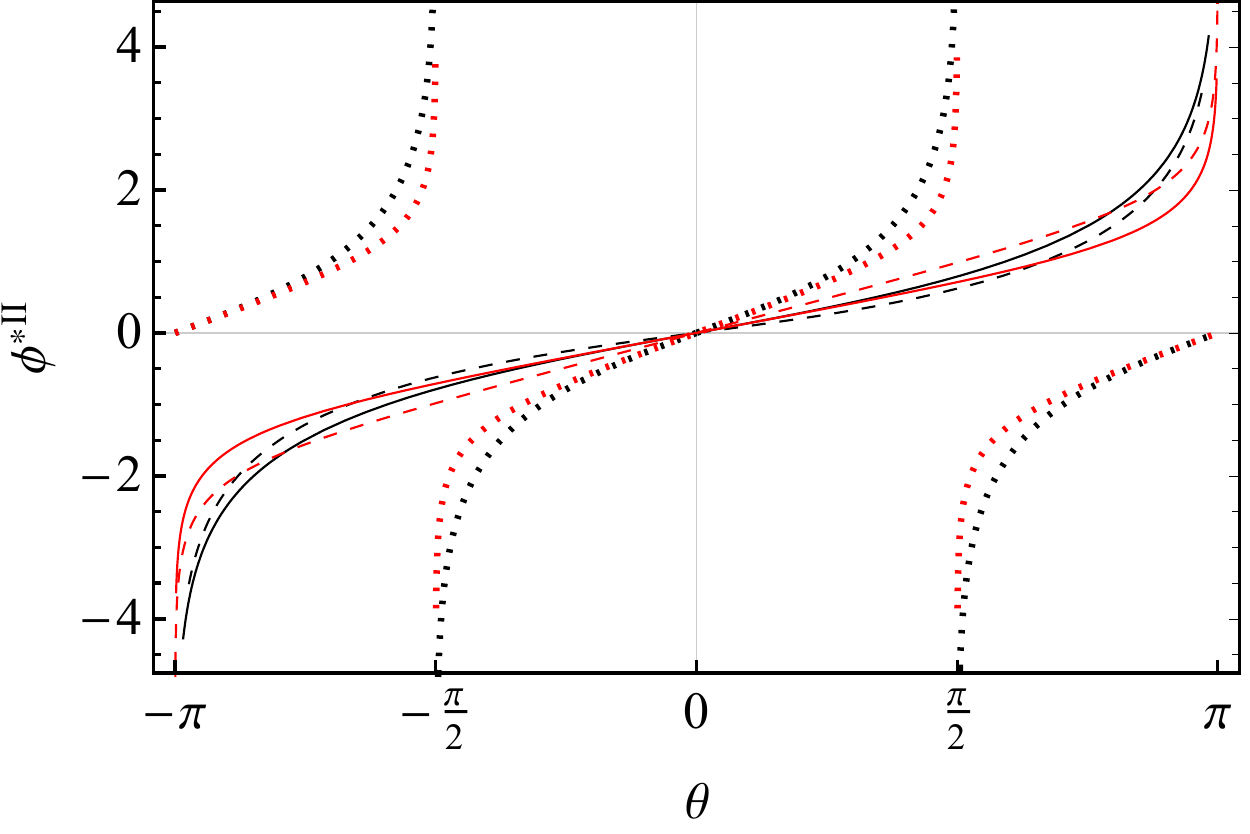}}$\;\;$
	\subfloat[]{\includegraphics[scale=0.69]{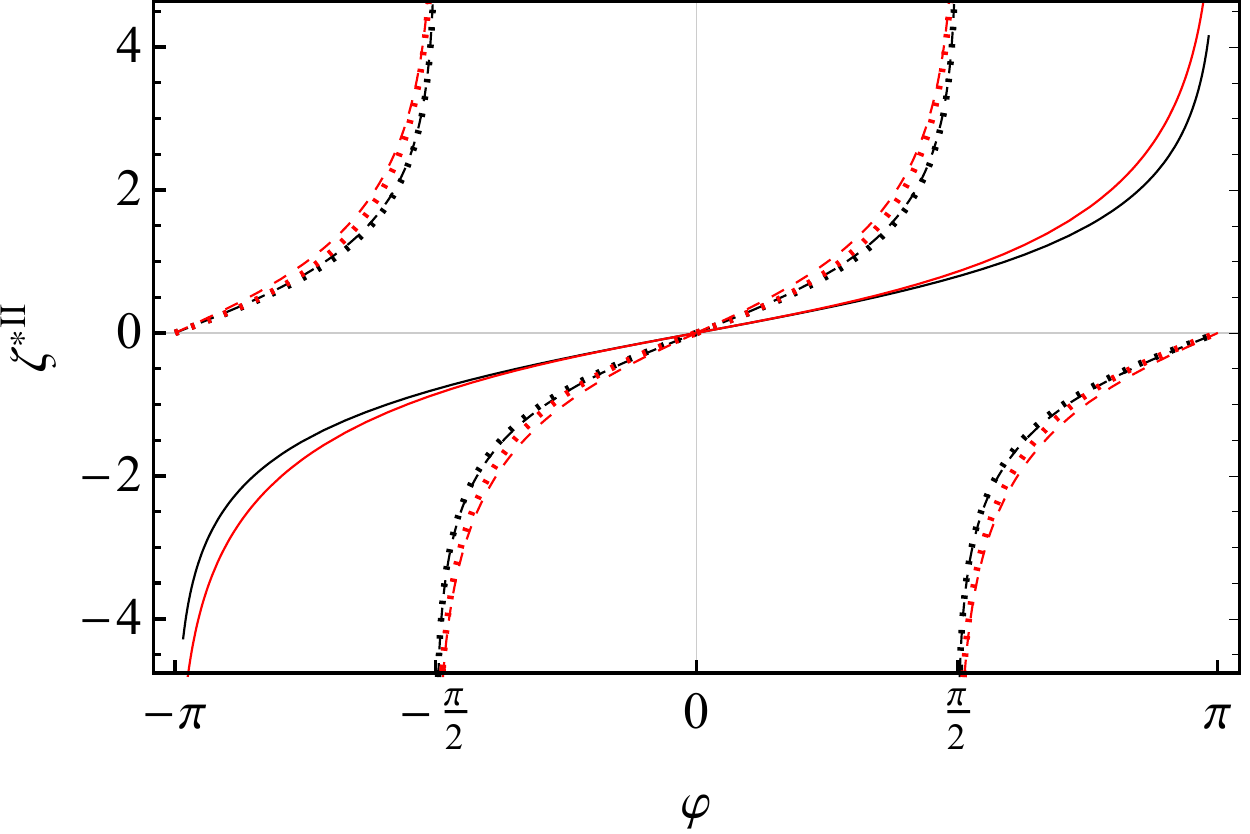}}
	\caption{(Color online) Scalar fields of model $II$ for $p=1/2$ (black) and $p=2$ (red): (a) $\phi^{*II}=\phi^{II}/ M^{2}$ as a function of $\theta$; (b) $\zeta^{*II}=\zeta^{II}/M^{2}$ as a function of $\varphi$. The plots are for $n=l=1$ (solid lines), $n=2$, $l=1$ (dashed lines) and $n=l=2$ (dotted lines).}\label{scalarII}
\end{figure}

\begin{figure}[!htb]
	\subfloat[]{\includegraphics[scale=0.58]{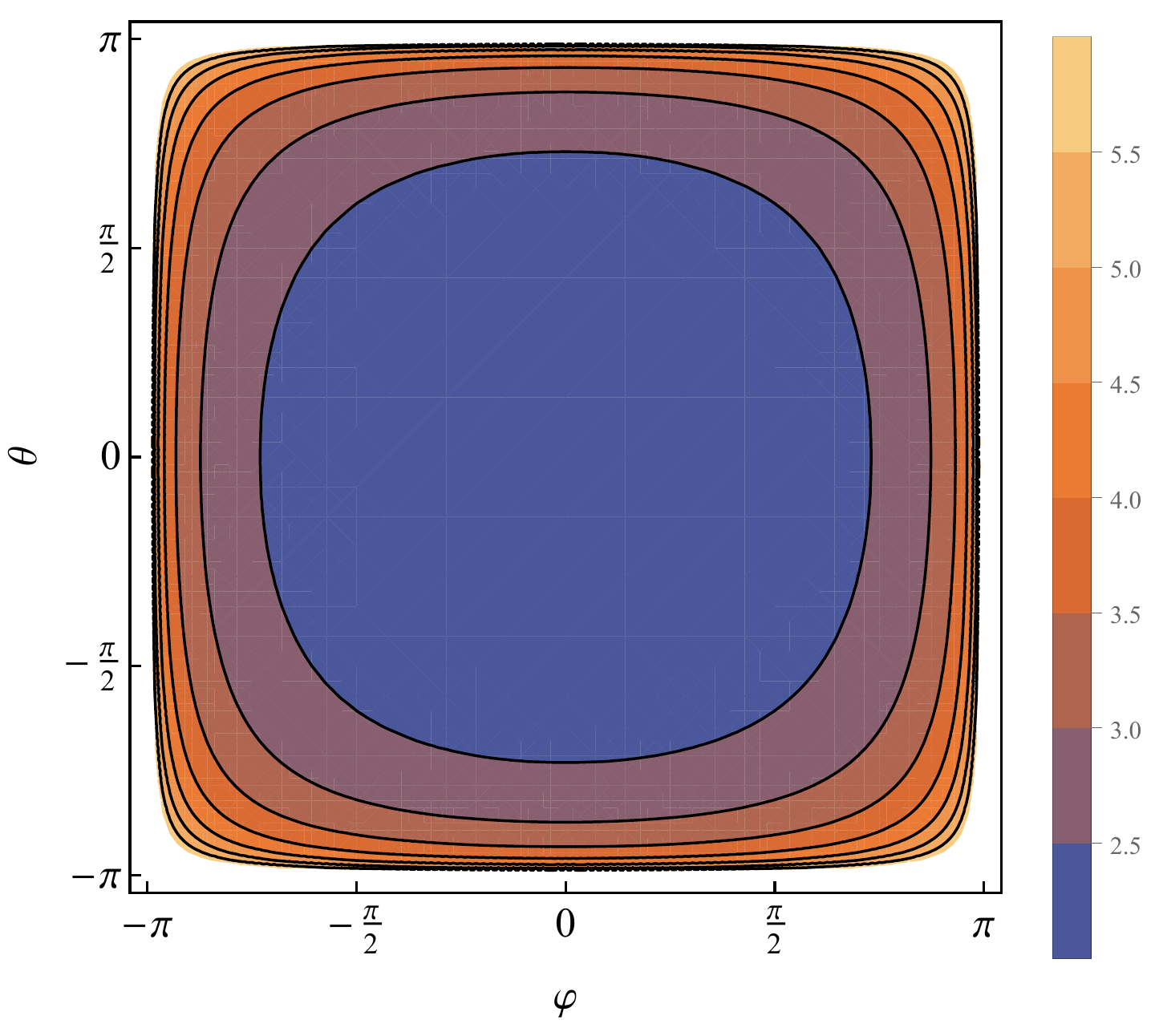}}$\;\;$
	\subfloat[]{\includegraphics[scale=0.58]{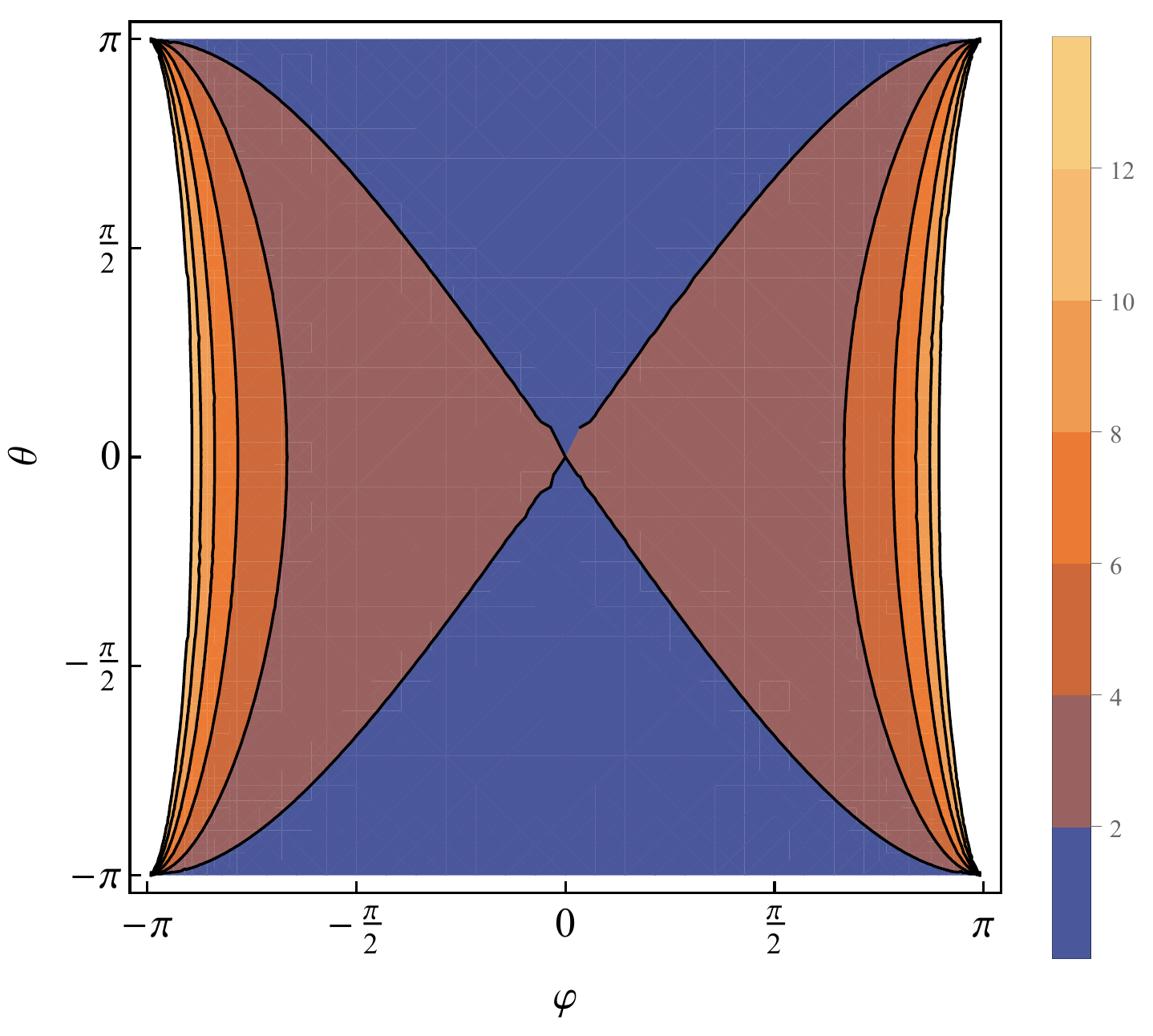}}
	\caption{(Color online) Potential $\mathcal{V}^{II}$ of model $II$ as a function of $\left(\theta,\varphi\right)$, for $p=1/2$ (a) and $p=2$ (b).}\label{figpotentialII}
\end{figure}

Model $II$ can be summarized by expressions \eqref{p<3}, \eqref{potentialII}, \eqref{scalarphiII} and \eqref{scalarzetaII}. The soft shortcoming of model $II$ is concerned with not being possible to rewrite $\mathcal{V}$ as function of $\phi$ and $\zeta$, since the expressions for \eqref{scalarphiII} and \eqref{scalarzetaII} are not invertible.
It would be advisable since one had started with the assumption that $\mathcal{V}=\mathcal{V}(\phi,\,\zeta)$. The shape of the scalar fields, $\phi$ and $\zeta$, and of the potential $\mathcal{V}$ are presented in Figs.~\ref{scalarII} and \ref{figpotentialII}, respectively.

\begin{figure}[!htb]
	\subfloat[]{\includegraphics[scale=0.58]{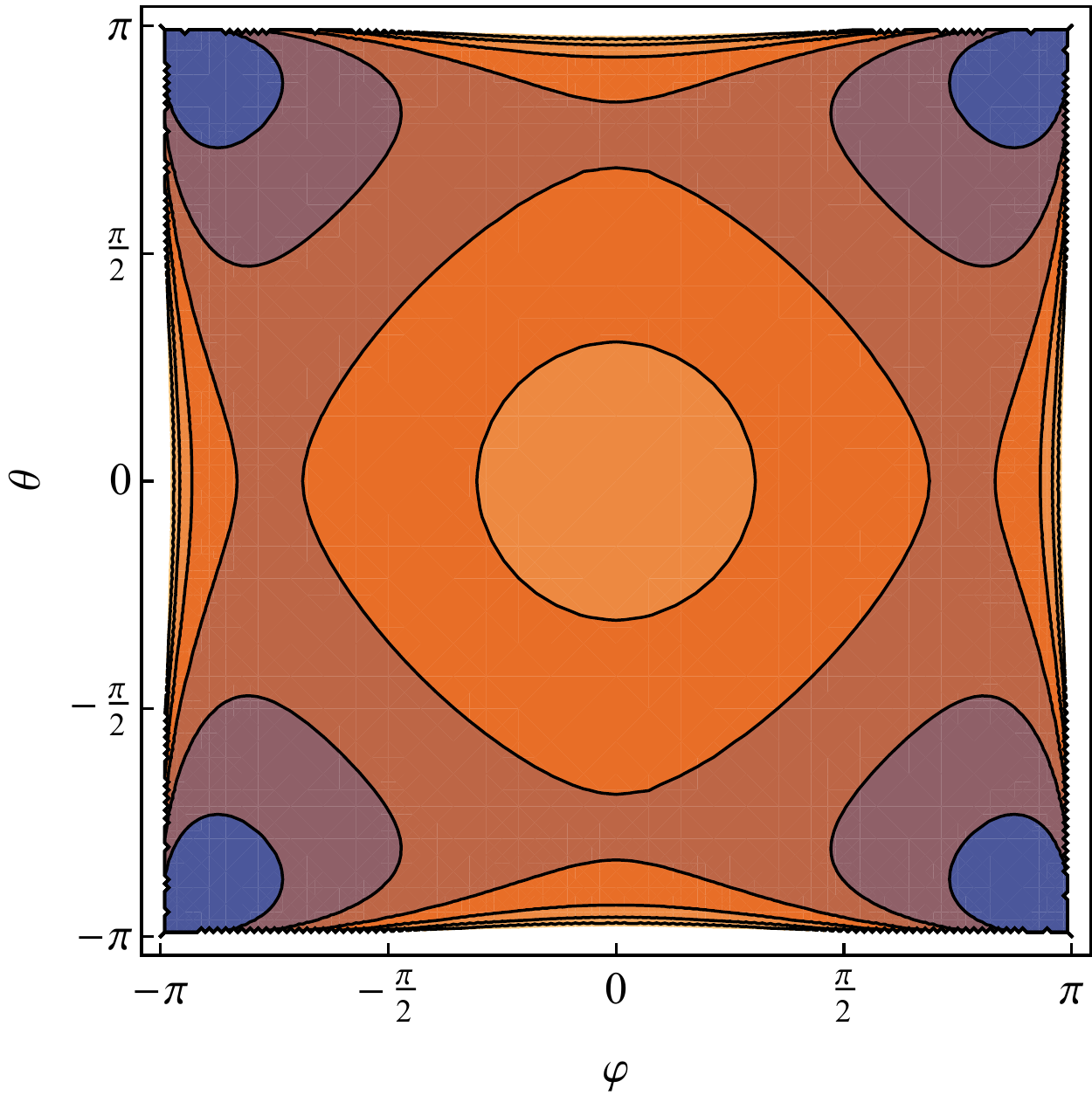}}$\;\;$
	\subfloat[]{\includegraphics[scale=0.58]{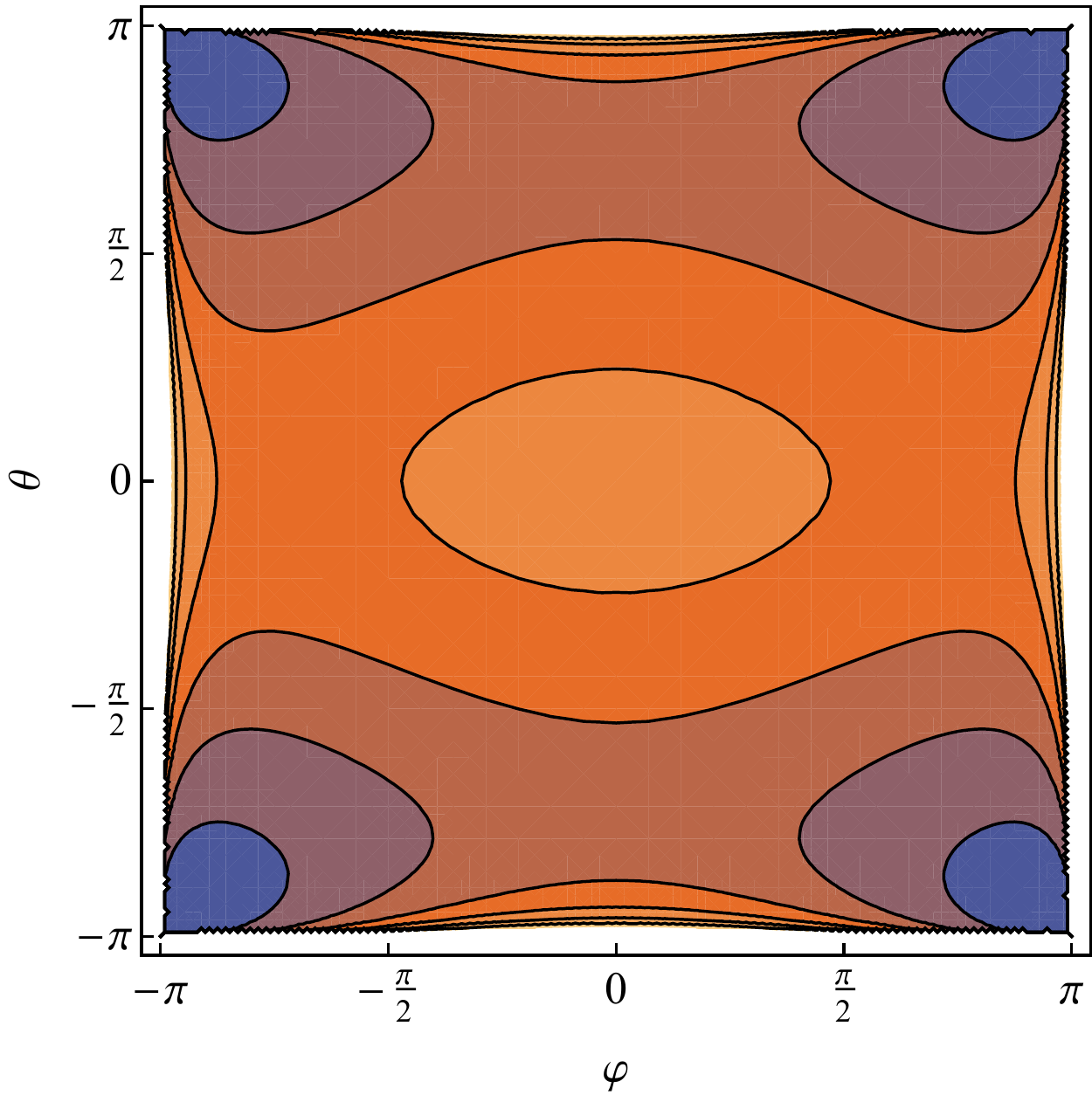}}$\;\;$
	\includegraphics[scale=0.57]{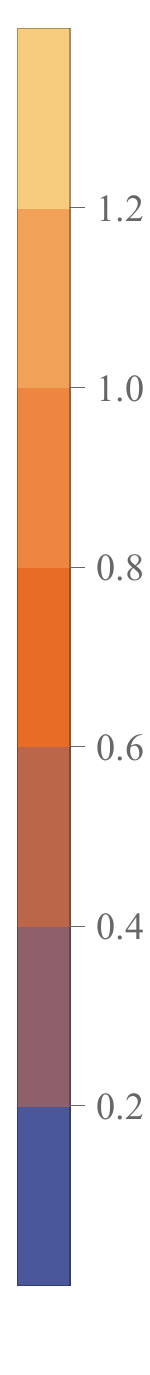}
	\caption{(Color online) Energy density $-T^{II}_{\mu\nu}\sqrt{-\mathrm{g}}$ of model $II$ as a function of $(\theta,\varphi)$, for $p=1/2$ (a) and $p=2$ (b).}\label{densityII}
\end{figure}

Analogously, the stress energy tensor is expressed by
\begin{equation*}
	T^{II}_{\mu\nu}=-2M^{4}\eta_{\mu\nu}\sqrt{\frac{\left|\cos \left(\frac{l\theta}{2}\right)\right|^{p}}{\left|\cos \left(\frac{n\varphi}{2}\right)\right|^{p-1} }}\left\{a_{\phi}\left[1-b_{\phi}\tan^{2}\left(\frac{l\theta}{2}\right)\right]+a_{\zeta}\left[1-b_{\zeta}\tan^{2}\left(\frac{n\varphi}{2}\right)\right]+\sqrt{\frac{\left|{\cos} \left(\frac{l\theta}{2}\right)\right|^{p-1}}{\left|{\cos}
			\left(\frac{n\varphi}{2}\right)\right|^{p}}}\left(\frac{{l}^{2}}{4{\rho}^{2}}+\frac{{n}^{2}}{4{r}^{2}}\right)\right\}.
\end{equation*}
which exhibits several singularities, depending on the values for $n$ and $l$. In fact, for $n=1$ and $l=1$ it has 2 singularities, one at $\varphi=\pi$ (or $-\pi$) and another one at $\theta=\pi$ (or $-\pi$). These singularities explains the number of cusps in the warp factor. In order to realize physically consistent solutions, the required energy to achieve their internal structure must be finite. From the perspective of the bulk, such a required energy is given by
\begin{align*}
	{E}^{II}_{\mu\nu}=\int_{\mathbb{E}^{6}}{T}^{II}_{\mu\nu}\sqrt{-\mathrm{g}}\mathrm{d}^{6}x\propto \int^{\pi}_{-\pi}\int^{\pi}_{-\pi}\frac{a_{\phi}\left[1-b_{\phi}\tan^{2}\left(\frac{l\theta}{2}\right)\right]+a_{\zeta}\left[1-b_{\zeta}\tan^{2}\left(\frac{n\varphi}{2}\right)\right]+\sqrt{\frac{\left|{\cos} \left(\frac{l\theta}{2}\right)\right|^{p-1}}{\left|{\cos}
				\left(\frac{n\varphi}{2}\right)\right|^{p}}}\left(\frac{{l}^{2}}{4{\rho}^{2}}+\frac{{n}^{2}}{4{r}^{2}}\right)}{\left|\sec \left(\frac{l\theta}{2}\right)\right|^{3/2}\left|\sec \left(\frac{n\varphi}{2}\right)\right|^{3/2}}\mathrm{d}\theta\mathrm{d}\varphi,
\end{align*}
with last integral converging for several values of $n$, $l$ and $p$\footnote{For instance, when $n=l=1$ and $p=1/2$, it integrates to
	$$
	\frac{2 \pi \Gamma \left(\frac{9}{8}\right)^2}{\Gamma \left(\frac{13}{8}\right)^2}+\frac{33 \pi ^3}{2^{8}\Gamma \left(\frac{7}{4}\right)^4}.
	$$}.
Therefore, even though model $II$ exhibits several singularities as depicted by the stress energy tensor, the total energy necessary to accomplish model $II$ is finite. This is an evinced advantage with respect to the model $I$. The form of the energy density for model $II$ ($T^{II}_{\mu\nu}\sqrt{-\mathrm{g}^{II}}$) is depicted in Fig.~\ref{densityII}. Although model $II$ has finite total energy, one may still argue against its physical significance, due to its recurrent singularities, a shortcoming that must be considered in the following model issues.

\section{The Kretschmann scalar ansatz (\ref{intersectingmetric})}\label{Kretschmann}
For model $IV$ one has reached the conclusion that the stress energy tensor is non-singular, but one may still find curvature singularities. From evaluating the Ricci scalar (or $R^{MN}R_{MN}$) one does not realize singularities, thus another parameter, the Kretschmann scalar, must be assessed in order to recognize the singularities of the model. The following calculations thus present the path for the Kretschmann scalar determination.
	
One once again rescales the metric by $\boldsymbol{g}=e^{-2A}\boldsymbol{\hat{g}}$, and determines the corresponding Riemann tensor,
\begin{align*}
{R^{I}}_{KJM}&={\hat{R}^{I}}{}_{KJM}-\hat{\nabla}_{J}\hat{\nabla}_{K}A\delta^{I}_{M}+\hat{\nabla}_{J}\hat{\nabla}_{M}A\delta^{I}_{K}+\hat{g}^{IS}\hat{g}_{JM}\hat{\nabla}_{K}\hat{\nabla}_{S}A-\hat{g}^{IS}\hat{g}_{JK}\hat{\nabla}_{M}\hat{\nabla}_{S}A
\\
&\qquad\qquad\qquad\qquad+A_{,J}A_{,M}\delta^{I}_{K}-A_{,K}A_{,J}\delta^{I}_{M}+A_{,K}A_{,L}\hat{g}^{LI}\hat{g}_{JM}-A_{,M}A_{,L}\hat{g}_{JK}\hat{g}^{LI}
\\
&\qquad\qquad\qquad\qquad\qquad\qquad\qquad\qquad\qquad\qquad+\hat{g}^{PS}A_{,P}A_{,S}\left(\hat{g}_{JK}\delta^{I}_{M}-\hat{g}_{JM}\delta^{I}_{K}\right),
\end{align*}
where ${\hat{R}^{I}}{}_{KJM}$ is obtained solely from $\boldsymbol{\hat{g}}$. The Kretschmann scalar is written in terms of the warp factor and the metric $\boldsymbol{\hat{g}}$ as
\begin{align*}
K={R}^{FKJM}{R}_{FKJM}=& e^{4A}\bigg\{\mathcal{K}+\hat{\mathfrak{K}}+4{\hat{R}}{}^{LN}\hat{\nabla}_{N}\hat{\nabla}_{L}A+4\hat{G}^{LN}A_{,L}A_{,N}+4\hat{g}^{HS}\hat{\nabla}_{H}\hat{\nabla}_{S}A\hat{g}^{KJ}\hat{\nabla}_{J}\hat{\nabla}_{K}A
\\
&+4\left(d-2\right)\hat{g}^{JN}\hat{g}^{KH}\hat{\nabla}_{H}\hat{\nabla}_{N}A\hat{\nabla}_{K}\hat{\nabla}_{J}A+8\left(d-2\right)\hat{g}^{JN}\hat{g}^{KH}A_{,K}A_{,J}\hat{\nabla}_{H}\hat{\nabla}_{N}A
\\
&-8\left(d-2\right)\hat{g}^{PS}A_{,P}A_{,S}\hat{g}^{KJ}\hat{\nabla}_{J}\hat{\nabla}_{K}A+2\left(d-2\right)\left(d-1\right)\hat{g}^{PS}A_{,P}A_{,S}\hat{g}^{NH}A_{,N}A_{,H}\bigg\},
\end{align*}
where one has defined,
$$
\hat{K}=\hat{R}^{FKJM}\hat{R}_{FKJM}=\hat{R}^{\mu\nu\kappa\rho}\hat{R}_{\mu\nu\kappa\rho}+\hat{R}^{ijkl}\hat{R}_{ijkl}=\mathcal{R}^{\mu\nu\kappa\rho}\mathcal{R}_{\mu\nu\kappa\rho}+\hat{\Sigma}^{ijkl}\hat{\Sigma}_{ijkl}=\mathcal{K}+\hat{\mathfrak{K}},
$$
$$
\mathcal{K}(x^{\mu})=\hat{R}^{\mu\nu\kappa\rho}\hat{R}_{\mu\nu\kappa\rho}=\mathcal{R}^{\mu\nu\kappa\rho}\mathcal{R}_{\mu\nu\kappa\rho},
$$
$$
\hat{\mathfrak{K}}(u,v)=\hat{R}^{ijkl}\hat{R}_{ijkl}=\hat{\Sigma}^{ijkl}\hat{\Sigma}_{ijkl}.
$$

After noticing that $A=A(u,v)$, for six dimensional space, in terms of $\boldsymbol{\omega}$, $A$ and $\boldsymbol{\sigma}$, one finds,
$$
K=e^{4A}\mathcal{K}(x^{\mu})+\mathfrak{K}(u,v)+16\sigma^{mh}\sigma^{jn}{\triangle}_{n}{\triangle}_{h}A{\triangle}_{j}{\triangle}_{m}A-32\sigma^{mh}\sigma^{jn}A_{,n}A_{,h}{\triangle}_{j}{\triangle}_{m}A+40\sigma^{mj}A_{,m}A_{,j}\sigma^{ps}A_{,p}A_{,s}.
$$
In particular, for the $p=0$ solutions ($III$, $IV$ and $V$), for which the metric is given by
$$
\boldsymbol{g}=e^{-2\hat{A}}e^{-2\tilde{A}}{\omega}_{\mu\nu}\mathrm{d}x^{\mu}\otimes\mathrm{d}x^{\nu}+ e^{-2\tilde{A}}\left(\mathrm{d}u\otimes\mathrm{d}u+\mathrm{d}v\otimes\mathrm{d}v\right)\implies \boldsymbol{\sigma}=e^{-2\tilde{A}}\boldsymbol{\gamma},
$$
it follows that
\begin{equation}\label{Kretschmannp=0}
K=e^{4\tilde{A}}\bigg( e^{4\hat{A}}\mathcal{K}(x^{\mu})+16\hat{A}_{,vv}{}^{2}-32\hat{A}_{,vv}\hat{A}_{,v}{}^{2}+40\hat{A}_{,v}{}^{4}+80\hat{A}_{,v}{}^{2}\tilde{A}_{,u}{}^{2}-32\hat{A}_{,vv}\tilde{A}_{,u}{}^{2}+e^{-4\tilde{A}}\mathfrak{K}(u)+16\tilde{A}_{,uu}{}^{2}+40\tilde{A}_{,u}{}^{4}\bigg),
\end{equation}
where $\mathfrak{K}(u)$ is the Kretschmann scalar out of $\boldsymbol{\sigma}=e^{-2\tilde{A}}\boldsymbol{\gamma}$, i.e. a single function of $u$.

\section{Determining the geometry for Sec.~\ref{predetermined}}

In Sec.~\ref{predetermined} one uses a predetermined geometry for the space $\mathbb{B}^{2}$. Here it is shown how the aforesaid is achieved for $\mathbb{B}^{2}=\mathbb{S}^{2}$. One starts by first considering a general geometry for $\mathbb{S}^{2}$, namely an ellipsoid. In this context an ellipsoid is $\mathbb{S}^{2}$, but with a particularly distorted metric. This metric can be determined from its immersion in $\left(\mathbb{R}^{(3)},\boldsymbol{\gamma}\right)$, where $\boldsymbol{\gamma}$ is the usual euclidean metric. For a triaxial ellipsoid, the immersion is a map
$$
\Phi:\mathbb{S}^{2}\rightarrow \mathbb{R}^{3},
$$
where it is defined by
$$
(\theta,\varphi)\mapsto\Phi(\theta,\varphi):=\left(a\cos\left(\varphi\right)\,\sin\left(\theta\right),b\sin\left(\varphi\right)\,\sin\left(\theta\right),c\cos\left(\theta\right)\right),
$$
with $a$, $b$ and $c$ being the three radii that define ellipsoid.

Then, from $\Phi$, it is possible to define a pull-back $\Phi^*$ of $T^{(0,2)}\mathbb{R}^{3}$:
$$
\Phi^*:T^{(0,2)}\mathbb{R}^{(3)}\rightarrow T^{(0,2)}\mathbb{S}^{2},
$$
according to
$$
\boldsymbol{\gamma}\mapsto \boldsymbol{\varepsilon}\left(X,Y\right):=\left(\Phi^{*}\boldsymbol{\gamma}\right)\left(X,Y\right)\equiv\boldsymbol{\gamma}\left(\Phi_{*}X,\Phi_{*}Y\right),
$$
where $\Phi_{*}$ is the push-forward on $T M$ induced by $\Phi$, which is defined by
$$
\left(\Phi_{*}X\right)^{i}=X^{a}\frac{\partial\left(x^{i}\circ\Phi\right)}{\partial y^{a}},
$$
where, finally, $x$ and $y$ are the coordinate chart maps for $\mathbb{R}^{3}$ and $\mathbb{S}^{2}$, respectively, also with $i\in\{1,2,3\}$ and $a\in\{1,2\}$. Substituting the above expressions back, one finds the components of the metric $\varepsilon$ of $\mathbb{S}^{2}$,
$$
\varepsilon_{ab}=\gamma_{ij}\frac{\partial\left(x^{i}\circ\Phi\right)}{\partial y^{a}}\frac{\partial\left(x^{j}\circ\Phi\right)}{\partial y^{b}},
$$
which leads to the metric,
\begin{widetext}
\begin{align*}
	\boldsymbol{\varepsilon}=&\left[a^{2}\cos^{2}\left(\varphi\right)\,\cos^{2}\left(\theta\right)+b^{2}\sin^{2}\left(\varphi\right)\,\cos^{2}\left(\theta\right)+c^{2}\sin^{2}\left(\theta\right)\right]\mathrm{d}\theta\otimes\mathrm{d}\theta
	\\
	&\qquad\qquad\qquad\qquad\qquad\qquad\qquad\quad+\frac{a-b}{2}\sin\left(2\varphi\right)\sin\left(2\theta\right)\mathrm{d}\varphi\otimes\mathrm{d}\theta+\left[a^{2}\sin^{2}\left(\varphi\right)+b^{2}\cos^{2}\left(\varphi\right)\right]\sin^{2}\left(\theta\right)\mathrm{d}\varphi\otimes\mathrm{d}\varphi.
\end{align*}
\end{widetext}
This is a triaxial set-up. It is a bit complicated, and even if one could find coordinates such that the off diagonal terms are null, which is always possible, one would end up with an extremely enhanced metric which can be simplified by making assumptions about the radii $a$ and $b$ into a spheroid configuration or a diaxial ellipsoid, i.e. $a=b=r$ and $c=\rho$, which would return the metric
\begin{equation}
\boldsymbol{\epsilon}=\left[r^{2}\cos^{2}\left(\theta\right)+\rho^{2}\sin^{2}\left(\theta\right)\right]\mathrm{d}\theta\otimes\mathrm{d}\theta+r^{2}\sin^{2}\left(\theta\right)\,\mathrm{d}\varphi\otimes\mathrm{d}\varphi.\label{spheroid}
\end{equation}
This result removes the off-diagonal terms, which turn the procedure into simpler analytical calculations, which should be still simpler for $r=\rho$, as it returns
\begin{equation}
\boldsymbol{\varsigma}=r^{2}\mathrm{d}\theta\otimes\mathrm{d}\theta+r^{2}\sin^{2}\left(\theta\right)\,\mathrm{d}\varphi\otimes\mathrm{d}\varphi.\label{sphere}
\end{equation}
Metrics from \eqref{spheroid} and \eqref{sphere} are exactly the metrics used in sec.~\ref{predetermined}.

	\footnotesize
	

\end{document}